\documentclass[10pt]{article}

\usepackage{graphicx}
\usepackage{graphicx,psfrag}
\usepackage{amsmath,amssymb,amsthm}
 
\def \vs {\vskip 0.2cm}

\def \n {\noindent}

\def \E {{\bf  \, E  }}

\def \P {{\bf P}}

\def \Tr {{\mathrm{\, Tr}}}

\def \bE {{\mathbb E}}

\def \bG {{\mathbb G}}
\def \bI {{\mathbb I}}

\def \bM {{\mathbb M}}

\def \bT {{\mathbb T}}

\def \bU {{\mathbb U}}

\def \bV {{\mathbb V}}

\def \bW {{\mathbb W}}

\def \sbW {\stackrel{*}{{\mathbb W}}}

\def \bY {{\mathbb Y}}

\def \CA {{\cal A}}

\def \CE {{\cal E}}
\def \CC {{\cal C}}

\def \CG {{\cal G}}
\def \CH {{\cal H}}

\def \CT {{\cal T}}

\def \CM{{\cal M}}
\def \CW {{\cal W}}
\def \CD {{\cal D}}
\def \CI {{\cal I}}

\def \CQ {{\cal Q}}
\def \CP {{\cal P}}
\def \CR {{\cal R}}

\def \CS {{\cal S}}

\def \CV {{\cal V}}

\def \D {\Delta}

\def \G{{\Gamma}}

\def \a {\alpha}

\def \b {\beta}
\def \g {\gamma}
\def \l {\lambda}
\def \r {\rho}

\def \vk {\varkappa}

\def \k {\kappa}

\def \s {\sigma}
\def \t {\tau}
\def \S {\Sigma}
\def \d {\delta}
\def \U {\Upsilon}
\def \u {\upsilon}
\def \th {\theta}

\def \Th {\Theta}

\def \vep {\varepsilon}

\def \ep {\epsilon}

\def \vp {\varphi}

\def \vsi {\varsigma}

\def \z {\zeta}

\def \Lam \Lambda

\def \S {\Sigma}

\def \o {\omega}

\def \la {\langle}
\def \ra {\rangle}

\def \bb {{\breve \beta}}

\def \bv {{\breve v}}

\def \brW {{\breve \CW}}

\def \bt {{\breve t}}

\def \fe {{\mathfrak e}}

\def \fv {{\mathfrak v}}

\def \fA {{\mathfrak A}}

\def \fB {{\mathfrak B}}

\def \fR {{\mathfrak R}}

\def \fs {{\frak s}}

\def \fS {{\frak S}}

\def \fI {{\mathfrak I}}

\def \fu {{\mathfrak u}}

\def \fa {{\mathfrak a} }

\def \fo {{\mathfrak o}}

\def\fO {{\mathfrak O}}

\def \fL {{\mathfrak L}}

\def \fM {{\mathfrak M}}

\def \fP {{\mathfrak P}}

\def \fm {{\mathfrak m}}

\def \fii {{\mathfrak i}}

\def \fF {{\mathfrak F}}

\def \rD {{\mathrm D}}

\def \rL {{\mathrm L}}

\def \RH {{\mathrm H}}

\def \RZ {{\mathrm Z}}

\def \RB {{\mathrm B}}

\def \rb {{\mathrm b}}

\def \rg {{\mathrm g}}

\def \rt {{\mathrm t}}

\def \RN {{\mathrm N}}

\def \rf {{\mathrm f}}

\def \rM {{\mathrm M}}

\def \dt {{\dot T}}

\def \ddt {{\ddot T}}

\def \dT {{\dot{ \hbox {T}}}} 

\def \ddT {{\ddot {\hbox{T}}}}

\begin{document}
\title{ On High Moments and the Spectral Norm
of Large Dilute Wigner Random Matrices\ \footnote{ The financial support of the research grant ANR-08-BLAN-0311-11 "Grandes Matrices
Al\'eatoires" (France)  is gratefully acknowledged}\ 
}

\author{O. Khorunzhiy\\ Universit\'e de Versailles - Saint-Quentin, Versailles, FRANCE\\
{\it e-mail:} oleksiy.khorunzhiy@uvsq.fr}
\maketitle
\begin{abstract}
 We consider a dilute version of the Wigner ensemble of $n\times n$ random real symmetric  \mbox{matrices}
$H^{(n,\rho )}$, where  $\rho$ denotes an average number of non-zero  {elements}
per row.
We study  the
asymptotic properties of the spectral norm $\Vert H^{(n,\r_n)}\Vert$ in the limit
 of infinite $n$ with
 \mbox{$\r_n = n^{2/3(1+\vep)}$, $\vep>0$. }
Our main result is that
the probability $\P\left\{ \Vert H^{(n,\r_n)} \Vert > 1+x n^{-2/3}\right\}$, $x>0$ is bounded
for any $\vep \in (\vep_0, 1/2]$, $\vep_0>0$
by an expression
that  does not depend on the particular values of the first several  moments \mbox{$V_{2l},  2\le l\le 6$} and 
$V_{12+2\phi_0}$,
\mbox{$\phi_0=\phi(\vep_0)$  }
of the matrix elements of $H^{(n,\rho)}$ provided they exist
and the probability distribution of the matrix elements is symmetric.
The proof is based on the study of the upper bound of the averaged moments
of random matrices with truncated random variables  $ \E\{ \Tr (\hat H^{(n,\r_n)})^{2s_n}\}$, 
$s_n = \lfloor \chi n^{2/3}\rfloor$ with $\chi>0$, in the limit $n\to\infty$.

 We also  consider the lower bound of  $\E\{ \Tr ( H^{(n,\r_n)})^{2s_n}\}$ and show that in the complementary asymptotic regime,
 when $\r_n = n^{\ep}$ with  $ \ep\in(0, 2/3]$ and  $n\to\infty$,
 the fourth moment $V_4$ enters  the estimates from below and the scaling variable $n^{-2/3}$ at the border of the
 limiting spectrum is to  be replaced by a variable related with  $\r_n^{-1}$.

\end{abstract}
\vskip 0.3cm

{\it Running title:} High Moments of Large Dilute Wigner Matrices

\section{Introduction}

The spectral theory of random matrices of high dimensions
has been started   by E. Wigner in the middle of fifties, when the 
eigenvalue distribution 
of  the ensemble $\{A^{(n)}\}$ of real symmetric 
matrices of the form
$$
\left (A^{(n)}\right)_{ij} = {1\over \sqrt n} a_{ij}, \quad 1\le i\le j\le n,
\eqno (1.1)
$$
where $\{a_{ij}, i\le j\}$ are jointly independent random variables with zero mean value and the variance $v^2$
has been studied \cite{W}. 
It was proved by E. Wigner that the normalized eigenvalue counting function of $A_n$ 
converges in the limit $n\to\infty$ to a non-random limiting distribution that has the density of the 
semicircle form.  It is common
to refer to (1.1) as to the Wigner ensemble of random matrices. The limiting eigenvalue distribution of $A_n$
is often referred as to the semicircle (or Wigner) law. 

The semicircle law 
has been then  improved and generalized in various aspects, in particular by relaxing the Wigner's conditions on the
 probability distribution
of the matrix elements $a_{ij}$
\cite{Gi,Pa},
 by the studies of different random matrix ensembles generalizing the form of (1.1)
\cite{MP},
 by the studies of  the extremal eigenvalues of $A^{(n)}$ and related
ensembles {\cite{BY,FK,G}, and others.

Later on, being motivated by  the universality conjecture
of the level repulsion in the spectra of heavy atomic nuclei
\cite{P},
a strong interest to the local properties  of the eigenvalue distribution of $A^{(n)}$
at the bulk and at the border of the limiting spectrum 
has lead to a number of powerful and deep results  (see  monographs \cite{AGZ} and  \cite{M} and references therein).
These results were mostly related with the ensembles of the form (1.1) with  the probability distribution
of $A^{(n)}$ that belong to a certain class of laws.

In the general situation of the Wigner ensemble (1.1), 
the breakthrough results in the studies of the local properties of the eigenvalue distribution
at the  border of the limiting spectrum  of the random matrices 
have been obtained  
in papers \cite{SS1,SS2} and  \cite{S}, where the eigenvalue distribution
of random matrices (1.1) has been studied for the first time on the local scale,
i.e. when the mean distance between the eigenvalues 
is of the order $n^{-2/3}$. 
These results were obtained with the help of  a deep improvement  of the moment method
initially proposed by E. Wigner. The local asymptotic regime at the border of the spectrum
is attaint in the limit $n\to\infty$  when the  order of the moments is proportional to $n^{2/3}$.
\vskip 0.1cm

One of generalizations of the Wigner ensemble (1.1) is given by an ensemble of
$n\times n$ real random matrices such that each row contains a random  number of non-zero elements
and the mean value of this number $\r_n$ is a function of  $n$.
Following the statistical mechanics terminology, where this kind of  models have been first considered,
it is natural to refer to this class of random matrices as to the sparse or dilute random matrices  \cite{MF,RB}. 
The limiting
eigenvalue distribution of  dilute random matrices or related ensembles is studied in a number of publications,
where, in particular, the semicircle law has been proved to be valid in the limit
$n,\r_n\to\infty$ \cite{KKPS,RB}.
The spectral properties at the edge of the limiting spectra has been studied in papers
\cite{K01,KKM}, however the local asymptotic regime has not been reached there.

In the present work, we consider the dilute version of Wigner random matrices 
and study its spectral properties on the  local regime at the border of the limiting spectra.
The paper is organized as follows. In Section 2, we describe the random matrix ensemble that we study 
and formulate our main theorems. The main technical result on the upper bound of the high moments of dilute 
Wigner random matrices  is also presented there; 
the general technique of  proof of this bound  is described. In Section 3, we introduce  necessary definitions
and formulate a basic principles of the estimates we obtain. Section 4 is devoted to  the proof
of our main technical result and the proof of main theorems as well.

In Sections 5 and 6, we prove
the auxiliary statements used in \mbox{Sections 3 and  4.} In Section 7, we 
obtain  the estimates from below for the high moments of dilute Wigner random matrices;
these estimates are related with certain generalizations of Catalan numbers. 
In Section 8, we discuss the results obtained. 

\section{Main results}

Let us consider a  family of real symmetric
random matrices $\{H^{(n,\r)}\}$ whose elements are determined by equality
$$
\left( H^{(n,\r)}\right)_{ij} = a_{ij}\, b_{ij}^{(n,\r)}, \quad 1\le i\le j \le n,
\eqno (2.1)
$$
where $\fA= \{a_{ij}, 1\le i\le j\}$ is an infinite family
of jointly independent identically distributed random variables and
  $\fB_n= \{ b^{(n,\r)}_{ij}, 1\le i\le j  \le n\}$
is a family of jointly independent between themselves random variables that are also independent
from $\fA$.
We denote by \mbox{$\E = \E_n$} the mathematical expectation with respect to the measure
$\P= \P_n$ generated by random variables 
$\{\fA,\fB_n\}$.

We assume that the probability distribution of random variables $a_{ij}$ is symmetric and denote their even moments by
$V_{2l}= \E (a_{ij})^{2l}, l\ge 1$ with  $V_2 = v^2 = 1/4$.

Random variables $b_{ij}^{(n,\r)}$ are proportional to the Bernoulli ones,
$$
b_{ij}^{(n,\r)}= {1\over \sqrt \r}
 \begin{cases}
  1-\delta_{ij} , & \text{with probability $\r/n$}, \\
0, & \text {with probability $1-\r/n$,}
\end{cases}
\eqno (2.2)
$$
where $\d_{ij} $ is the Kronecker $\d$-symbol.

\vskip 0.2cm

Our main result is related with the asymptotic behavior of the maximal in the
absolute value eigenvalue of $H^{(n,\r)}$,
$$
\l_{\max} ^{(n,\r)} = \Vert H^{(n,\r)}\Vert = \max_{1\le k\le n} \vert \l_k(H^{(n,\r)})\vert ,
$$
in the limit when $n$ and $\r$ tend to infinity.

 \vskip 0.2cm
{\bf Theorem 2.1.} {\it Let  the probability distribution of $a_{ij}$ be such that
the moment  $V_{12 +2\phi} = \E \vert a_{ij}\vert ^{12+2\phi}$ exists with some $\phi>0$.
If $\r_n = n^{2/3(1+\vep)}$ with any given $\vep > {3\over 6+\phi}$, then 
the  limiting probability
$$
\limsup_{n\to\infty} \P\left\{ \l_{\max} ^{(n,\r_n)} \ge \left( 1+ {x\over n^{2/3}} \right) \right\}
\le \fP(x), \quad x>0
\eqno (2.3)
$$
admits the universal bound in the sense that $\fP(x)$
 does not depend on the  values
of $V_{2l}$ with $2\le l\le 6$ and $V_{12+2\phi}$.
}

\vskip 0.1cm
Assuming more about the probability distributions of $a_{ij}$, one can relax the restriction  on $\vep$ of Theorem 2.1.
The following statement is true.

\vskip 0.2cm
{\bf Theorem 2.2.} {\it Let $\tilde a_{ij}, 1\le i\le j$ be independent identically distributed  bounded random variables,
$\vert \tilde a_{ij} \vert \le U$ such that their probability distribution is symmetric. Then the maximal eigenvalue
$\tilde \l_{\max}^{(n,\r)} = \l_{\max}( \tilde H^{(n,\r)})$ of 
the random matrix with elements $\tilde H^{(n,\r)}_{ij} = \tilde a_{ij} b^{(n,\r)}_{ij}$ 
admits the same asymptotic bound as (2.3)
$$
\limsup_{n\to\infty} \P\left\{\tilde  \l_{\max} ^{(n,\r_n)} \ge \left( 1+ {x\over n^{2/3}} \right) \right\}
\le \fP(x), \quad x>0
\eqno (2.4)
$$
in the limit $n\to\infty$, $\r_n= n^{2/3(1+\tilde \vep)}$ with any given $\tilde \vep\in (0,1/2]$.
}

\vs

{\it Remarks.}

1.  Theorems 2.1 and 2.2 remain true in the case when the random matrices
$H^{(n,\r_n)}$ (2.1) are hermitian, where  $a_{ij}$ are complex jointly independent random variables
 and 
$b_{ij}^{(n,\r_n)}$ are still determined by  (2.2).
In this case the upper bound $\fP(x)$ can be slightly diminished.
This difference  should disappear in the asymptotic regime of infinite $n$ and $\r_n$ such that  $\r_n\ll n^{2/3}$.
We  discuss  this topic   in more details  in Section 7.
\vskip 0.3 cm
2. Theorem 2.1 is in agreement with  the   statements  of work \cite{K}, where the existence of
the moment $V_{12+2\phi}$ with any positive $\phi>0$ is proved to be sufficient
for the upper bound (2.3) to hold for $\r_n = n$
when the dilute random matrices coincide with those of the Wigner ensemble (1.1).
Moreover, the proofs of theorems given in the present paper 
hold in the case of $\rho_n = n$ without any change. Thus
we obtain once more  the results of \cite{K} by the technique developed here.

\vskip 0.3cm
The proofs of  Theorems 2.1 and 2.2 are related with   the detailed study of the averaged moments
$
\rM_{2s}^{(n,\r)} = \E \rL_{2s}^{(n,\r)},
$
$$
\rL_{2s}^{(n,\r)} = 
\Tr \left( H^{(n,\r)}\right)^{2s} =  \,
\sum_{i_0, i_1, i_2,\dots , i_{2s-1} =1}^n \
H^{(n,\r)} _{i_0i_1}
H^{(n,\r)} _{i_1i_2} \cdots
H^{(n,\r)} _{i_{2s-1}i_0}
\eqno (2.5)
$$
in the limit of infinite $n,\r$ and $s$. To study the case described by Theorem 2.1,
we consider  random matrices with  truncated random variables $a_{ij}$.
This makes the proofs of Theorems 2.1 and 2.2 almost identical up to  the final stages.

\vs

Given a sequence $U_n>0, n\ge 1$, we introduce   truncated random variables
$$
\hat a_{ij}= \hat a^{(n)}_{ij} =
\begin{cases}
a_{ij}, & \text{  if $  \vert a_{ij}\vert \le U_n$}, \\
0, & \text{ otherwise.}
\end{cases}
$$
Theorems 2.1 and 2.2 will follow from our main technical result  related with
the moments  (2.5) of  random matrices $\hat H^{(n,\r)}_{ij} = \hat a_{ij}^{(n)} \, b_{ij}^{(n,\rho)}$.

\vs
{\bf Theorem 2.3}. {\it Under conditions of Theorem 2.1,
the following inequality holds,}
$$
\limsup_{n\to\infty} \E  \Tr \left( \hat H^{(n,\r_n )}\right)^{2s_n} \le \fM(\chi)<\infty ,
\eqno (2.6)
$$
{\it where $s_n = \lfloor \chi n^{2/3}\rfloor$ with $\chi>0$, $\r_n = n^{2/3(1+\vep)}$ and $U_n= n^\delta$ 
with a proper choice of $\delta>0$. The upper bound
does not depend on the values of $V_{2l}$, $2\le l\le 6$ and $V_{12+2\phi}$. }

\vs
Theorem 2.3 is proved in Section 4.  
\vs

Following the original E. Wigner's idea, it is natural to consider the right-hand side of (2.5) as the weighted sum
over paths of $2s$ steps. In particular, we can write that 
$$
\hat \rM^{(n,\r)}_{2s} = 
\E  \Tr  \left( \hat H^{(n,\r )}\right)^{2s}  = \sum_{\CI_{2s} \in \bI_{2s}(n)}
\hat \Pi_a(\CI_{2s}) \, \Pi_b(\CI_{2s}),
\eqno (2.7)
$$
where the sequence $\CI_{2s} = (i_0,i_1,\dots, i_{2s-1},i_0), i_k\in \{1,2,\dots, n\} $ is
regarded as a closed path  of $2s$ steps $(i_{t-1},i_{t})$ with the discrete time $t\in [0,2s]$. 
We will also say that $\CI_{2s}$
is a trajectory of $2s$ steps. The set of all possible trajectories of $2s$ steps over $\{1,\dots, n\}$ is denoted by $\bI_{2s}(n)$.

The weights
$
\hat \Pi_a(\CI_{2s}) $ and $\Pi_b(\CI_{2s})$ are naturally determined
as the mathematical expectations  of the products of corresponding random variables,
$$
\hat \Pi_a(\CI_{2s}) = \E \left( \hat a_{i_0i_1}\cdots \hat a_{i_{2s-1}i_0}\right), \quad
\Pi_b(\CI_{2s}) = \E \left(b_{i_0i_1}\cdots b_{i_{2s-1}i_0}\right).
\eqno (2.8)
$$
Here and below, we omit the superscripts in $b^{(n,\r)}_{ij}$ when no confusion can arise.

\vs 
In a series of papers by Ya. Sinai and A. Soshnikov \cite{SS1,SS2,S}, a powerful and deep approach 
 to study the moments $\rM_{2s}^{(n)}$ 
of Wigner random matrices 
in the limit $n,s\to\infty$
has been developed. It is based on the classification of the family  of trajectories $\bI_{2s}(n)$
according to the number of their self-intersections. Corresponding classes of equivalence 
depend also on the types of these self-intersections (simple self-intersections, simple open self-intersections, 
simple self-intersections with multiple edges). In paper \cite{KV}, this approach has been completed 
by the notions  of the instants of broken tree structure and proper and imported cells that are important in the studies 
of the trajectories $\CI_{2s}$ that have many steps with common starting point. 
The complete version of the Sinai-Soshnikov description 
has been further developed in \cite{K} to generalize the results of \cite{S} to the case of Wigner random matrices
whose elements have a finite number of moments. 

In the present paper, we propose a new
improvement of the approach  of \cite{K} that enables us to study high moments of dilute random matrices
in the asymptotic regime that describes the local properties of their  spectra at the 
border of the limiting eigenvalue distribution.
The method proposed here,  when applied to the ensemble of Wigner random matrices,  
essentially   simplifies  the technique of \cite{K}.
Our results can be regarded as the first completely rigorous proofs of the statements 
obtained
in previous works in this direction.


\section{ Even walks and classes of equivalence}

Given a trajectory
$\CI_{2s}$, we write that $\CI_{2s}(t) = i_t$, $t\in [0,\dots, 2s]$ and consider a subset
$\bU(\CI_{2s};t) = \{ \CI_{2s}(t'), \, 0\le t'\le t\}\subseteq \{1,2,\dots,n\}$.  We denote by
$ \vert \bU(\CI_{2s};t)\vert$ its cardinality.
Each $\CI_{2s}$ generates a {\it walk} $\CW_{2s}= \CW^{(\CI_{2s})}_{2s}= \{\CW(t), 0\le t\le 2s\}$ that we determine as a sequence
of $2s+1$ symbols  (or equivalently, letters)
 from an ordered  alphabet, say
$ {\cal A} = \{\a_1, \a_2, \dots\}$. The walk $\CW^{(\CI_{2s})}_{2s}$ is
 constructed with the help of the following recurrence rules \cite{KSV}:
\vskip 0.1cm
1) $\CW_{2s}(0)= \a_1$;

2) if $\CI_{2s}(t+1) \notin \bU(\CI_{2s};t)$, then $\CW_{2s}(t+1) = \a_{\vert \bU(\CI_{2s};t)\vert +1}$;

\hskip 0.44cm if there exists $t'\le t$ such that $\CI_{2s}(t+1) = \CI_{2s}(t')$, then $\CW_{2s}(t+1) = \CW_{2s}(t')$.

\vskip 0.1cm

\noindent For example,
$
\CI_{16} = (5,2,7,9,7,1,2,7,9,7,2,7,2,1,7,2,5)
$
produces the walk
$$
\CW_{16} =  (\a_1,\a_2,\a_3,\a_4,\a_3,\a_5,\a_2,\a_3,\a_4,\a_3,\a_2,\a_3,\a_2,\a_5,\a_3,\a_2,\a_1).
$$
One can say that the pair $(\CW_{2s}(t-1), \CW_{2s}(t))$ represents the $t$-th step of the  walk $\CW_{2s}$ and that
$\a_1$ represents the {\it root} of the walk $\CW_{2s}$. 

Given  two  trajectories $\CI'_{2s}$ and $ \CI''_{2s}$, we say that they  are equivalent
$\CI'_{2s}\sim \CI''_{2s}$ if $\CW^{(\CI'_{2s})}_{2s} = \CW^{(\CI''_{2s})}_{2s}$
and   denote by $\CC_\CW = \CC_{\CW_{2s}}$ the corresponding  class of equivalence.
It is clear that
$$
\vert \CC_\CW\vert = n(n-1)\cdots (n-  \vert \bU(\CI_{2s};2s)\vert+1).
\eqno (3.1)
$$

Given $\CW_{2s}$, one can draw a {\it graphical representation}  $ g(\CW_{2s}) = (\bV_g, \bE_g)$ that can be
considered as  a kind of multigraph
with the set $\bV_g$ of   vertices labelled
by $\a_1, \dots, \a_{\vert \bU(\CI_{2s};2s)\vert} $ and the set $\bE_g$ of $2s$ oriented edges
(or equivalently, arcs) labelled by $t\in [1, \dots, 2s]$;
the edge $e_t= (\a_i,\a_j)$ is present in  $\bE_g$ in the case when
 $\CW_{2s}(t-1)=\a_i$ and $ \CW_{2s}(t)= \a_j$, i.e. the $t$-th step $(\a_i,\a_j)$ is present in $\CW_{2s}$.
 To describe the properties of $g(\CW_{2s})$ in general situations, we will 
use greek letters $\a,\b,\gamma,\dots $ instead of the symbols from the ordered alphabet ${\cal A}$.
In this case the root of the walk  will be  denoted by  $\varrho$. Given vertex $\b $ such that 
$\CW_{2s}(t)=\b$, we will say that $\b$ is seen in $\CW_{2s}$ at the instant of time $t$.
By an   abuse of terminology, we refer to  $g(\CW_{2s})$ as to  the {\it graph} of the walk  $\CW_{2s}$.

 Let us  define the {\it current multiplicity} of the couple of vertices $\{\b,\g\}$, $\b,\g \in \bV_g$ up to the instant $t$ 
 by the following variable
$$
\fm_\CW^{(\{\b,\g\})}(t) = \# \{t' \in [1,t]: (\CW(t'-1),\CW(t'))= (\b,\g)\ \ {\hbox{or}} \ \  (\CW(t'-1),\CW(t'))= (\g,\b) \}
$$
and say that $\fm_\CW^{(\{\b,\g\})}(2s)$ represents the total multiplicity of the couple $\{\b,\g\}$.

The probability law of $\hat a_{ij}$ being symmetric, the weight of $\CI_{2s}$ (2.8)
is  non-zero, $\hat \Pi_a(\CI_{2s})\neq 0$
only in the case when 
 $\CI_{2s}$ is such that in the corresponding graph of the walk  $\CW^{(\CI_{2s})}_{2s}$ each couple 
 $\{\a,\b\}$ has an even multiplicity
 $\fm_\CW^{(\{\a,\b\})}(2s) = 0( \hbox{mod}\, 2)$.
 We refer to the walks of such trajectories as to  the {\it even closed  walks} \cite{SS1} and denote
 by $\bW_{2s}$ the set of all possible even closed walks of $2s$ steps. In what follows, we consider
 the even closed walks only and refer to them simply as to the walks.

\subsection{Marked steps and self-intersections}

Regarding an instant of time $t$, we say that the couple $(t-1,t)$
represents the {\it step of time} number $t$. It is natural to say that the pair $(\CW_{2s}(t-1), \CW_{2s}(t))= \fs_t$
represents the {\it step of the walk} number $t$.  
Given $\CW_{2s}\in \bW_{2s}$, we say that the instant of time $t$ is {\it marked} \cite{SS1} if the couple
$\{\a,\b\}= \{\CW_{2s}(t-1),\CW_{2s}(t)\}$
has an odd current multiplicity at the \mbox{instant $t$,}
$\fm_\CW^{(\{\a,\b\})}(t)= 1(\hbox{mod}\, 2)$. We also say that  the step of the walk $\fs_t$   and the corresponding
edge $e_t$
of $g(\CW_{2s})$
are   marked. All other steps and edges are called the {\it non-marked} ones.
Regarding  the collection of marked edges $\bar \bE_s$ of $g(\CW_{2s})$, we can consider the multigraph
$ \bar g_s = (\bar \bV_s, \bar \bE_s)$. Clearly, $\bar \bV_s = \bV_s$ and $\vert \bar \bE_s\vert=s$.
It is useful to keep the time labels of the edges $\bar \bE_s$ as they are  in  $\bE_s$.

Any even closed  walk $\CW_{2s}\in \bW_{2s}$
generates a sequence $\theta_{2s}$ of
$s$ marked and $s$ non-marked instants that can be regarded as a binary sequence of
$0$'s and $1$'s.
Such a  sequence $\theta_{2s}$ is known to encode a
Dyck path of $2s$ steps. We  denote by   $\theta_{2s}= \theta(\CW_{2s})$ the
Dyck path of $\CW_{2s}$  and 
say that $\th(\CW_{2s})$ represents the {\it Dyck structure} of $\CW_{2s}$.
\vs 
Let us denote by
 $\Theta_{2s}$ the set of all Dyck paths of $2s$ steps.
It is  known that $\Theta_{2s}$ is  in one-by-one correspondence with the set of all half-plane rooted trees
$\CT_s\in \bT_s$ constructed with the help of
$s$ edges. Sometimes we will also use the denotation ${\bf T}_{2s} = \CT_s$. 
The correspondence between 
$\Theta_{2s}$ and $\bT_s$ can be  established with the help of  the chronological run $\fR$ over the edges of $\CT_s$.
The cardinality of $\bT_s$ being given by the Catalan number that we denote by 
$$
\rt_s= {(2s)!\over s!\,  (s+1)!} ,
\eqno (3.2)
$$
we  refer to the elements of $\bT_s$ as to the {\it Catalan trees}.
We consider the edges of the tree $\CT_s$ as the oriented ones in the direction away from its root.  

Given a Catalan tree $\CT_s\in \bT_s$, one can label its vertices with the help of letters of $\CA$ according to 
$\CR_{\CT}$. The root vertex gets the label $\a_1$ and each new
vertex that has no label is labelled by the next in turn letter. 
We denote the walk obtained  by
${\stackrel{\circ}{\CW}}_{2s} [\CT_s]$ and the corresponding Dyck path $\th_{2s} = \th({\stackrel{\circ}{\CW}}_{2s})$ will be denoted also as
 $\theta_{2s} = \theta(\CT_s)$.

\vs
Given $\CW_{2s}$, we denote by $\th^*(\CW_{2s})$ the height of the corresponding Dyck path,
$$
\th^*(\CW_{2s}) = \max_{0\le t\le 2s} \th_{2s}(t), \quad \th_{2s} = \th(\CW_{2s}).
$$
This is also the height of the tree $\CT_s$, $\th^*(\CT_s) = \max_{0\le t\le 2s} \th_{2s}(t)$, $\th_{2s} = \th(\CT_s)$.

\begin{figure}[htbp]
\centerline{\includegraphics[width=12.5cm,height=3.5cm]{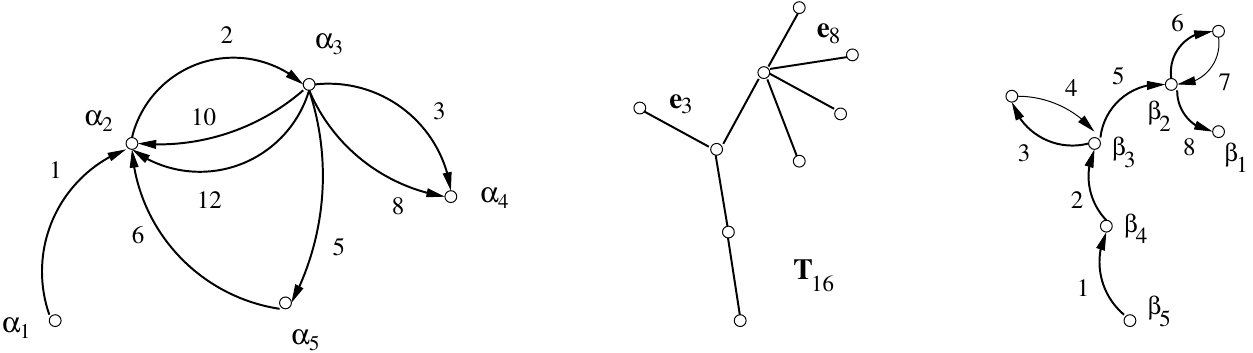}}
\caption{\footnotesize{ Graph   $\bar g(\CW_{16})$,  tree ${\bf T}_{16}= \CT_8= \CT(\CW_{16})$ and a part of the chronological run
over $\CT_8$
}}
\end{figure}

Any Dyck path $\theta_{2s}$ generates   a sequence $(\xi_1, \xi_2, \dots, \xi_s)$, $\xi_i\in [1,2s-1]$
such that each  step of the walk  $\fs_{ \xi_i}$, $1\le i\le s$  of ${\stackrel{\circ}{\CW}}_{2s}[\theta_{2s}]$ is marked.
We denote this sequence by $\Xi_s= \Xi(\th_{2s})$.  Given $\Xi_s$ and $\t\in [1,s]$,
one can uniquely reconstruct $\theta_{2s}$ and find corresponding instant of time $\xi_\t\in [1,2s-1]$.
We will  say that $\t$ represents the {\it $\t$-marked instants} or
{\it instants of marked time} that varies from $1$ to $s$; sometimes we will  simply say that
$\t$ is the {\it marked instant} when no confusion with the term "marked instant of time" can arise.

On Figure 1, we present the graph  $\bar g(\CW_{16})= (\bV_8,\bar \bE_8)$ 
of the walk $\CW_{16}$ 
as well as its Catalan tree ${\bf T_{16}} = \CT_8 = \CT(\CW_{16})$  
 and  a part of the chronological run over 
$\CT_8$ on the time interval $[0,8]= [0,\xi_6]$.
We have $\th^*(\CW_{16})=4$. The set of five vertices $\{ \b_1, \dots, \b_5\}$ represents the 
{\it descending part} of the tree $\CT_6$ that is a part of $\CT_8$.
\vs

Given a walk $\CW_{2s}$ and a letter $\b$ such that $\b\in \bV_g(\CW_{2s})$,
we say that the instant of time $t'$ such that $\CW_{2s}(t')=\b$ 
represents an {\it arrival } $\fa$ at $\b$ and that 
$t'$ 
is the {\it arrival instant of time}.
If $t'$ is marked, we will say that $\a$ is the marked arrival at $\b$.
In $\CW_{2s}$, there can be several marked arrival instants of time at $\b$ that we denote by 
$1\le t^{(\b)}_1< \cdots < t^{(\b)}_N$.
For any non-root vertex $\b$, we have $N=N_\b\ge 1$. 
The first arrival instant of time $\b$ is always the marked one.
We can say that $\b$ is created at this instant of time.
To unify the description, we assume that the
root vertex $\varrho$
is  created at the  zero instant of time $t^{(\rho)}_1=0$ and add the corresponding zero  marked instant to the list of the
marked arrival instants at $\varrho$.

If $N_\b\ge 2$, then we say that the $N$-plet $(t^{(\b)}_1,\dots, t^{(\b)}_N)$ of marked arrival instants of time represents the
 {\it self-intersection} of $\CW_{2s}$, $\b$ is the {\it vertex of self-intersection}, and this self-intersection is of the {\it degree}
 $N$ \cite{SS1}.
 We say that the self-intersection degree of $\b$ is equal to $N$ and denote this by  $\vk(\b)=N_\b$.
  Clearly, if $\b\neq \varrho$, then
the self-intersection degree $\vk(\b)$ indicates the number of marked edges of $g(\CW_{2s})$ that have  $\b$ as their tails.

If $\vk(\b)=2$, then we say that  $\b$ is  the vertex of  {\it simple self-intersection} \cite{SS1}.
If the walk is such that $\vk(\b)=2$ and 
at the second marked arrival instant
$t^{(\b)}_2$ there is at least one couple $\{\b,\g\}$ with an odd current multiplicity, 
$\fm_{\CW}^{(\{\b,\g\})} (t^{(\b)}_2-1)=1({\hbox {mod}}\, 2)$,
then $\b$ is referred to as to the vertex of {\it open simple self-intersection}
and $t^{(\b)}_2$ is the {\it instant of open simple self-intersection} \cite{SS2}.
We will also say that the vertex $\b$ is open at the instant of time $t'= t_2^{(\b)}-1$  or that 
$\b$ is a $t'$-open vertex.

\subsection{Vertices and edges of $g(\CW_{2s})$ and diagram $\CG(\CW_{2s})$}

Given a walk $\CW_{2s}$ and  an integer  $k_0\ge 1$, we
consider all vertices $\b\in \bV_g$ such that their self-intersection degree $\vk(\b)\le k_0$ and
say that they are the {\it $\mu$-vertices}. We will denote the collection of $\mu$-vertices
by $\bV_g^{(k_0,\mu)} = \bV^{(\mu)}_g$.
The vertices $\g$ with
 $\vk(\g)\ge k_0+1$ are
referred to as  the {\it $\nu$-vertices}, $\g\in \bV_g^{(\nu)}$. 

Regarding a $\nu$-vertex $\b$, we say that all oriented marked edges of $\bar  \bE_g$
of the from $(\g,\b)$ are the {\it $\nu$-edges.} We color the $\nu$-edges in black and denote 
by $\nu_k$ the number of vertices $\b$ such that $\varkappa(\b)=k$, $k\ge k_0+1$. 
We denote by $\bar \bE_g^{(\nu)}$ the collection of the $\nu$-edges and determine 
the subset $\bar \bE^{(k_0)}_g = \bar \bE_g \setminus \bar \bE^{(\nu)}_g$. 

\vs

The number of  $\mu$-vertices $\b$ such that $\vk(\b)=1$ will be denoted by $\mu_1$.
In this case, all marked edges of the form $(\g,\b)$ will be referred as the $\mu_1$-edges and colored in grey. 
\vs

Let us choose a  $\mu$-vertex $\b$ such that $\vk(\b)\ge 2$ and consider
the marked arrivals $\bar \fa_i$ at $\b$. 
Let  $e_2=(\g,\b)$ be the edge of $\bar \bE_g$ that corresponds to the second arrival $\bar \fa_2$ at $\b$. 
In our considerations, we do not assume that $\g$ is  different from $\b$.
We denote by  $\xi_{\t_2}$ the marked instant of time of $e_2$ 
and consider the the sub-walk $\CW_{[0,t']}$ of $\CW_{2s}$  with $t'=\xi_{\t_2}-1$.
We  classify the $\mu$-vertices and corresponding edges 
 according to the properties of the second and the third arrival at them.

First, let us consider   the edge $e_2=(\g,\b)$  of $\bar \bE_g$ that corresponds to the second arrival $\bar \fa_2$ at $\b$. 
We denote by  $\xi_{\t_2}$ the marked instant of time of $e_2$ 
and consider the the sub-walk $\CW_{[0,t']}$ of $\CW_{2s}$  with $t'=\xi_{\t_2}-1$.
We distinguish following properties with respect to $\bar \fa_2$:

\vs
\n {\bf (a)} the edge $e_2 = (\g,\b)$ is such that there exists a marked edge $e' =(\b,\g)$ such that 
$e'\in \bar \bE_g^{(k_0)}$ and $e'\in g(\CW_{[0,t']})$; in this case we say that 
$e_2$ is the $q$-edge;

\vs
\n {\bf (b)} if the edge $e_2$ does not verify condition (a) and there exists a marked edge
$e''=(\g,\b)$ such that $e''\in g(\CW_{[0,t']})$, then  we say that $e_2$ is the $p$-edge;

\vs 
\n {\bf (c)} if the edge $e_2$ does not verify (a) and does not verify (b) and the vertex $\b$ is $t'$-open,
then we say that $e_2$ is the $o$-edge. 

\vs 
We denote by $\bM_2'$ the collection of $\mu$-vertices $\b$ such that the second arrival $\bar \fa_2$ at $\b$
verifies one of the three conditions listed  above and denote its cardinality by 
$\mu_2'= \vert \bM_2'\vert$. 
We say that the first arrival $\fa_1$ at $\b$
represents an $f$-edge of the graph $\bar g_s$ and color it in red. 
The $o$, $p$ and $q$-edges are colored in blue color.
All  edges that correspond to the arrivals $\fa_i$, $i\ge 3$ at $\mu_2'$-vertex $\b$, if they exist,  
will be referred as the $u$-edges
and colored in green color. We denote by $u_2$ the total number of such edges of $\bar g_s$.

 \vs

If $\b$ is such that $\vk(\b)=2$ and
neither $(a)$ nor $(b)$ nor $(c)$ is verified, we say that $e_2$ is the $\mu$-edge at color it  in blue color. 
The collection  of  vertices of this kind 
will be denoted by $\bM_2''$ with the cardinality $\vert \bM_2''\vert = \mu_2''$. 

\vs 
Let us take a $\mu$-vertex $\b$ such that $\vk(\b)\ge 3$ that does not belong to $\bM_2'$. 
If the third arrival $\bar \fa_3$
verifies at least one of the two conditions,  either $(a)$ or $(b)$, 
with $\t_2$ and $e_2$  replaced by $\t_3$ and $e_3=e(\xi_{\t_3})= (\g,\b)$, respectively, 
  we say that $\b\in \bM_3'$. In this case, 
assuming the same ordering
of the  conditions (a) and (b) as before, we attribute to the marked edge $e_3$ of the third arrival at $\b$ one of two labels, either 
 $q$ or $p$. All edges that correspond to the arrivals $\bar \fa_i$, $i\ge 4$ at $\b$, if they exist, will 
be referred as the $u$-edges. We denote by $u_3''$ the total number of such edges of $\bar g_s$. The cardinality of 
$\bM_3'$ will be denoted by $\mu_3'=\vert \bM_3'\vert$. 

If $\b\in \bM_3'$, then we say that $e_3$ is the $\mu$-edge and color  it    in blue. 
We color the edges of the first and the second arrivals at $\b$
 in red and refer to them as to the $f$-edges.

\vs Finally, let us consider the vertices $\b$ such that $\vk(\b)\ge 3$ and $\b\notin \bM_3' \cup \bM_2'$.
The family of such vertices will be denoted by $\bM_3''$ with  $\vert \bM_3''\vert = \mu_3''$.
In this case, we say that all three arrival edges at $\b$ are the $\mu$-edges and color them in blue.
The edges of all subsequent arrivals at $\b$ are referred as the green  $u$-edges. The total number of such green  edges
will be denoted by $u_3'$.

\vs
\vs 
Summing up these considerations, 
we see that a given walk  $\CW_{2s}$ generates a kind of {\it graphical diagram} $\CG$ that describes the vertices of self-intersections
of $\CW_{2s}$ and the structure of the corresponding edges. More rigorously,
we define a diagram $\CG$ as a collection of vertices $\fv_i\in \CV(\CG)$ $\vert \CV(\CG)\vert = \vert \bV_g\vert$ 
and {\it half-edges} $\fe_j \in \CE(\CG)$ attached
to $\fv_i$. The half-edges have heads but have no tails. Instead of the tail, we
attach to the corresponding end of $\fe$ a circle that we refer to as  the {\it window} $\fo$. 
The windows can contain the numerical data; in this case we will say that these numbers 
represent a {\it realization} of the windows. In general, we will say that the numerical data in the windows
that comes from $\CW_{2s}$ 
represents a {\it realization  of the diagram} $\CG$ given by  $\CW$; we denote this realization by $\la\CG\ra_{\CW}$. 
In what follows, 
we will refer to the triplet $(\fv,\fe,\fo)$ either as to the {\it edge-window} or simply as to the edge of $\CG$. The edge-windows of 
$\CG$ are colored according to the colors of the corresponding edges of $g(\CW_{2s})$. 
Then we can determine the same classification of  elements of $\CV(\CG)$  as it is done for those of $\bV_g$. 

\vs 
Regarding the example walk $\CW_{16}$ of Figure 1, we see that the graph $g(\CW_{16})$
contains five vertices $\a_1,\dots, \a_ 5$. The vertices of self-intersections are represented by two of them, $\a_2, \a_4\in \bM_2'$.
 Therefore the realization of the diagram $\CG(\CW_{16})$ contains two vertices,
$\CV(\CG) = \{v_1, v_2\}$. There are four edge-windows at $v_1$ with  $\la(\fo_1,\fo_2,\fo_3,\fo_4)\ra_\CW = (1,6,10,12)$ 
and two at $v_2$, $\la(\fo_1,\fo_2)\ra_\CW = (3,8)$. We order the vertices $v_i$ according 
to the values in the blue windows of the second arrivals. 
If 
$k_0\ge 4$, then  $v_1$ has one red, one blue and two green edges attached, 
the vertex $v_2$ has one red edge and one blue edge. 

The vertices of $\CV(\CG)$ are not ordered but those of $\la \CG\ra$ are. In the general situation, 
we order the vertices of $\la \CG\ra$ according either to the instants of the second arrivals, if the vertices are from $\bM_2'\cup \bM_2''$,
or according to the instants of the third arrivals, if they are from $\bM_3'\cup \bM_3''$, or to the instants of the last arrivals,
if the corresponding vertices of $g(\CW_{2s})$ are the $\nu$-vertices.

\subsection{Classes of walks and trajectories}

Given $\th_{2s}$, an even closed walk $\CW_{2s}$  
is determined by its values at the marked and the non-marked instants of time.
The general estimation principle used in papers \cite{SS1,SS2} and \cite{S} 
is based on the observation that the knowledge of the instants of self-intersections determines all
values of $\CW_{2s}$ at the marked instants of time; this knowledge being  
added by a rule $\U$ of the non-marked passages determines completely the walk $\CW_{2s}$ (see Section 5 for the rigorous
definition of $\U$).

The vertices of self-intersections and the properties  of the corresponding edges
 are described by diagrams $\CG$. Any diagram is characterized by  the following  set of variables, 
$$
\CS= (r,p,q, \mu_2'',  u_2; \mu_3', \mu_3'', u_3, \bar \nu),
$$ 
where
 $\bar \nu = \bar \nu^{(k_0+1)} = (\nu_{k_0+1}, \dots , \nu_s)$ and $\nu_k$ denotes the number of
 vertices $v$ such that $\vk(v)=k$. Let us denote the set of such diagrams by 
 $ \bG(\CS)$. 
 The elements of $\bG(\CS)$ differ by the positions of green edges attached to the $\mu$-vertices. 
\vs 
We say that $\CW_{2s}$ belongs to the class
$\bW_{2s}(\CG)$, $\CG=\CG(\CS)$ if the graph $g_s = g(\CW_{2s})$ 
has a collection of $\mu$-vertices  $\bM_2', \bM_2'', \bM_3', \bM_3''$ with corresponding cardinalities,
where $r+p+q = \mu_2'$ and $r$ is the number of self-intersections determined by  $o$-edges,
$p$ is the number of self-intersections with $p$-edges, and $q$ is the number of self-intersections with  $q$-edges.
Then obviously, 
$$
\vert \CV(\CG)\vert   = \mu_1 + \mu_2 + \mu_3+ \sum_{k=k_0+1}^s \nu_k,
$$
where $\mu_2 =  \mu_2' +\mu_2'' $ and $\mu_3 =  \mu_3' +\mu_3''$
and 
$$
\vert \CE(\CG)\vert = \mu_1 + 2\mu_2 + 3 \mu_3 + u_2 + u_3 + \sum_{k=k_0+1} k\nu_k = s.
\eqno (3.3)
$$
Let us recall that $\vert \CV(\CG)\vert = \vert \bV_g\vert$ and $\vert \CE(\CG)\vert = \vert \bar \bE_g\vert $
and denote $
\Vert \bar \nu\Vert = \sum_{k=k_0+1}^s k\nu_k
$.

\vs

\vs
{\bf Lemma 3.1.} {\it If $\CI_{2s}$ is such that 
$\CW(\CI_{2s})\in \bW_{2s}(\CG)$ with  $\CG\in \bG(\CS)$, then
$$
\hat \Pi_a(\CI_{2s}) \, \Pi_b(\CI_{2s}) \le \ \left( {V_2\over n}\right)^{\mu_1 + \mu_2'+ r+ 2\mu_2''  + 2 \mu_3'+ 3\mu_3''}
\ \cdot \left( { U_n^2\over \rho} \right)^{p +  q + \mu_3'+ u_2 + u_3+ \Vert \bar \nu\Vert} \ .
\eqno (3.4)
$$
}

{\it Proof of Lemma 3.1.} Regarding the factor
$\hat \Pi_a(\CI_{2s})$ of (2.8) determined by  a given diagram $\CG$,  we replace by $U_n$ all random variables $\hat a_{ij}$
that correspond to the $\nu$-edges, $u$-edges, $p$-edges and $q$-edges
together with all their non-marked
counterparts. When doing this, only one case needs a special attention, namely the case
when the vertices $\g$ and $\b$ are joined by two $q$-edges of the form $(\b,\g)$ and $(\g,\b)$.
However, it is easy to consider  all possible configurations of the marked edges with heads $\g$ and $\b$
and  to show that (3.4) is valid 
for this class of diagrams.
$\Box$ 

\vs


Given a walk $\CW_{2s}$, we determine the {\it enter cluster } of $\b$ of its graph
$g(\CW_{2s})$
as the set of  all marked edges  of the form $(\a_i,\b)$,
$\Lambda(\b) = \Lambda(\b; \CW_{2s}) = \left\{ (\a_j,\b) \in \bar \bE_g\right\}$.
Similarly, we determine the
{\it exit cluster} of a vertex $\b$
as the set of  all edges  $(\b,\g_i)$,
 $
 {\D}(\b) = \D(\b; \CW_{2s}) = \left\{  (\b,\g_i) \in \bar \bE_g\right \}.
 $
 Sometimes, when no confusion can arise,  we will use the same denotations, $\Lambda$ and $\Delta$ for the 
 collections of vertices that are the heads (or tails, respectively)  of the  corresponding edges. 
 
Regarding $\CW_{2s}$, we find the exit cluster of maximal cardinality
and  say that it determines the {\it maximal exit degree} of the walk,
$$
{\CD}(\CW_{2s}) = \max_{\b \in \bV_g} \vert \D(\b;\CW_{2s})\vert.
$$
Given $\th=\th_{2s}$, let us consider a sub-class
$\bW^{[\th]}_{2s}(\rD;\CG)\subset \bW_{2s}(\CG)$
of walks $\CW_{2s}$ of Dyck structure $\th$ such that the maximal exit degree of $\CW_{2s}$ is equal to $\rD$,
$\CD(\CW_{2s}) = \rD$.
Given $\CG$, let us denote by $ \CG_\Diamond$ the collection of its blue, green and black edge-windows and
by $\CG_\circ $ the collection of its red edge-windows. 

\vs
Let  $\la \CG_\Diamond\ra_s$ be  a realization of  the corresponding edge-windows filled with the values 
from $[1,\dots, s]$. Given such a realization, we perform a run of the walk $\CW_{2s}$ with the self-intersections prescribed. 
If such walk exists, we denote   
 $\la \CG_\circ \ra= \la \CG_\circ\ra_{\CW}$ the realization of values in the red  edge-windows of $\CG$ recorded 
 during the run of $\CW_{2s}$. 

\vs


{\bf Lemma 3.2.} {\it Given a rule $\U$, denote by   $\bW^{[\th]}_{2s} {(\rD;\CG, \U)}$  the family  of walks  $\CW_{2s}$ such that
$\CW_{2s}\in \bW_{2s}^{[\th]}(\rD;\CG, \U)$, $\CG \in \bG(\CS)$.
 Then
 $$
 \vert \bW^{[\th]}_{2s} {(\rD;\CG, \U)}
\vert\  =  \ \sum_{ \la \CG_\circ\ra} \ \sum_{\la \CG_\Diamond\ra_s} \ 1,
$$
where 
$$
\sum_{  \la \CG_\Diamond\ra_s} \ 1 \le
{s^{r+p+q+u_2}\over r!\, p!\, q!} \cdot
{1\over \mu_2''!} \left( { s^2\over 2} \right)^{\mu_2''}
\cdot  
{ s^{\mu_3'+u_3}\over \mu_3'!} \cdot {1\over \mu_3''!} \left({ s^3\over 6}\right)^{\mu_3''} 
\cdot \prod_{k=k_0+1}^s {1\over \nu_k!} \left( {s^k\over k!}\right)^{\nu_k},
\eqno (3.5a)
$$ 
 and 
 $$
 \sum_{\la \CG_\circ\ra} 1 \le (2\th^*_{2s})^r\, \rD^p\, k_0^q\, \cdot (s(\rD +k_0))^{\mu_3'}
 \eqno (3.5b)
 $$
 and therefore
 $$
 \vert \bW^{[\th]}_{2s} {(\rD;\CG, \U)}
\vert\  \le \  
{(2s\th^*_{2s})^r\over r!}\cdot {(s\rD)^p\over p!} \cdot {(sk_0)^q\over q!} \cdot
{1\over \mu_2''!} \left( { s^2\over 2} \right)^{\mu_2''}\cdot s^{u_2+u_3}
$$ 
$$
\times \ 
{ (s^2(\rD+k_0))^{\mu_3'}\over \mu_3'!} \cdot {1\over \mu_3''!} \left({ s^3\over 6}\right)^{\mu_3''} 
\cdot \prod_{k=k_0+1}^s {1\over \nu_k!} \left( {s^k\over k!}\right)^{\nu_k},
$$
where $\th^*_{2s}$ is the height of the Dyck path $\th_{2s}$. }

\vs
We prove Lemma 3.2 in Section 5.
\vs
{\it Corollary of Lemma 3.2. } Let us consider the family of walks,
$$
\bW^{[\th]}_{2s} (\rD; \CS)\,  = \, \bigsqcup_{\CG \in \bG(\CS)} \ \bigsqcup_{\U\in \bY}
\bW_{2s}^{[\th]}(\rD; \CG,\U).
$$
Then 
$$
\vert \bW^{[\th]}_{2s} (\rD;\CS)\vert 
\le \  
{(6s\th^*_{2s})^r\over r!}\cdot {(3s\rD)^p\over p!} \cdot {(3sk_0)^q\over q!} \cdot
{1\over \mu_2''!} \left( { s^2\over 2} \right)^{\mu_2''}\cdot  { (8k_0^4 s\mu_2')^{u_2} \over u_2!}
$$ 
$$
\times \ 
{ (s^2(\rD+k_0))^{\mu_3'}\over \mu_3'!} 
\cdot {1\over \mu_3''!} \left({ s^3\over 6}\right)^{\mu_3''} 
\cdot { (16k_0^5 s\mu_3)^{u_3}\over u_3!} 
\cdot \prod_{k=k_0+1}^s {1\over \nu_k!} \left( {(2ks)^k\over k!}\right)^{\nu_k}.
\eqno (3.6)
$$

\vs

Let us introduce  a class of walks 
$\bW^{(u)}_{2s}(\rD; \CS)$ such that the height of their Dyck structures  is equal to $u$.
Combining the results of Lemma 3.1 and Corollary of Lemma 3.2, it is not difficult to prove the following statement.
\vs 

{\bf Lemma 3.3.} 
{\it Let us denote by $\CC(\bW^{(u)}_{2s}(\rD; \CS))$ the family of trajectories $\CI_{2s}$ such that 
their walks belong to $\bW^{(u)}_{2s}(\rD; \CS)$. 
 Then 
$$
\sum_{\CI_{2s} \in \CC(\bW_{2s}^{(u)} {(\rD;\CS)})}
\hat \Pi_a(\CI_{2s}) \ \Pi_b (\CI_{2s})\le
 V_2^{s} \ \vert \Th^{(u)}_{2s}\vert  \ \exp\left\{ - {(s-\s)^2\over 2n}\right\}
$$
$$
\times\  {1\over \mu_2''!} \left( { s^2\over 2n}\right)^{\mu_2''} 
\cdot   \RH^{(u,\rD,k_0)}_{(\CS;2)}(1) \cdot 
\RH_{(\CS;3)}^{(\rD,k_0)}(1) 
\cdot \RH_{(\CS;\bar \nu)}^{(k_0+1)}(1) \, ,
\eqno (3.7)
$$
where \ $\Th_{2s}^{(u)} = \left\{ \th_{2s}\in \Th_s, \ \th^*_{2s} = u\right\}$, \ $\s = \mu_2+\mu_3 + u_2 + u_3 + \vert \bar \nu\vert _1$ 
with 
$$
\vert \bar \nu\vert_1 = \sum_{ k=k_0+1}^s (k-1) \nu_k,
$$
and 
$$
\RH^{(u,\rD,k_0)}_{(\CS;2)}(h)=  
{1\over r!} \left( { 6h su\over n}\right)^r \cdot {1\over p!} \left( { 3h sD\hat U_n^2\over \rho}\right)^p
$$
$$
 \times \ {1\over q!} \left( { 3hsk_0\hat U_n^2\over \rho}\right)^q \cdot  
 {1\over u_2!} \left( { 8 h k_0^4 s \mu_2'  \hat U_n^2\over \rho}\right)^{u_2},
 \eqno (3.8a)
$$
$$
 \RH_{(\CS;3)}^{(\rD, k_0)}(h) = {1\over \mu_3'!} \left( { 9h (D+k_0) s^2 \hat U_n^2\over n\rho}\right)^{\mu_3'} \cdot 
{1\over \mu_3''!} \left( {3h s^3\over 2 n^2} \right)^{\mu_3''} \cdot 
{1\over u_3!} \left( { 16h k_0^5 s\mu_3 \hat U_n^2\over \rho} \right)^{u_3},
\eqno (3.8b)
$$
and 
$$
\RH_{(\CS;\bar\nu)}^{(k_0+1) }(h)=  \prod_{k=k_0+1}^s
{1\over \nu_k!} \left(  {n  (2h ks)^k \, \hat U_n^{2k}\over k!\,  \rho^k} \right)^{\nu_k}
\eqno  (3.8c)
$$
with $h\ge 1$ and $\hat U_n^2 = U_n^2/ V_2$.
}

\vs
The upper bound (3.7) represents  a natural generalization of the estimates obtained for the first time   in
papers \cite{SS1,SS2,S} and \cite{R}. Further modifications of these estimates were presented in \cite{K}. 
The form of (3.6) together with expressions (3.7), (3.8), (3.9) and (3.10)
is based on  a new description related with the form and the structure of diagrams $\CG$.
It gives 
a considerable simplification and powerful improvement of the general approach used in
\cite{K}.  The rigorous proof of Lemma 3.3 answers 
a number of  questions that arise when reading papers \cite{R,S}. 
We prove Lemma  3.3  in Section 5.

 The results of Lemma 3.3 are sufficient to get the upper bound of  the leading term of (2.5) and to show that
the contribution of those  of walks that have multiple edges and that have bounded maximal exit degree
$\CD (\CW_{2s} ) \le n^{\epsilon}$ with certain $\epsilon$ vanishes  in the limit  $n\to\infty$ (see Section 4). 
To study the family of walks such that $\CD(\CW_{2s})> n^{\epsilon}$, we need 
to consider  the properties of corresponding graphs in more details.


\subsection{Vertex of maximal exit degree}

Let us consider a walk $\CW_{2s}$ and find the first letter  $\a_i = \bb$  such that
$
 \vert \D(\bb)\vert = \max_{\b\in \CV(\CW_{2s})}\  \vert \D(\b)\vert.
$
We will refer to $\bb$  as to the {\it vertex of maximal exit degree} and
 denote $\CD(\bb)= \vert \D(\bb)\vert$.
In this section we study the properties of even closed walks related with
the vertex of maximal exit degree $\bb$. To classify the arrival edges at $\bb$, we 
need to determine
 reduction procedures related with $\bb$. These procedures are similar to those considered in  \cite{KV}.


\subsubsection{Reduction procedures and reduced sub-walks}

Any walk $\CW_{2s}$ can be considered as an ordered set 
of its steps $\fs_t$, $1\le t\le 2s$, where  $\fs_t = (\CW_{2s}(t-1),\CW_{2s}(t))$. Inversely, each pair of letters
$\a, \b$ such that $\CW_{2s}(t-1)= \a$ and $\CW_{2s}(t)=\b$ with some $t$ represents an element of the ordered set
of steps of $\CW_{2s}$  that we denote by 
$\fS = \fS(\CW_{2s})$. To each element $\fs_i\in \fS$ we attribute the label $i$
that is simply the number of step in $\CW_{2s}$. 
These labels order in natural way  the elements of $\fS$. We do not change these labels during reduction
procedures we consider below.
\vs 

 Given $\CW_{2s}$, 
let  $t' $ be the minimal instant of time such that 
\vs 
\hskip 0.4cm i) the step $\fs_{t'}$ is the marked step of $\CW_{2s}$;

\hskip 0.4cm ii) the consecutive to $\fs_{t'} $ step $\fs_{t'+1}$ is non-marked;

\hskip 0.4cm iii) $\CW_{2s} (t'-1) = \CW_{2s}(t'+1)$.
\vs
\noindent 
If such $t'$ exists, , we can apply to $\fS$
a reduction procedure $\hat \CR$ that removes from $\fS$ two consecutive elements $\fs_{t'}$ and $\fs_{t'+1}$; 
we denote $\hat \CR(\fS) = \fS'$. The ordering labels of elements of $\fS'$ are inherited from 
those of $\fS$.

\vs 

 It is clear that the ordered set $\fS'$ can be considered as a new walk
$\CW'_{2s-2}$ that is again an even closed walk. 
We denote $\CW'_{2s-2} = \hat \CR(\CW_{2s})$ and say that 
 $\hat \CR$ is the {\it strong reduction} procedure  of the walk $\CW_{2s}$.
Then we can apply $\hat \CR$ to $\CW'_{2s-2}$ and get an even closed walk
$\CW''_{2s-4} = \hat \CR(\CW_{2s-2})$. Repeating this action maximally possible number of times $m$,
 we get the walk
$$
\hat \CW_{2\hat s} = ( \hat \CR)^{m}
(\CW_{2s}), \quad \hat s = s-m,
$$
that we refer to as  the {\it strongly reduced walk.} We denote $\hat \fS = (\hat \CR)^m(\fS)$.

We introduce  a {\it weak reduction} procedure $\breve R$ of $\fS$ that removes
from   $\CW_{2s}$
the pair of consecutive steps,  $\fs_{t'}, \fs_{t'+1}$ such that the conditions (i)-(iii) are verified
and

iv) $\CW_{2s}(t')\neq \bb$.

\noindent We denote by 
$$
\brW_{2\breve s}= (\breve \CR)^{l}(\CW_{2s}), \quad  \breve s = s-l
\eqno (3.9)
$$
 the result of the action
of maximally possible number of consecutive weak reductions $\breve \CR$ and denote $\breve \fS = (\breve \CR)^l (\fS)$. 
In what follows, we sometimes omit the subscripts $2\hat s$ and $2\breve s$.

\vs

Let us consider the example walk $\CW_{16}$ (3.1). We see that  $\CD(\CW_{2s})=5$ and the vertex of maximal exit degree
is $\a_3$. The strongly reduced walk $(\hat \CR)^3(\CW_{16}) = \hat \CW_{10}$ is as follows,
$$
\hat \CW_{10} = (\a_1, \a_2, \a_3, \a_5, \a_2, \a_3, \a_2, \a_5, \a_3, \a_2, \a_1)
$$
and the corresponding reduced set of steps  $\hat \fS = \hat \fS(\hat \CW_{10})$ is
$$
\hat \fS = \{ \fs_1, \fs_2, \fs_5, \fs_6, \fs_7, \fs_{12}, \fs_{13}, \fs_{14}, \fs_{15}, \fs_{16}\}.
$$
For this example walk, we have $\brW_{10}=\hat \CW_{10}$.

 \vs
Regarding the difference $\breve \fS\setminus \hat \fS = \check \fS$, 
 we see that it represents  a collection
of sub-walks, $\check W = \cup_j \check \CW^{(j)}$.
 Each sub-walk $\check \CW^{(j)}$  can be reduced by a sequence of  the strong reduction procedures $\hat \CR$
 to an empty walk. We say that $\check \CW^{(j)}$ is of the {\it tree-type} structure,
or that $\check \CW^{(j)}$   is a tree-type sub-walk. It is easy to see that 
any $\check \CW^{(j)}$ starts by a marked step and ends by a non-marked steps and there is
no steps of $\hat W$ between these two steps of $\check \CW^{(j)}$.  
We say that $\check \CW^{(j)}$ is {\it non-split}.

\vs
It is not hard to see that the remaining part $\tilde \fS = \fS\setminus \breve \fS$ is given by a collection of 
subsets $\tilde \fS = \cup_k \tilde \fS^{(k)}$, each of $\fS^{(k)}$ represents a  non-split tree-type sub-walk
$\tilde \CW^{(k)}$, 
$$
\tilde W = \cup_k \tilde \CW^{(k)}.
\eqno (3.10)
$$
In this definition we assume that each sub-walk   $\tilde\CW^{(k)}$ is maximal by its length.

\subsubsection{Tree-type sub-walks attached to  $\bb$}

Given $\CW_{2s}$, let us consider the  instants of time $0\le t_1< t_2<\dots t_L\le 2s$
such that for all $i=1,\dots, L$ the walk  arrives at $\bb$ by the steps
of  $\brW_{2\breve s }$,
 $$ 
\CW_{2s}( t_i) = \bb \quad {\hbox{and}} \quad
\fs_{t_i}\in \brW_{2\breve s}.
\eqno (3.11)
$$
We say that  $t_i$ are  the {\it $\breve t$-arrival} instants of time of  $\CW_{2s}$.
Let us consider   a sub-walk 
that corresponds to the subset $\fS_{[t_i+1,t_{i+1}]} = \{\fs_t, t_i+1\le t\le t_{i+1}\}  \subseteq\fS$; we denote this sub-walk by 
$\CW_{[t_i, t_{i+1}]}$.  In general situation, we also denote by $\CW_{[t',t'']}$ a sub-walk that is not necessary even and/or closed.

\vs Let us consider  the interval of time $[t_i+1, t_{i+1}-1]$ between two consecutive $\breve t$-arrivals at $\bb$.
It can happen that $\CW_{2s}$ arrives at $\bb$ at some instants of time $t' \in [t_i+1,t_{i+1}-1]$, $\CW_{2s}(t') = \bb$.   
We denote by  $\tilde  t_{(i)}$ the maximal value of such $t'$. 

\vs
{\bf Lemma 3.4.} {\it The sub-walk $\CW_{[t_i, \tilde t_{(i)}]}$ is of the tree structure and coincides 
with one of the maximal tree-type
sub-walks $\tilde \CW^{(k')}$ of (3.10). }

\vs
This  statement means   that the walk $\CW_{2s}$ is such that after an arrival at $\bb$ by a step
of $\brW$, it creates a tree-type sub-walk $\CW^{(k')}$  that is not interrupted by the steps of $\brW$,
and that all   steps performed after $\CW^{(k')}$  
 belong again to $\breve \CW$,
 $$
\{ \fs_{t}, \ t_i+1\le t\le \tilde t_{(i)}\} \subseteq \tilde \fS, \quad  \{\fs_t, \ \tilde t_{(i)}+1\le t\le t_{i+1}\} \subseteq \breve \fS.
 $$

\vs {\it Proof of Lemma 3.4}. 
 It is clear that the step $\fs_{ \tilde t_{(i)} }$ does not belong to $\breve \fS$ and is 
non-marked. Then this step makes a part of a non-split tree-type sub-walk 
$\tilde \CW^{(k')} = \CW_{[t'', \tilde t_{(i)}]}$
such that $\tilde \CW^{(k')}(t'')=\bb$. Then the previous step $\fs_{t''}\in \breve \fS$ is such that
$\CW_{2s} (t'')=\bb$. Then $t'' = t_i$. Lemma is proved.

\vs
\vs

As a consequence of Lemma 3.4, we see that the sub-walk  $\CW_{[t_i, \tilde t_{(i)}]}= \tilde \CW^{(k')}$
is of the tree-type structure that ends at $\bb$ by a non-marked step
and therefore starts at $\bb$ by the marked step. Let us consider
the family of all marked exits edges from $\bb$ performed  by the marked steps on the interval
of time 
$[t_i, \tilde t_{(i)}]$ and denote these edges of 
$\bar \bE = \bar \bE(\CW_{2s})$ by $\tilde \D_i$.
Regarding the non-empty subsets $\tilde \D_j$, we
say that $\tilde \D_j$ represents the {\it exit sub-clusters of tree type} attached to  $\bb$ and denote
by $d_j = \vert \tilde \D_j\vert$ its  cardinality, $d_j\ge 1$, $j=1,\dots, L'$.
These tree-type sub-clusters are numerated in natural chronological order.
Clearly, any non-empty tree-type sub-cluster is attributed to  a uniquely determined $\bt$-arrival instant at $\bb$.

\vs
Regarding the even walk $\brW_{2\breve s}$, we can determine the corresponding
Dyck path $\breve \theta_{2\breve s} = \theta (\brW_{2\breve s})$ and the tree
 $\breve \CT_{\breve s}
= \CT(\breve  \theta)$.
It is clear that $\breve  \CT_{\breve s}$
can be considered as a subtree of  $\CT_s = \CT( \theta(\CW_{2s}))$. One can determine the difference
$\tilde \CT = \CT_s \diagdown
 \breve \CT_{\breve s}$ such that it  is represented by a collection of
sub-trees $\tilde \CT^{(j)}$. Not to overload the paper, we do not present rigorous definitions here.

\vs 
Regarding the Catalan tree $\CT(\th_{2s})$,  we  say that the chronological run
$\fR_{\CT}$  is represented by $2s$ directed arcs $\varpi_i$, $1\le i\le 2s$
drawn over $\CT_s$ (see Figure 1).
This chronological run uniquely determines $L$ arcs $\breve  \varpi_l$, $1\le l\le L$ that 
 correspond
to the arrival instants of time $\breve t_l$ at $\bb$.  Also the corresponding vertices $\u_l$ of the tree $\CT_s$ are determined.
It is clear  that $\u_l$ are not necessarily different for different $l$. 

The sub-trees $\tilde \CT^{(l)}$ are attached at  $\u_l$
and the chronological run over $\tilde \CT^{(l)}$ starts immediately after
the arc $\breve \varpi_l$ is drawn.
We will say that these arcs $\breve \varpi_l$ represent   
the {\it nest cells} from where  the sub-trees
$\tilde \CT^{(l)}$, $1\le l\le L$ grow. It is clear that the sub-tree $\CT_l$ has
$d_l\ge 0$ edges attached to its root $\rho_l$ represented by  the vertex $\u_l$.
Returning to $\CW_{2s}$, we will say that the arrival instants of time $\breve t_l$
represent the {\it arrival cells} at $\bb$.
In the next sub-section, we describe a classification of the arrival cells at $\bb$ 
that represents a natural improvement of the approach proposed in \cite{KV}.

\subsubsection{Classification of arrival cells at $\bb$ and BTS-instants}

Let us consider a $\bt$-arrival cell $t_i$ (3.11). 
If the step  $\fs_{t_i}$ of $\CW_{2s}$ is marked,
then we say that $t_i$ represents  a {\it  proper cell} at $\bb$.
If the step  $\fs_{t_i}$ is non-marked and  $\fs_{t_i}\in \check \CW= \breve \CW\setminus \hat \CW$,
then we say that $t_i$ represents a {\it mirror cell} at $\bb$.
If  the step $\fs_{t_i}\in \hat \CW$ is non-marked,
then we say that $t_i$ represents an {\it imported cell } at $\bb$.

\vs
Let  us consider $I$ proper cells  $\check t_i$ 
such that the corresponding step $\fs_{\check t_i}$ belongs to $\check \fS$. We denote by 
$ x_i$ corresponding to $\check t_i$ marked instants, $ x_i = \xi_{\check t_i}$,
$1\le i\le  I$ and write that $\bar x_I = (x_1, \dots, x_I)$.
 Each proper cell $x_i$ is associated with  a number  of corresponding mirror cells.
 We denote this number by $m_i\ge 0$ and write that $M= \sum_{i=1}^{I} m_i$ and 
 $\bar m_I = (m_1, \dots, m_I)$.
 \vs

Regarding the strongly reduced walk $\hat \CW_{2\hat s}$, we denote by 
 $\hat t_k$ the proper cells such that the steps $\fs_{\hat t_k} \in \hat \fS$. Corresponding
 to $\hat t_k$ marked instants will be denoted by $z_k$, $1\le k\le  K$. Then 
 $\bar z_K = (z_1,\dots, z_K)$ and 
 clearly $\vk(\bb) = I+K$.

 \vs

\vs

Regarding any walk $\CW_{2s}$, we observe that  
if the set  $\hat \fS$ is non-empty, then there exists at least one pair of elements of $\hat \fS$,
$(\fs', \fs'')$ such that $\fs'$ is a marked step of $\hat \CW_{2\hat s}$, $\fs''$ is the non-marked one 
and $\fs''$ follows immediately after $\fs'$ in $\hat \fS$. We refer to each pair of this kind
as to  the pair of {\it broken tree structure} of $\CW_{2s}$ or in abbreviated form, 
 the BTS-pair of $\CW_{2s}$. 
If $\tau' $ is the marked instant that corresponds to $\fs'$, 
we will simply say that $\tau'$ is the {\it BTS-instant}  of $\CW_{2s}$ \cite{KV}. We will refer to the 
vertex $\g = \CW_{2s} (\xi_{\t'})$ as to the {\it BTS-vertex} of the walk.

\vs

Regarding the strongly reduced walk $\hat \CW$, let us  consider 
   a non-marked arrival step at $\bb$ that we denote by $\bar \fs = \fs_ {\bar t}$.
   Then one can uniquely determine the marked instant 
    $\t'$ such that all steps  $ \fs_t\in \hat \fS$
with $\xi_{\t'}+1\le t \le \bar t$ are the non-marked ones.
Let us denote by $ t''$ the instant of time of the first non-marked step $\fs_{\bar t''} \in \hat \fS$
of this series of non-marked steps. 
Then $(\fs_{t'}, \fs_{ t''})$ with  $t' = {\xi_{\t'}}$  is the BTS-pair of $\CW_{2s}$ that corresponds to $\bar t$.
We will say that $\bar t$ is attributed to  the corresponding BTS-instant $\t'$. 
It can happen that several arrival instants $\bar t_i$ are attributed to  the same BTS-instant $\t'$.
We will also say  that the BTS-instant $\t'$ {\it generates}  the imported cells that are attributed to it.

\vs  

Let us consider a BTS-instant $\t$   such that 
$\CW_{2s}(\xi_{\t}) =  \bb$. As it is said above, we  denote such marked instants by $z_k$, $1\le k\le K$. 
Assuming that a marked BTS-instant $z_k$ generates $f_k'\ge 0$ imported cells, we denote
by $\vp^{(k)}_1, \dots, \vp^{(k)}_{f'_k}$ the positive numbers   such that 
$$
\CW_{2s}( \xi_{z_k} + \sum_{j=1}^l \vp^{(k)}_j) = \bb
\quad {\hbox{for all}} \quad 1\le l\le f'_k.
\eqno (3.12)
$$ 
If for some $\tilde k$ we have $f'_{\tilde k}=0$,
then we will say that $z_{\tilde k}$ does not generate any imported cell at $\bb$.
We denote $\bar \vp^{(k)} = (\vp^{(k)}_1, \dots, \vp^{(k)}_{f'_k})$. 
\vs 

Let us consider a BTS-instant $\t$ that generates imported cells at $\bb$ and such that 
$\CW_{2s} (\xi_\t)\neq \bb$. We denote such BTS-instants by $y_j$, $1\le j\le J$. 
Assuming that a marked BTS-instant $y_j$ generates $f''_j+1$ imported cells, $f_j''\ge 0$,
we denote
by $\ell_j, \psi^{(j)}_1, \dots, \psi^{(j)}_{f''_j}$ the positive numbers   such that 
$\CW_{2s} (\xi_{y_j} + \ell_j) = \bb$ and 
$$
\CW_{2s}\left( \xi_{y_j} +\ell_j+  \sum_{l=1}^k \psi^{(j)}_l\right) = \bb \quad {\hbox{for all}} \quad 1\le k\le f''_j.
\eqno (3.13)
$$ 
In this case we will say that the first arrival at $\bb$ given by the instant of time
$\xi_{y_j} + \ell_j$ represents the {\it principal} imported cell at $\bb$.
All subsequent arrivals at $\bb$ given by (3.13) represent the {\it secondary} imported cells at $\bb$. 
We will say that $y_j$ is the {\it remote} BTS-instant with respect to $\bb$ and will use denotations  
$\bar y_J= (y_1, \dots, y_J)$ and $\bar \ell_J = (\ell_1, \dots, \ell_J)$.
We also denote 
$\bar \psi^{(j)}= (\psi^{(j)}_1, \dots, \psi^{(j)}_{f''_j})$. 

All arrivals determined  by (3.12) will be also referred to as the secondary imported cells at $\bb$. 
We will say that corresponding BTS-instant $z_k$ is the {\it local} one with respect to $\bb$. 
\vs

We see that for a given walk $\CW_{2s}$, the proper, mirror and imported cells 
at its vertex of maximal exit degree
are characterized by  a set of parameters, $(\bar x, \bar m)_{ N}, $, 
$(\bar z, \Phi,\bar f')_{K}$, where 
$\Phi_K= (\bar \vp^{(1)}, \dots \bar \vp^{(K)})$, $\bar f'_{K}  = (f'_1, \dots, f_{K}')$ and 
$(\bar y, \bar \ell,  \Psi, \bar f'')_J$, where 
$\Psi_J= (\bar \psi^{(1)},\dots, \bar \psi^{(J)})$, 
$\bar f''_J = (f''_1, \dots, f''_J)$. We also denote
$
F' = \sum_{k=1}^{K} f'_k $ and  $
F'' = \sum_{j=1}^{J} f''_j
$.

\vs
Summing up, we observe that 
the  vertex  $\bb$ with  $\varkappa(\bb) = I+K$
has the  total number of cells  given by $R = I+M+K+ 2J+F$,  where 
$F=F'+F''$. In what follows, we will use the following denotation for the set of parameters described above,
$\left\{ (\bar x,\bar m)_I, (\bar z, \Phi,\bar f')_{K}, (\bar y, \bar \ell,  \Psi, \bar f'')_J\right\}= \langle\CP_R\rangle$.
It would be an instructive exercise to consider the example walk $\CW_{16}$ of Figure 1 and to determine
its  $\la\CP_R\ra$.





\section{Proof of main results}

Remembering that  $\vep > {\displaystyle 3\over \displaystyle 6+\phi}= \vep_0$, let us choose
$$
k_0= \lfloor {3 \over \vep - \vep_0} \rfloor +2  \  , \ \d = {\vep_0+\vep \over 6}\ , \  {\hbox{and   }}\ U_n= n^\d.
\eqno (4.1)
$$
We assume $k_0$ to be an even number.

\vs 

Our aim is to show that the averaged trace ${\hat {\mathrm L}}_{2s_n}^{(n,\rho)}$ (2.7) admits an upper bound
in the limit $n, s_n\to\infty $, $s_n = \lfloor\chi n^{2/3}\rfloor, \ \chi >0$ that we denote by $(n,s)_\chi\to\infty$. 
We also prove that 
 the trajectories $\CI_{2s}$ such that the graphs of their walks have multiple edges vanish in this limit.
Then Theorem 2.3 will follow. 

\vs 
In the spirit of \cite{R,S}, we consider the following   partition of the sum (2.7),
$$
\hat \rM_{2s}^{(n,\rho)} = \E  \Tr \left( \hat H^{(n,\r_n )}\right)^{2s_n}= Z_{2s}^{(1)} + Z_{2s}^{(2)} + Z_{2s}^{(3)},
\eqno (4.2)
$$
where 

- $Z_{2s}^{(1)}$ is the sum over the trajectories $\CI_{2s}\in \CC(\CW_{2s})$
such that the  graphs $g(\CW_{2s})$ have neither  $p$-edges nor $\mu_3'$-vertices;

- $Z_{2s}^{(2)}$ is the sum over the trajectories $\CI_{2s}\in \CC(\CW_{2s})$
such that the  graph $g(\CW_{2s})$ have at least  one $p$-edge or one $\mu_3'$-vertex
and the maximal exit degree  $g(\CW_{2s})$
is bounded, $\CD(\CW_{2s})\le n^{\ep}$ and 

-   $Z_{2s}^{(3)}$ is the sum over the trajectories $\CI_{2s}$
such that the  graph $g(\CW_{2s})$  has   the maximal exit degree
 $\CD(\CW_{2s})> n^{\ep}$.

As we will see, the choice of  $\ep = {( \vep -\vep_0)/ 12}$ will be  sufficient for our purposes.
The sub-sum $Z^{(1)}_{2s}$ will be also represented as a sum of  two parts in dependence of the
presence of $q$-edges, or $u$-edges, or $\nu$-vertices.


\subsection{Estimate of $Z^{(1)}_{2s}$}

Following the definitions of Section 3, we can write that
$$
Z^{(1)}_{2s}  \ = \ \sum_{u=1}^s\ \sum_{\la \CS^{(1)}\ra } \ \sum_{ \CG \in \bG(\la \CS^{(1)}\ra )}\
\sum_{\la \CG_\Diamond\ra_s}\ \ \sum_{\CW_{2s} \in \bW_{2s}^{(u)}\left(\la \CG_\Diamond\ra_s\right)}\  \ 
\sum _{ \CI_{2s} \in \CC ( \CW_{2s} )} \ \hat \Pi_a(\CI_{2s})\cdot \Pi_b(\CI_{2s}),
\eqno (4.3)
$$
where $\CS^{(1)}= (r, 0, q, \mu_2'', u_2; 0, \mu_3'', u_3; \bar \nu)$ represents a particular case of the set of variables $\CS$ (see sub-section 3.3)
and the sum over $\la \CS^{(1)}\ra$ runs over all non-negative integer values of its parameters 
such that the condition (3.3) is verified. 
Let us represent   $Z^{(1)}_{2s}$ as   a sum of two terms,
$$
Z^{(1)}_{2s} = Z^{(1,1)}_{2s} + Z^{(1,2)}_{2s},
$$
where $Z^{(1,1)}_{2s}$ is the sum over the realizations of $ \CS^{(1,1)} = (r,0,0,\mu_2'',0;0,\mu_3'', 0;0)$. Then the
corresponding walks do not have neither $q$-edges, nor $u$-edges, nor $\nu$-vertices. 

\vs 
To estimate the right-hand side of (4.3) from above, we use Lemma 3.3 of \mbox{Section 3} 
with $p=q=u_2=\mu_3'=u_3=\vert \bar \nu\vert_1= 0$. Then $\s = r+\mu_2''+ \mu_3''$ and we can write that 
 $$
 \exp\left\{ - {(s-\s)^2\over 2n}\right\} \le \exp\left\{ - {s^2\over 2n}\right\}\cdot 
 \left( e^{{s\over n}} \right)^{ \mu_2'' + r + \mu''_3 }. 
 \eqno (4.4)
 $$
 Applying the upper bound (3.7) to the right-hand side of (4.3), 
 replacing the sum over $\la \CS^{(1)}\ra$ by the sum over all possible values of  variables $r$, $\mu_2''$,  and $\mu_3''$
 and using (4.4), 
we can write the following inequality,
$$
Z^{(1;1)}_{2s} \le 
n  V_2^s \, \exp\left\{ - { s^2\over 2n} \right\} \  \sum_{u=1}^s \, \vert  \Th_{2s}^{(u)} \vert 
\ \sum_{r=0}^s \ {1\over r!} \left( { 6 su\over n} e^{s/n} \right)^r 
 $$
$$
\times \ \sum_{ \mu_2'' = 0}^s\ 
{1\over \mu''_2!}
\left( { s^2\over 2n} e^{s/n} \right)^{\mu''_2}
\cdot  \sum_{\mu''_3=0}^s \  {1\over \mu''_3!} \left( { 3 s^3\over 2 n^2} e^{s/n}\right)^{\mu''_3},
$$
Taking into account that $e^{s/n}\le 2$ for large values of $n$, we get the bound 
$$
Z^{(1,1)}_{2s} \le
{n V^s_2\,  \rt_s} \,\RB_s(12 \chi^{3/2})\,
\cdot  \exp\left\{{s^2\over 2n}\left(e^{s/n}-1\right) + {3 \chi ^3}   \right\},
$$
where we denoted 
$$
\RB_s(x)\, = {1\over \rt_s} \,  \sum_{u=1}^s \vert \Theta^{(u)}_{2s}\vert \exp\left\{  {xu\over \sqrt s}\right\}
\ = \ 
{\bf E}_{s} \left( \exp\{ x \th^*/\sqrt s\}\right).
\eqno (4.5)
$$
In (4.5),  
 ${\bf E}_s(\cdot)$ denotes the mathematical expectation with respect to  the uniform measure on the set of Catalan trees $\bT_s$.  
It is proved in \cite{KM} that
  $\RB_s(x)$ in the limit of infinite $s$   is given by  the exponential moment of the maximum of 
the normalized Brownian excursion.
Using  convergence $\RB(x) = \lim_{s\to\infty} \RB_s(x)$ \cite{KM}  and elementary relation 
$$
{n \rt_{s_n} \over 4^{s_n}} =  {n\, (2s_n)!\over 4^{s_n} \, s_n! \, (s_n +1)!} = {1\over \sqrt {\pi \chi^3}} (1+ o(1)), \quad s_n = \chi n^{2/3},
$$
that follows from the Stirling formula, we conclude that 
$$
\limsup_{n\to\infty} \RZ^{(1,1)}_{2s_n} \le {1\over \sqrt {\pi \chi^3}} \ \RB(12\chi^{3/2}) \ e^{4\chi^3}.
$$

Let us consider the sub-sum $Z^{(1,2)}_{2s}$. 
 Applying the upper bound (3.7) to the right-hand side of (4.3), using  the analog of (4.4)
 and replacing the sum over $\la \CS^{(1,2)}\ra$ by the sum over all possible values of  its variables,
we can write the following inequality,
$$
Z^{(1,2)}_{2s} \le
n V_2^s \, \sum_{u=1}^s \, \vert  \Th_{2s}^{(u)} \vert  \sum_{r=0}^s \ {1\over r!} \left( { 12 su\over n} \right)^r 
\  \sum_{ \mu_2'' = 0}^s\ 
\exp\left\{ - { s^2\over 2n} \right\} \ {1\over \mu''_2!}
\left( { s^2\over 2n} e^{s/n} \right)^{\mu''_2}
$$
$$
\times\  \sum_{\mu''_3=0}^s \  {1\over \mu''_3!} \left( { 3 s^3\over  n^2} \right)^{\mu''_3} \cdot 
\ \ \sum_{u_2+u_3+ \vert \bar \nu \vert_1 \ge 1}  \  
{1\over u_2!} \left( { 16k_0^4 sr  \hat U_n^2\over \rho}\right)^{u_2} 
$$
$$
\times \
 \ {1\over u_3!} \left( { 32k_0^5s \mu''_3 \hat U_n^2\over \rho} \right)^{u_3}
\cdot  \   \ \prod_{k=k_0+ 1}^s  \,
{1\over \nu_k!} \left(   n \left(  {C_1 s \hat U_n^{2}\over  \rho}\right)^{k}\,  \right)^{\nu_k},
\eqno (4.6)
$$
where we denoted $C_1 =\, \sup_{k\ge 2}\,  { 2k\over  (k!)^{1/k}}$ and  used relation
 $e^{s/n}\le 2$.

Regarding the last product of (4.6), we can write that 
$$
n \left(  {C_1 s \hat U_n^{2}\over  \rho}\right)^{k} = 
 A_n^{(k_0)}  \left(  {C_1 s \hat U_n^{2}\over  \rho}\right)^{k-k_0},
\quad 
 A_n^{(k_0)}  =
 n \left( {C_1 s  \hat U_n^{2} \over  \rho}\right)^{k_0}.
  $$ 
It follows from (4.1) that 
$
A_n^{(k_0)} \le  
\left( {C_1\chi \over  V_2}\right)^{k_0} \, n^{-2\ep}
$ with \mbox{$\ep=(\vep-\vep_0)/12$}
and therefore
$$
n \left(  {C_1 s \hat U_n^{2}\over  \rho}\right)^{k} \le 
\left( {C_1\chi \over  V_2}\right)^{k_0} \, n^{-2\ep} \cdot  \left( { C_1\chi\over V_2}n^{-4\ep}\right)^{k-k_0}\ \, 
\eqno (4.7)
$$
where we have taken into account that $n^{2/3} U_n^2/\rho \le n^{-4\ep}$. 
Then the following relation is true,
$$
\sum_{\vert \bar \nu \vert_1 = \jmath_1}^s 
 \ \prod_{k'=1}^s  \,
{1\over \nu_{k_0+k'}!} \left( A_n^{(k_0)}\left(  {C_1 s \hat U_n^{2}\over  \rho}\right)^{k'}\,  \right)^{\nu_{k_0+k'}} 
$$
$$
\le
\exp\left\{ \left( { C_1\chi\over V_2}\right)^{k_0} n^{-2\ep} 
\sum_{k'=1}^\infty \left( { C_1\chi\over V_2}n^{-4\ep}\right)^{k'} \right\}
-\jmath_1
\le 
\exp\left\{ \left({ C_1\chi\over V_2}\right)^{k_0} n^{-2\ep}  \right\}
-\jmath_1,
$$
where $\jmath_1$ takes values $0$ or $1$.
Denoting 
$C_2 = 64k_0^5 \chi/ V_2$, we can also write that 
$$
 \sum_{u_3=\jmath_2}^s \ {1\over u_3!} \left( { 32k_0^5s \mu''_3 \hat U_n^2\over \rho} \right)^{u_3}
 \le 
  \begin{cases}
 \exp\{ C_2\,  \mu_3''\,  n^{-4\ep}\} , & \text{if  $\jmath_2 =0$}, \\
C_2 \, n^{-4\ep}\,  e^{\mu''_3}, & \text {if $\jmath_2=1$}
\end{cases}
$$
for sufficiently large values of $n$.
Similar computation shows that
$$
\sum_{u_2=\jmath_3}^s {1\over u_2!} \left( { 16k_0^4 sr  \hat U_n^2\over \rho}\right)^{u_2} \le 
  \begin{cases}
 \exp\{ C_2\,  r\,  n^{-4\ep}\} , & \text{if  $\jmath_3 =0$}, \\
C_2 \, n^{-4\ep}\,  e^{r}, & \text {if $\jmath_3=1$.}
\end{cases}
$$

\vs 
Using elementary inequality
$$
\sum_{\jmath_1 + \jmath_2 +\jmath_3\ge 1}\left\{ \cdot\right\}
\le 
\sum_{\jmath_1\ge 1,  \jmath_2\ge 0, \jmath_3\ge 0}\left\{ \cdot\right\}
+\sum_{\jmath_1\ge 0,  \jmath_2 \ge 1, \jmath_3\ge 0}\left\{ \cdot\right\}
+\sum_{\jmath_1\ge 0, \jmath_2 \ge 0, \jmath_3\ge 1}\left\{ \cdot\right\}
$$
and  accepting that $n$ is such that $\exp\{s/n + C_2 n^{-4\ep}\}\le 2$, we see that 
relation (4.6) implies the following  inequality,
$$
Z^{(1,2)}_{2s} \le
n  V_2^s \, \sum_{u=1}^s \vert \Theta^{(u)}_{2s}\vert \exp\left\{ 12e\chi^{3/2} {u\over \sqrt s}\right\}
$$
$$
\times
\exp\left\{{s^2\over 2n}\left(e^{s/n}-1\right) + 3e \chi ^3  + 2C_3 n^{-2\ep} \right\} \left( 2C_2n^{-4\ep} +  2C_3 n^{-2\ep} \right) ,
$$
where $C_3 = 2C_1 \chi/V_2$. 
Now it is clear that 
$
Z^{(1,2)}_{2s} = o(1), \quad (n,s)_\chi\to\infty
$
and therefore 
$$
\limsup_{(n,s)_\chi \to\infty} \, Z^{(1)}_{2s} 
\le {1\over \sqrt {\pi \chi^3}} \ \RB(12\chi^{3/2}) \ e^{4\chi^3}.
\eqno (4.8)
$$

\vs 

\subsection{Estimate of $Z^{(2)}_{2s}$}

\vs 
Rewriting relation (4.3) with $\CS^{(1)}$ replaced by $\CS^{(2)} = ( r,p,q,\mu_2'',u_2; \mu_3', \mu_3'', u_3; \bar \nu)$
and using the result of Lemma 3.3 together with  (4.4),
we can write that   
$$
Z^{(2)}_{2s} \le
n  V_2^s\, \sum_{\rD=1}^{n^\ep}\   \sum_{u=1}^s \, \vert  \Th_{2s}^{(u)} \vert 
\  \sum_{ \mu_2'' = 0}^s\ 
\exp\left\{ - { s^2\over 2n} \right\} \ {1\over \mu''_2!}
\left( { s^2\over 2n} e^{s/n} \right)^{\mu''_2} \cdot  \sum_{r=0}^s \ {1\over r!} \left( { 12 su\over n} \right)^r 
$$
$$
\times\  \sum_{ \stackrel{p,\mu_3':  }
{p+\mu_3'  \ge 1 }}
\ {1\over p!} \left( { 6s\rD\hat U_n^2\over \rho}\right)^p
 \cdot  {1\over \mu_3'!} \left( { 9(\rD+k_0) s^2 \hat U_n^2\over n\rho}\right)^{\mu_3'}
\ 
\cdot \sum_{u_2=0}^s {1\over u_2!} \left( { 32k_0^4 sr  \hat U_n^2\over \rho}\right)^{u_2} 
$$
$$
\times \ \sum_{q=0}^s \,   {1\over q!} \left( { 6sk_0\hat U_n^2\over \rho}\right)^q 
 \cdot 
  \sum_{\mu''_3=0}^s \  {1\over \mu_3!} \left( { 3 s^3\over n^2} \right)^{\mu''_3}
\cdot  \ \sum_{u_3=0}^s\,   {1\over u_3!} \left( { 64k_0^5s \mu_3 \hat U_n^2\over \rho} \right)^{u_3}
$$
$$
\times  \ \sum_{\vert \bar \nu\vert\ge 0}  \ \prod_{k'=1}^s  \,
{1\over \nu_k!} \left( A_n^{(k_0)}\left(  {C_1 s \hat U_n^{2}\over  \rho}\right)^{k'}\,  \right)^{\nu_k},
\eqno (4.9)
$$

Taking into account that 
$$
{s \rD U_n^2\over  V_2 \rho} \le {\chi\over V_2} n^{-3\epsilon} \quad {\hbox{and}} \quad
{ \rD s^2 U_n^2\over  V_2 n\rho} \le {\chi^2\over V_2} n^{-3\epsilon} \cdot 6\chi^2 n^{-1/3} 
\le {\chi^2\over V_2} n^{-3\epsilon},
\eqno (4.10)
$$
and repeating computations of the previous subsection, we 
deduce from (4.9) the following estimate,
$$
Z^{(2)}_{2s} \ \le \  {48\chi\over V_2} n^{-2\epsilon} \ 
n  V^s_2 \, \sum_{u=1}^s \vert \Theta^{(u)}_{2s}\vert
 \exp\left\{ 24\chi^{3/2} {u\over \sqrt s}+ 8 \chi ^3  + 2A_n^{(k_0)} \right\}.
$$
Then 
$$
Z^{(2)}_{2s} = o(1), \quad \hbox{ as} \quad (n,s)_\chi \to\infty.
\eqno (4.11)
$$




\subsection{Estimate of $Z^{(3)}$}

In this subsection we estimate the sub-sum of (2.5) that corresponds to the  walks that have 
a vertex $\bb$ of 
large exit degree $\rD$. 
In previous works (see e.g. \cite{K01,K,S}), it is observed that  in the case of $L$
arrival cells at $\bb$, the underlying Dyck path and the corresponding tree $\CT_s$
has to have at least one vertex $\u$ whose exit degree is not less than $D/L$. Then the collection
of such trees has an  exponentially small cardinality with respect to the
total number of trees,  with the exponential factor determined by  a value proportional to $ D/L$.

In the case of dilute random matrices, this observation is not sufficient to get the needed estimates of
the cardinality of the set of   such walks. Roughly speaking,
our aim is to get the upper bound related with the exponential factor determined by values proportional to $D$.
Let us briefly outline the main idea of the proof of corresponding bounds. 

We consider the families of walks such that their vertex of maximal exit degree
$\bb$ is characterized by a collection of certain parameters $\CP$. This set of parameters 
described in Section 3 involves the 
marked BTS-instants $\t_i$ and corresponding descending lengths $\ell, \varphi, \psi$. 
Given $\CP$,
 the positions of the  nest cells in the tree $\CT_s$ with given exit sub-clusters are uniquely determined.
This implies fairly strong exponential estimate for the 
number of corresponding trees (see D-lemma of Section 6).
From another hand, it appears that the collection of parameters $\CP$ can be  naturally inserted into  
the diagrams $\CG$ that we use to estimate the number of walks in the corresponding classes.

\subsubsection{Diagrams and classes of walks}

Given $u$, $\rD$ and $\CS$, 
 we consider a family  
$\bW^{(u)}_{2s}(\rD; \CS,   \la  \CP_R\ra)$ given by the walks
such that their vertex of maximal exit degree $\bb$ is characterized
by the following set of numerical data (see subsection 3.4),
$$
\la\CP_R \ra= \left\{ (\bar x,\bar m)_I, (\bar z, \Phi,\bar f')_{K}, (\bar y, \bar \ell,  \Psi, \bar f'')_J\right\}, \quad R= I+M+K+2J+F.
\eqno (4.12)
$$
 It is convenient to  write that $\la\CP_R\ra= (\la \CQ_R\ra, \la\CH_R\ra)$ 
 with  $\la \CQ_R \ra= \{\bar x_I, \bar z_K, \bar y_J\}$
and $\la \CH_R\ra = \left\{\bar m_I, (\Phi, \bar f')_K, (\bar \ell, \Psi, \bar f'')_J\right\}$. 
In what follows, when no confusion can arise, 
we omit the angles in the denotations of $\CP_R$ (4.12), as well as in $\CQ_R$ and $\CH_R$, that denote  numerical realizations
of the sets of parameters $\CP_R$, $\CQ_N$ and $\CH_R$, respectively. 
In denotations
$\CQ_R$ and $\CH_R$, we keep the subscript $R$ to indicate that these values are taken from 
one common set (4.12). 
\vs

Let us describe the construction of the family $\bW^{(u)}_{2s}(\rD; \CS,    \CP_R)$. Given $\CS$, we 
build  a diagram $\CG(\CS)$ as it is done in  Lemma 3.2 (see Section 3).
Regarding the set of vertices $\CV(\CG)$, we add to it  a vertex $\breve v$ of the self-intersection degree
$\vk(\bv) = N = I+K$, i.e. a vertex with $I+K$ edge-windows attached. 
 We denote  the diagram obtained by $\CG^*(\CS)$.

The next step is to distribute $J$ labels over the edge-windows of $\CG$ that will be filled by
the values $\bar y_J$. 
We refer to these labels as to the {\it $y$-labels}.
The windows with $y$-labels will represent the marked instants of the remote BTS-pairs.
Therefore  one cannot use the first arrival edges attached to the vertices of $\CG$.  
Also the  edge-windows attached to the vertices of $\bM_2''$ cannot be used.
Thus, we have $\mu_2' + u_2 + \mu_3 +u_3 +\vert \bar \nu\vert_1$ windows of $\CG$ 
in our disposition 
and then the number of ways to distribute $J$ labels is  bounded from above by 
$
{\mu_2' + u_2 + \mu_3 +u_3 +\vert \bar \nu\vert_1}\choose {J}$. The following upper bound
will be useful for us,
$$
{ {\mu_2' + u_2 + \mu_3 +u_3 +\vert \bar \nu\vert_1}\choose {J}}\le 
{ (\mu_2' + u_2 + \mu_3 +u_3 +\vert \bar \nu\vert_1)^J\over J!} 
$$
$$
\le\  {1\over h^J}\  \exp\{ h(\mu_2' + u_2 + \mu_3 +u_3 +\vert \bar \nu\vert_1)\}, \quad h\ge h_0 >1.
\eqno (4.13)
$$

We denote by $\CG_{\la\CQ\ra}(\CS) = \CG_\CQ(\CS)$ a diagram obtained
from $\CG(\CS)$ by insertion of the  marked instants $\la\CQ_R\ra$ into the chosen  $J$ edge-windows 
of $\CG(\CS)$.

\vs 
The collection of blue, green and black edge-windows of $\CG_{\CQ}(\CS)$ that remain free 
will be denoted by $\stackrel{\diamond}{ \CG}_{\CQ}(\CS) $.
We denote by 
$$
\langle \stackrel{\diamond}{ \CG}_{\CQ}(\CS)\rangle_s = \langle \CG_\triangleleft \rangle_s
$$ 
a realization of the values
in these blue, green and black windows of $\CG_{\la \CQ\ra}(\CS)$. 
This gives the realization of the diagram $\CG^*(\CS)$ that we denote as follows,
$$
\la \CG^\star\ra_s = \la \bv \ra\uplus \la \CG_{\triangleleft} \rangle_s,
$$
where $\la \bv\ra$ denote the realization of the edge-windows attached to $\bv$
given by the values $(\bar x_I,\bar z_K)$. 
The red $f$-edges of $\CG_\CQ(S)$
are  denoted by $ \CG_\circ$ as before. 
\vs 

The exit maximal exit degree of a walk  $\CW_{2s}\in \bW^{(u)}_{2s}(\rD; \CS,  \bb,  \CP_R) $
can be represented as follows,  $\rD = \breve D + \tilde D$, where $\breve D$ is the number of marked edges
of the form $(\bb, \g)$ that belong to the reduced walk $\breve \CW$ (3.9). 
It is known that the number of marked exits from $\bb$ is equal to the number of non-marked
arrival steps  at $\bb$, $\breve D = M+F+J$ and that $F\le K$ (see \cite{KV} and also Lemma 5.1 of \cite{K}). 
It is not hard to see  that $M\le I-1$. 
Then we can write that 
$$
\tilde D = \rD - M-F-J \ge \rD - I-K-J.
\eqno (4.14)
$$

\vs
We see that in a particular $\CW_{2s}$,  
 $\tilde D$ edges are attributed to  $R$ proper and imported cells at $\bb$
 and  $R \le 2(I+J+K) $. Let us denote by $\bar d_R = (d_1, \dots , d_R)$
 a particular distribution of $\tilde D $ balls over $R$ ordered boxes, $\sum_{i=1}^d d_i = \tilde D = D_R$. 
 Then the total number of different sets $\bar d_R$ is bounded by 
 $$
 {D_R + R-1 \choose R-1} \le { D_R +L-1\choose L- 1}, \quad L= 2(I+J+K).
 \eqno (4.15) 
 $$
 
 Having determined all of the values described above, we denote by  
 $$
 \bW^{(u)}_{2s}(\rD, \bar d_R; \la \CG^\star\ra_s, \la \CH_R\ra, \U) = 
 \bW^{(u)}_{2s}( \bar d_R; \la \CG^\star\ra_s,  \CH_R, \U) , 
 \eqno (4.16)
 $$
the family of walks $\CW_{2s}$ with $\th^*(\CW_{2s})=u$ and such that $\CW_{2s} $ follow
 the given  rule $\U$ 
of non-marked continuations.

 

\subsubsection{Exponential estimate and the upper bound} 

It is clear  that the number of ways to attribute  $F'$  imported cells 
to   $K$ marked instants and $F''$ imported cells  to $J$ marked instants  are bounded by 
$$
{ F' +K-1\choose K-1} \le 2^{F'+K} \quad {\hbox{ and}} \quad 
{F''+ J-1\over J-1} \le 2^{F''+J},
$$
  respectively,
and the number of ways to distribute $M\le I$ mirror cells over $I$ marked arrivals at $\bb$ 
is bounded by $2^{2I}$.  
We see that the sub-sum $Z^{(3)}_{2s}$ can be bounded by the following expression (cf. (4.3)), 
$$
Z^{(3)}_{2s} \ \le \  
\sum_{\rD = \lfloor n^\epsilon\rfloor}^s\ \sum_{u=1}^s \ \sum_{I,K: \, I+K\ge 1} \ 
\sum_{M=0}^I\ \sum_{J=0}^s\ \ 
\sum_{F', F''\ge 0}\  \sum_{\bar x_I, \bar y_J, \bar z_K} \ 2^{F+K+J+2I}\ 
$$
$$
\times 
\ \sum_{\la \CS\ra} \ \sum_{\CG(\CS)}  \ \sum_{\la \CQ_R\ra \in \CG(\CS)}\ \ 
\sum_{\la \CG_\triangleleft \ra_s}  \ \ \sum_{\bar d_R} \ \ \sum_{\U} \  \sum_{\la \CH_R\ra} 
\ \Pi(\bb) \ \vert \U_\bb\vert 
$$
$$
\times\  \sum_{\CW_{2s} \in \bW_{2s}^\star}  \ \sum_{ \CT_{2s} \in \CC(\CW_{2s})} \hat \Pi_a(\CI_{2s})\, \Pi _b(I_{2s}),  
\eqno (4.17)
$$
where 
$\Pi(\bb)$ is the weight of the edges attached to  $\bb$ and $\vert \U_\bb\vert $ stands for  the estimate of the local rule of non-marked passages.  
In (4.17),  we also introduce a denotation  $\bW_{2s}^\star = \bW^{(u)}_{2s}( \bar d_R; \la \CG^\star\ra_s, \CH_R, \U)$.
Assuming that $\CG^\star$ is such that $\mu_2' = r+p+q$, we use 
the following upper bound  (see Lemma 6.3 of Section 6),
$$
\sum_{u=1}^s\ \sum_{\CH_R} 
\vert  \bW^{(u)}_{2s}(\rD, \bar d_R; \la \CG^\star\ra_s, \CH_R, \U)\vert \le 
{s^{r/2}} \,   B_r \ 2^r\, 
{\rD^p}\, k_0^q\, (s(\rD+k_0))^{\mu_3'} 
$$
$$
\times\   4^R\, (D_R + 1 )\, e^{-\eta D_R}\, \rt_s,
\eqno (4.18)
$$
where $B_r = \sup_{s\ge 1} {\bf E}_s( (\th^*/\sqrt s)^r)$ (see (4.5)) and  $\D_R = \tilde D = \sum_{r=1}^R \, d_R$. 
Let us note that (4.14) implies that
$$
4^R\, (D_R+1)\, e^{-\eta \rD_R} \le s \, 4^L \, e^{\eta (I+J+K)} \, e^{-\eta \rD}.
\eqno (4.19) 
$$ 
It follows from (4.15) that 
$
\sum_{\bar d_R} 1 \le {\rD+L-1 \choose L-1} 
$
and elementary analysis shows that
$$
\sup_{ L\ge 1} \ {4^L\over h_0^L} \cdot {(\rD + L-1)!\over \rD! \, (L-1)!} 
\le  \exp\left\{ 4\rD\over h_0\right\}.
\eqno (4.20)
$$
The product $\Pi(\bb)\, \vert  \U_{\bb}\vert $ can be bounded with the help of following inequalities that are true for $h\ge 1$ and 
sufficiently large values of $n$ (see Section 5),  
$$
e^{h(I+K)}\, \Pi(\bb)\, \vert \U_\bb\vert   \le \, e^{h(I+K)}
{s^{I+K} (C_1 U^2)^{I+K}\over \rho^{I+K-1}}
$$
$$
 \le 
{1 \over h^{I+K} }
\begin{cases} ( hs e^h )^{k_0} , 
 & \text{if $I+K\le k_0$}, \\
 e^{hk_0}\, n^{-\epsilon} (h e^h n^{-\epsilon/2})^{I+K-k_0}, 
 & \text {if $I+K\ge k_0+1$.}
\end{cases}  
\eqno (4.21)
$$
Finally, it is not difficult to see  that 
$$
\sum_{\bar y_J} \sum_{ \la \CG_\triangleleft\ra} \ 1 = \sum_{ \la \CG_\Diamond\ra} \ 1 \, ,  
\eqno (4.22)
$$
where the last sum is estimated by the right-hand side of inequality (3.5a).

\vs 
Taking into account the upper bound (3.5a)  and relations  (4.4), (4.13), (4.18)-(4.22) 
and accepting that 
$n^\epsilon \ge e^{hk_0}$, 
we deduce from (4.17) the following inequality,
$$
Z^{(3)}_{2s} \le \ (hse^h)^{k_0+3} \ 
\sum_{\rD = \lfloor n^\epsilon\rfloor}^s\ \ \sum_{I,K: \, I+K\ge 1} \ \sum_{J=0}^s\ \ 
{(8h_0)^{2(I+J+K)}\over h^{I+J+K}} 
\, \exp\left\{ - \eta \rD + {4 \rD\over h_0}\right\}
$$
$$
\times \, 
\exp\left\{ {s^2\over 2n} \left( e^{s/n}-1\right)\right\}   \ \sum_{r,p,q\ge 0} \, 
B_r
\cdot   \RH_{(\CS,2)}^{(\sqrt s, \rD,k_0)}(e^h)
$$
$$
\times 
\sum_{\mu_3', \mu_3'', u_3\ge 0} \ \RH_{(\CS,3)}^{(\rD, k_0)}(e^h) \ 
 \ \sum_{\vert \bar \nu \vert_1\ge 0} \ \RH_{(\CS, \bar \nu)}^{(k_0+1)}(e^h),
\eqno (4.23)
$$
where we have used denotations (3.7).
\vs 
Assuming that 
$$
\lim_{s\to\infty}\  \sum_{r=0}^s \ {(2\chi^{3/2} e^{h})^r \over r!} \ B_r 
= \RB(2\chi^{3/2} e^h ),
\eqno (4.24)
$$
we see that 
 the choice of 
 $h= (8h_0+1)^2+1$ and  $h_0 = 8/\eta$ in the right-hand side of (4.23) is sufficient for the relation 
 $$
Z^{(3)}_{2s} = O\left( s^{k_0+4}\,  \exp\{- \eta n^{\ep}/2\} \right)= o(1)
\eqno (4.25)
$$
to hold in the limit $(n,s)_\chi\to\infty$. 

\vs 
Relation  (4.24) can be proved provided 
$B_r = \sup_{s\ge 1} B_r^{(s)} = \lim_{s\to\infty} B_r^{(s)}$. This statement would follow from the monotonicity of 
$B^{(s)}_r$ with respect to $s$. We did not find any related reference in the literature but assume that this monotonicity is true. 
We do not study this question in the present paper. In fact,  the existence of an  upper bound of  
the left-hand side of (4.24) 
would be  sufficient for us.

\subsection{Proof of Theorem 2.1 and Theorem 2.2}

Using the standard arguments of the probability theory, we can write that
$$
\P\left( \left\{\o:  \hat \l_{\max}^{(n,\r_n)} > 2v \left( 1 + {x\over n^{2/3} } \right) \right\}\right)  \le 
{ \E \Tr ( \hat H^{(n,\r_n)})^{2s_n}
\over \left( 2v (1+x n^{-2/3})\right)^{2s_n} },
\eqno (4.26)
$$
where $\hat \l_{\max} ^{(n,\r_n)}$ is the maximal in absolute value eigenvalue of $\hat H^{(n,\r_n)}$. 

Relations (4.2), (4.8), (4.11) and (4.25) show that
$$
\limsup_{n\to\infty}  \E \Tr ( \hat H^{(n,\r_n)})^{2s_n} 
= \limsup_{n\to\infty} \left( Z_{2s_n}^{(1,1)} + Z_{2s_n}^{(1,2)}+Z_{2s_n}^{(2)} +Z_{2s_n}^{(3)} \right)
$$
$$
\le  {1\over \sqrt {\pi \chi^3}} \, e^{4\chi^3} \, 
\RB(6\chi^{3/2}) = \fM(\chi).
\eqno (4.27)
 $$
 It follows from (4.26) and (4.27) that 
$$
\limsup_{n\to\infty} \P\left\{ \hat \l_{\max}^{(n)} > 2v \left( 1 + {x\over n^{2/3} } \right) \right\} \le 
\inf_{\chi>0} \fM(\chi) e^{-x \chi} = {\mathfrak P}(x).
\eqno (4.28)
$$
Let us consider the subset 
$\Lambda_n \subseteq \Omega$, $ \Lambda_n = \cap_{1\le i< j\le n} \left\{ \omega: \vert a_{ij}(\omega) \vert \le U_n\right\}$. 
Denoting $\hat \lambda_{\max}^{(n,\r_n)} = \l_{\max}(\hat H^{(n,\r_n)})$, we can write that 
$$
\P\left( \left\{\o:   \l_{\max}^{(n,\r_n)} >y \right\}\right) = 
\P\left( \left\{\o:   \hat \l_{\max}^{(n,\r_n)} >y \right\}\right)  + 
\P\left( \left\{ \o:  \l_{\max}^{(n,\r_n)} >y \right\} \cap \bar \Lambda_n\right), 
\eqno (4.29)
$$
where $\bar \Lambda_n = \Omega\setminus \Lambda_n$. 
Clearly,
$$
\P \left\{ \bar \Lambda_n \right\} \le \sum_{1\le i<j\le n} \P\left\{ \vert a_{ij} \vert > U_n\right\}
\le n^2 { \E \vert a_{ij} \vert ^{12+2\phi} \over U_n^{12+2\phi} }.
\eqno (4.30)
$$
The choice of $\d$ given by  (4.1)  is sufficient 
for the right-hand side of (4.30) to vanish in the limit $n\to\infty$.
Then we can  deduce from (4.28) and (4.29) the following estimate,
$$
\limsup_{n\to\infty} \P\left\{  \l_{\max}^{(n,\r_n)} > 2v \left( 1 + {x\over n^{2/3} } \right) \right\} \le 
{\mathfrak P}(x)
\eqno (4.31)
$$
that proves (2.3). 
Theorem 2.1 is proved. $\Box$

\vs
Returning to (4.27), we observe that 
$Z_{2s_n}^{(1,2)}+Z_{2s_n}^{(2)} +Z_{2s_n}^{(3)} = o(1)$ as $n\to\infty$. This means that
the upper bound $\fM(\chi)$ is the same as if one considers the moments of Wigner random matrices
$H^{(n,n)}$ (2.1) provided
$V_{12+2\phi}< \infty$. The same is true for the random matrices  $A_n$ of Gaussian Orthogonal Ensemble
 \cite{M}. 
 The limit of the corresponding sub-sum $Z^{(1,1)}_{2s_n}(A_n)$ exists \cite{S},
 $
\lim_{n\to\infty}  Z_{2s_n}^{(1,1)}(A_n) = \CM_{\hbox{\tiny{GOE}}}(\chi)
$
and therefore we conclude about the existence of the limit 
$$
 \lim_{n\to \infty} 
\E \Tr ( \hat H^{(n,\r_n)})^{2s_n} = \CM_{\hbox{\tiny{GOE}}}(\chi).
\eqno (4.32)
$$

\vs 
\vs 

Let us prove Theorem 2.3. Regarding the moments of random matrices $\tilde H^{(n,\r_n)}$,
we can use the same representation as before and write that
$$
\tilde \rM_{2s_n}^{(n,\r_n)} = \E \Tr \left( \tilde H^{(n,\r_n)}\right)^{2s_n} = \tilde Z_{2s_n}^{(1,1)} + \tilde Z_{2s_n}^{(1,2)}+ 
\tilde Z_{2s_n}^{(2)}+ \tilde Z_{2s_n}^{(3)},
\eqno (4.33)
$$
where $\tilde Z_{2s_n}$ are determined in the same way  as it is done in (4.2).
It is easy to  see that the estimate of $\tilde Z_{2s_n}^{(1,1)}$ is the same as that  of
$Z_{2s_n}^{(1,1)}$.

To get the estimate of $\tilde \RZ_{2s_n}^{(1,2)}$, we repeat the computations used to estimate $Z_{2s_n}^{(1,2)}$
with the only difference that (4.7) is replaced by expression
$$
n{ (C_1U^2 s)^k \over (v^2 \r)^k} = \left({ C_1 U^2\over v^2} \right)^{k} n^{1- 2\vep k_0/3 - 2\vep (k-k_0)/3},
\quad k\ge k_0+1.
$$
Then the choice of the technical value $k_0 = \lfloor { 3\over 2\vep} \rfloor +1$ (cf. (4.1)) is sufficient to conclude that 
$\tilde Z_{2s_n}^{(1,2)} = o(1)$ as $n\to\infty$. The same concerns the estimate of $\tilde Z_{2s_n}^{(2)}$ and $\tilde Z_{2s_n}^{(3)}$.

These arguments show that
$$
\limsup_{n\to\infty} \tilde \rM^{(n,\r_n)}_{2s_n} \le
{1\over \sqrt {\pi \chi^3}} \, \RB(6\chi^{3/2}) \, e^{4\chi^3}
\eqno (4.34)
$$
for any positive $\vep\in  (0,1/2]$.
It is easy to see  that (4.34) implies (2.4). Theorem 2.2 is proved. $\Box$

\vs
In conclusion, let us say that the upper bounds (4.27) and (4.34) we obtain are not optimal. 
Regarding the estimate of $Z^{(1,1)}_{2s}$ (see Section 4), we observe that the leading contribution to this sum
comes from the walks with simple self-intersections (the open and not open ones) 
and the triple self-intersections that are not open.
The las group of intersections gives the factor $\exp\{ \chi^3 \}$ instead of $\exp\{4\chi^3\}$. 
 
The simple open self-intersections  with no BTS-pairs should be included into the term $\mu_2''$
and their contribution  is normalized by the term $\exp\{-s^2/(2n)\}$. Thus the factor $\RB$ of (4.27) and (4.3)
should take into account two kind of BTS-pairs and is to be replaced by $\RB(2\chi^{3/2})$. 
Therefore one could expect that the explicit form of relation like (4.32) should be as follows,
$$
 \lim_{n\to \infty} 
\E \Tr ( \tilde H^{(n,\r_n)})^{2s_n} = \CM_{\hbox{\tiny{GOE}}}(\chi) =
{1\over \sqrt{\pi\chi^3}} \, \RB(2\chi^{3/2})\, s e^{2\chi^3}.
\eqno (4.35)
$$

It is not hard to observe that in the case of hermitian random matrices, 
only one type of BTS-pair is present in the simple open self-intersections. Therefore 
we expect that the following limit holds,
$$
 \lim_{n\to \infty} 
\E \Tr ( \tilde H^{(n,\r_n)})^{2s_n} = \CM_{\hbox{\tiny{GUE}}}(\chi) =
{1\over \sqrt{\pi\chi^3}} \, \RB(\chi^{3/2})\,  e^{2\chi^3}.
\eqno (4.36)
$$

\vs
Let us note that the limiting expressions $\CM_{\hbox{\tiny{GOE}}}(\chi)$ and $\CM_{\hbox{\tiny{GUE}}}(\chi)$
are well known in the spectral theory of random matrices.
 They can be computed with the help of orthogonal polynomial technique (see  \cite{S} and references therein).
By using the Laplace transform  method, it can be shown that these values
are related with the Tracy-Widom distributions $F(t) = \P( \lambda^{(n)}_{\max} \le 1 + t/n^{2/3})$ of GOE and GUE, respectively
\cite{AGZ,S,TW}. Then one conclude also about the convergence of the left-hand side of (2.3) to the Tracy-Widom law.
This discussion  of these questions goes  out of the range of the present paper. 


\section{Auxiliary statements} 

\subsection{Proof of Lemma 3.2}

Let us introduce the following denotations. A  self-intersection with 
the marked instants $ \t_1< \dots< \t_k\le s$ will be denoted
by $\fii^{(k)}(\t_1, \dots, \t_k)$.  We denote the vertex of self-intersection $\fii^{(k)}$
 by $v = v(\fii^{(k)})$.
  A collection of self-intersections $\{ \fii^{(k_j)}\}, j\ge 1$  with given marked instants $\t_i\in [1,\dots, s]$ 
 will be denoted by 
$\langle \fI\rangle _s$. 
Given $\th= \th_{2s}$, $\langle \fI\rangle_s$ and a rule of non-marked passages  $\U$, 
the corresponding walk $\CW_{2s}$, if it exists,  is unique and 
 completely determined by the chronological run along $\theta$. 

\vs
We will  also use  the following denotation of the self-intersection with a number of free instants,
$\tilde \fii^{(k)}_l= \tilde \fii^{(k)}(\fO_l, \t_{k}^{(l)})$, where $\fO_l= (\fo_1,\dots, \fo_l)$, 
$\t_k^{(l)} = (\t_{l+1}, \dots \t_k)$ and 
$$
\tilde \fii^{(k)}(\fo_1, \dots , \fo_l,\t_{l+1},\dots, \t_{k} ) = \sqcup_{x_1< \dots < x_l< \t_{l+1}}\,  
\fii^{(k)}(x_1, \dots, x_l, \t_{l+1},\dots, \t_k).
$$
We say that $\fo_j$ are the {\it windows} of self-intersection. In what follows, we will omit the tilde sign and the subscript $l$ 
 in
$ \tilde \fii^{(k)}_l(\fO_l, \t_k^{(l)})$, when no confusion can arise.

\vs 
Let us denote by 
$\bW_{2s}^{[\th]}(\langle \fI\rangle_s, \tilde  \fii^{(k)}_l; \U)$ a family of walks that have a Dyck structure $\th$, 
the set of given self-intersections $\langle \fI\rangle_s$, follow  the  rule $\U$, 
and have an additional  $k$-fold self-intersection
$\tilde \fii^{(k)}_l$.
We consider all possible walks of this kind with all possible values of the free instants of self-intersection $\tilde \fii^{(k)}$.
Given  $(\th, \langle \fI\rangle_s,\U)$, we omit these 
variables and indicate 
their presence by an asterisk   in the denotation 
$\stackrel{*}{\bW}_{2s}$.
The next sub-section represents a rigorous formulation of the observations  used in \cite{SS1} and \cite{S}. 
The tools developed will give us means to proceed in  more complicated cases.

\subsubsection{ Walks with $\b\in \bM_2'$}

Let us consider a family of walks 
$
\sbW_{2s}\left(\fii^{(2)}(\fo_1,\t_2)\right) 
$
 such that the  instant of the second marked arrival 
 $\bar \fa_2= \fa_2$  at $v= v(\fii^{(2)})$ is given by $\t_2$. 
The set  $\sbW_{2s} \left(\fii^{(2)}(\fo_1,\t_2)\right)$ can be constructed as follows. 
Regarding  a   sub-walk
$\CW_{[0,\xi_{\t_2}-1]}$ that is completely determined by $(\th, \langle \fI\rangle_s,\U)$,  we  set 
$\CW_{2s}(\xi_{\t_2})= \b$, $\b\in g(\CW_{[0,\xi_{\t_2}-1]})$.  Then, respecting 
the rules of $\th, \langle \fI\rangle_s$ and $\U$, we  
continue $\CW$  to the time interval $[\xi_{\t_2}+1,2s]$. If such a walk exists, we add it to the 
list of elements of $\sbW_{2s} \left(\fii^{(2)}(\fo_1,\t_2)\right)$. 
In this case, the value $\t'$
that is the first arrival instant at $\b$ fills the window $\fo_1$. We say that 
 the value $\t_1=\t'$ of the simple self-intersection is a 
realization of the window $\fo_1$ given by the walk $\CW_{2s}$. 
We  denote  this realization  by 
$\t_1=\langle \fo_1\rangle_{\CW_{2s}}$.
\vs 

In this construction, we see that the only condition on $\t'$
at the instant $\t_2$ of the second arrival $\fa_2$ 
is such that there is no other marked arrivals at $\b$ than the first one, $\t'$.
Let us  say that this situation describes the {\it unconstraint} simple self-intersection,
when there is no other special condition
imposed on  $\fa_2$.

\vs 

\vs
Let us denote by  
$\fii^{(2)}\left( \fo_1, \left[{\stackrel{\t_2}{\fu_2}}\right] \right) $ 
a  simple self-intersection $\fii^{(2)(\fo_1, \t_2)}$
 with a certain condition $\fu_2$ 
imposed on the second arrival instant $\fa_2$ at   $v(\fii^{(2)})$.

\vs
We  consider first the construction of walks 
$\CW_{2s}\in\  \sbW_{2s} \left(\fii^{(2)}\left(\fo_1,\left[ { \stackrel{\t_2}{\fu_2}} \right]\right)\right)$ with
open simple self-intersection $\fii^{(2)}$. 
that corresponds to the condition (c)  of subsection 3.2 with denotation $\fu_2=(o)$. 
Starting with a uniquely determined  sub-walk $\CW_{[0,\xi_{\t_2}-1]}$ that follows the rules of $\th$, 
$\langle \fI\rangle_s$ and $\U$, 
we see that 
the vertex $v=\g$ of the open simple self-intersection $\fii^{(2)}$ can be chosen from the subset of vertices
${\bV}^{(o)}_{t'} \in \bV(g)$,  $g(\CW_{[0,\xi_{\t_2}-1]})$ that are open at the instant of time $t'= \xi_{\t_2}-1$,
 i.e. that have at least one
open non-oriented edge attached to $\g$. We will say that $\bV^{(o)}_{t'}$ is the set of {\it $t'$-open} vertices.
Choosing one of the admissible vertices, we continue the run of the walk according to the rules 
$\th$, $\langle\fI\rangle_s$ and $\U$, if it is possible.

A simple but  important  observation is that for  any instant of time $\xi_{\t_2}$, 
the following upper bound holds,
$\max_t \vert {\bV}'(\CW_{[0,t]})\vert \le 2\th^*
$, where  $\th^*$ is the  height of the Dyck path $\th_{2s}$
  \cite{SS2}.
Therefore we can write that
$$
\# \left\{ \t_1: \ \t_1= \langle \fo_1\rangle_{\CW_{2s}}, \ 
\CW_{2s}\in\ 
\sbW_{2s} \left(
\fii^{(2)}
\left(
\fo_1,
\left[ {\stackrel{\t_ 2}
{\fu_2}}\right] 
 \right) \right)
\right\} \le 2\th^*, \quad \fu_2 = (o).
\eqno (5.1) 
$$
Clearly, 
$$
\sbW_{2s} \left(
\fii^{(2)}
\left( \fo_1,\left[ {\stackrel{\t_ 2} {\fu_2}}\right] \right) \right)\subseteq \ 
 \sbW_{2s} \left(\fii^{(2)}\left(\fo_1,{{\t_ 2}} \right) \right)
 $$
and
we can say  that the set $ \sbW_{2s} \left(
\fii^{(2)}
\left(
\fo_1,
{{\t_ 2}} 
 \right) \right)$ is {\it filtered} by the condition $\fu_2$.
 With a certain  abuse of language, we can say that 
 the possible values $\t'$ at the \mbox{window $\fo_1$} 
 are {\it restricted} or {\it filtered} by the condition $\fu_2$ imposed on the second arrival $\fa_2$. 
 In a brief form, we will say that the first arrival $\fa_1$ is filtered by the property $\fu_2$ of the second arrival $\fa_2$.

\vs

The next case of simple  self-intersection with filtered first
arrival is given by  condition $(b)$ of subsection 3.2. 
Corresponding self-intersection  produces a $p$-edge of the form $(\g,v)$, $v= v(\fii^{(2)})$. 
In this case,  a necessary condition
is that $v$ belongs to the exit cluster  $\Delta(\alpha)$ of $g(\CW_{[0,\xi_{\t_2}-1]})$ with 
 $\a= \CW(\xi_{\t_2}-1)$  that is uniquely determined.  
We denote this condition by $\fu_2 = (\D)$.

Condition $\fu_2 = (\D)$ means that we can continue the sub-walk $\CW_{[0,\xi_{\t_2}-1]}$
to the instant of time $\xi_{\t_2}$ with a letter (a vertex) that belongs to $\D(\a)$. 
To choose this letter means also to choose the marked instant of the first arrival at this vertex. 
Therefore, if one considers the family of walks 
$\sbW_{2s} \left(\rD;
\fii^{(2)}\left(\fo_1,  \left[ {\stackrel{\t_ 2} {\fu_2}}\right] \right) \right)$
such that $\vert \CD(\CW_{2s})\vert =  \rD$,  then 
$$
\# \left\{ \t_1: \ \t_1=\langle \fo_1\rangle_{\CW_{2s}}, \ \CW_{2s}\in
\ \sbW_{2s} \left(
\rD; \fii^{(2)}\left(\fo_1,
\left[ {\stackrel{\t_ 2}
{\fu_2}}\right]  \right) \right)\right\}  \le \rD, \quad \fu_2 = (\D).
\eqno (5.2)
$$
Thus, $\sbW_{2s}(\fii^{(2)}\left(\fo_1,{{\t_ 2}} \right))$ is filtered by the condition $\fu_2 = (\D)$. 
We will say that $\fii^{(2)}$ of this type produces a multiple edge in $\bar \bE_g^{(k_0)}$.

\vs

The second type of the filtering related with the multiple edges in $\bar \bE_g^{(k_0)}$ 
is given by  condition (a) of subsection 3.2. 
In this case, a necessary condition on $\langle \fo_1\rangle_{\CW_{2s}}=\t_1$  is  that
the marked edge $(v(\fii^{(2)}),\a)= ( \CW_{2s}(\xi_{\t_1}),
\CW_{2s}(\xi_{\t_2}-1) )$ 
exists and belongs to $\bar \bE_g^{(k_0)}$. The latter condition implies that 
$v(\fii^{(2)})\in \Lambda( \a)$ and means than the vertex $v(\fii^{(2)})$ is the $\mu$-vertex. 
We denote this necessary condition by $\fu_2 = (\Lambda^{(k_0)})$.
The number of vertices in $\Lambda(\a)$ bounded by $k_0$, we can write that 
$$
\# \left\{ \t_1: \ \t_1=\langle \fo_1\rangle_{\CW_{2s}}, 
\ \CW_{2s}\in
\  \sbW_{2s} \left(\fii^{(2)}\left(\fo_1,  \left[ {\stackrel{\t_ 2}
{\fu_2}}\right]  \right)\right) 
\right\} \le k_0, \quad \fu_2 = (\Lambda^{(k_0)}).
\eqno (5.3)
$$
In this case, we will say  that the values $\t'$ at the  window $\fo_1$ are filtered by  the 
\mbox{$\Lambda^{(k_0)}$-condition}. 

\vs
It is obvious that the upper bounds (5.1), (5.2) and (5.3) are true in the  case when $\b \in \bM_2'$, i.e. when
$\kappa(\b)= m \ge 2$. To prove this, it is sufficient to consider
the walks with the self-intersection $\tilde \fii^{(m)}_1 = \tilde \fii^{(m)}(\fo_1, \t_2,\t_3,\dots, \t_m)$. 

\vs 
Let us note that  in the reasonings above, we did not make any difference between two procedures,
the first one related with
 the continuation of the sub-walk $\CW_{[0, \xi_{\t_2}-1]}$ to the instant of time $\xi_{\t_2}$ with some letter
 that verifies one or another condition
 and the second one given by the 
 filtration of the set of  walks $\sbW_{2s}(\fii^{(i)}(\fo_1,\t_2))$ with respect to one or another filtering condition. 
 It is easy to see that these two procedures are similar and lead to the same results, i. e. to the same estimates
 of the cardinalities of families of walks.
We will use this observation in the sub-sections that follow. 
\vs
The last remark is that in the reasoning  above the cases when $v(\fii^{(2)})=\a$ are not excluded.
This means that the estimates presented are valid in the cases of loops. 
This is also true with respect 
to the proofs of estimates that  follow. 


\subsubsection{Walks with $\b\in \bM_3'$}

Using the tools of the previous subsection, we can  describe the properties of walks that belong to 
$\sbW_{2s}(\fii^{(3)}(\fo_1,\fo_2, \t_3))$, where the vertex of self-intersection $ \fii^{(3)}$ is  
such that $\b = v(\fii^{(3)})\in \bM_3'$. If the edge $e_3 = e(\xi_{\t_3})$ of the third arrival $\bar \fa_3=\fa_3$ 
 is the $p$-edge, we will write that $e(\xi_{\t_3})\in \bV^{(p)}(g)$. 
If $e_3= e(\xi_{\t_3})$ is the $q$-edge, we will write that $ e(\xi_{\t_3})\in \bV^{(q)}(g)$.

\vs 
{\bf Lemma 5.1.} {\it Let us denote by $\sbW_{2s}(\rD; \fii^{(3)})$ a family of walks 
with $k_j\le k_0$, $j\ge 1$ and such that $\CD(\CW_{2s}) = \rD$. Then  
 $$
 \# \left\{ (\t_1,\t_2)= \langle (\fo_1,\fo_2)\rangle_{\CW},  \CW \in 
 \sbW_{2s}\left(\rD; \fii^{(3)}\left(\fo_1, \fo_2, \left[ { \stackrel{\t_3}{\fu_3}} \right] \right)\right)  \right\} 
 \le 
  \begin{cases}
 s \rD,&\text{if  $\fu_3 = (\D)$}, \\
s k_0,&\text {if $\fu_3 = (\Lambda^{(k_0)})$.}
\end{cases}
 \eqno (5.4)
 $$
 }
 
 \vs 
 The statement of Lemma 5.1 seems to be a simple generalization of  relations (5.2) and (5.3) 
 of the previous sub-section.  However, the proof of Lemma 5.1 crucially depends on   the classification of vertices 
 of $\bM_2'$ and 
 $\bM_3'$ we introduced 
and therefore represents a key point in  the   general method of the proof of Lemmas 3.2 and 3.3.
Let us stress that the estimates like (5.4) were not considered in papers \cite{SS1, SS2} and \cite{S}. 
 \vs

{\it Proof of Lemma 5.1.} 
Let us start with the case $e_3\in \bV^{(p)}(g)$. Then the  edge $e_3 = (\a,\b)$ is such that 
the marked  edge $e'=(\a,\b)$ exists in the graph of the sub-walk $\CW_{[0, \xi_{\t_3}-1]}$
and $\a = \CW(\xi_{\t_3}-1)$. 
By  the definition, $\b\notin \bM_2'$. This means that the edge $e'$ represents either the first 
marked arrival at $\b$, $e'= e(\fa_1)$ or the second one, $e'= e(\fa_2)$, 
and that the situation when $e(\fa_1) = e(\fa_2) = (\a,\b)$
is excluded. 
\vs 
We are going to   construct   the walks of $\sbW_{2s}(\rD; \fii^{(3)}(\fo_1,\fo_2, \t_3))$ 
with  \mbox{$v(\fii^{(3)})\in \bM_3'$}  such that
 $e_3= e(\xi_{\t_3})$ is a $p$-edge. 
We denote the exit cluster
  of  \mbox{$\a = \CW(\xi_{\t_3}-1)$} 
  by  $\D(\a) = \{\b_1,\dots \b_l\}$ with $l\le \rD$.
 Let us assume for the moment that  $\vk(\b_i)=1$ for all $i\in [1,\dots, l]$. 
 We choose one of the vertices $\b_i$ and take the marked instant $\t'$ such that $\CW_{[0, \xi_{\t_3}-1]}(\xi_{\t'})=\b_i$. 
 The sub-walk $\CW_{[0, \xi_{\t_3}-1]}$  will be continued at the instant of time $\xi_{\t_3}$ with the letter $\b_i$.

 Our aim is  to construct a sub-walk $\bar  \CW_{[0, \xi_{\t_3}-1]}$ that has a  self-intersection $(\cdot, \cdot)_1$ with 
 the participation 
 of $\t'$. The subscript $1$
 indicates the fact that this simple self-intersection is not present in $ \CW_{[0, \xi_{\t_3}-1]}$.
 We  proceed as follows: we take any marked instant of time $\t''$ that is not involved into the self-intersections
 already present and consider the letter $\g''= \CW_{[0, \xi_{\t_3}-1]}(\xi_{\t''})$. If $\t''<\t'$, then we replace $\b_i$ by $\g''$.
 \vs 
 
 The walk $\bar \CW_{[0, \xi_{\t_3}-1]}$ performs 
a self-intersection $(\t'',\t')$  and in general situation, 
this fact  changes the run of the walk after $\xi_{\t'}$ 
with respect to the run of the initial  sub-walk $  \CW_{[0, \xi_{\t_3}-1]}$.
Therefore the exit cluster $ \tilde \D(\a)$ can differ from  $\D(\a)$. 
Indeed, one can easily find examples of sub-walks $\bar \CW$ and $ \CW$  such that 
 $\tilde  \D(\a)\cap \D(\a) = \emptyset$. 

As a consequence, it can happen that $e(\t')\notin \bar \D(\a)$. Moreover, if one considers another
value $\tilde \t'$ such that $e(\tilde \t')\notin \D(\a)$, one cannot guarantee that 
$e(\tilde \t') \notin \bar \D'(\a)$. 
Thus, without  additional considerations,
we cannot say  that given $\t_3$ and $\t''$, the set of all possible values of $\t'$ has a cardinality bounded by 
the maximal exit degree $\rD$. 

\vs 
An important observation here is that the changes of the exit cluster of $\a$ are possible
only in the cases when the sub-walk performs a BTS-couple  $(\xi_{\t'}, \xi_{\t'}+1)$. If it is not the case,
all cells at $\a$ created up to the instant $\xi_{\t_3}-1$ remain the same and therefore the edges of $\D(\a)$
are determined uniquely by $\theta_{2s}$. The key point is that relation  $\b\in \bM_3'$ implies
that $\b\notin \bM_2'$. This means that the second arrival $\fa_2$ does not verify the condition (c)
mentioned above and we cast off 
those walks
that  arrive by $\fa_2$  at an open  vertex. 
 Therefore $\fa_2$ cannot represent   the BTS-pair and  
 the set $\D(\a)$ is determined uniquely by parameters 
 $\th$, $\langle \fI\rangle_s$ and $\U$. The exit cluster $\D(\a)$ 
 does  not change  when the values   $\t''$ and $\t'$ vary.
 Then the edge $e(\t') = (\a,\b_i) = (\a,\g'')$ always belongs to $\D(\a)$.
 \vs
  If $\t'<\t''$, then we replace the letter $\g''$ by $\b_i$. Again, the fact that $(\a,\b_i)\in \Lambda(\a)$ remains true.
  We see that the number of walks that verify conditions imposed is bounded by $(s-1)D$. 
Thus the following upper bound holds,
 $$
 \# \left\{ (\t_1,\t_2)_1 = \langle (\fo_1,\fo_2)\rangle_{\CW} ,  \CW \in\,  
 \sbW_{2s}\left(D; \fii^{(3)}\left(\fo_1, \fo_2, \left[ { \stackrel{\t_3}
 {{{\fu_3}}}} \right] \right)\right) 
 \right\} \le (s-1)\rD, \ \fu_3 = (\D).
 $$
 Let us consider the case when $\vk(\b_i)=2$ in the graph of the sub-walk $\CW_{[0, \xi_{\t_3}-1]}$. 
Then we can write that 
$$
 \# \left\{ (\t_1,\t_2)_1 = \langle (\fo_1,\fo_2)\rangle_{\CW} ,  \CW \in\,  
 \sbW_{2s}\left(D; \fii^{(3)}\left(\fo_1, \fo_2, \left[ { \stackrel{\t_3}
 {{{\fu_3}}}} \right] \right)\right) 
 \right\} \le \rD, \ \fu_3 = (\D).
 $$
 These two estimates give (5.4) with $\fu_3 = (\D)$.

 \vs 
 Let us  construct   the walks of $\sbW_{2s}(\fii^{(3)}(\fo_1,\fo_2, \t_3))$ with  $v(\fii^{(3)})\in \bM_3'$  such that
 $e_3= e(\xi_{\t_3})\in \bV^{(q)}(g)$. 
 Given $\th$, $\langle \fI\rangle_s$ and $\U$, the enter cluster
  of  \mbox{$\a = \CW_{[0, \xi_{\t_3}-1]}(\xi_{\t_3}-1)$} 
  that is $\Lambda(\a) = \{\b_1,\dots \b_l\}$, $l\le k_0$ is uniquely determined.
 Let us assume for the moment that  $\vk(\b_i)=1$ for all $i\in [1,\dots, l]$. 
 We choose one of the vertices $\b_i$ and take the marked instant $\t'$ such that $\CW_{[0, \xi_{\t_3}-1]}(\xi_{\t'})=\b_i$. 
 The sub-walk will be continued at the instant of time $\xi_{\t_3}$ with the letter $\b_i$.
 
 We are going to construct a sub-walk $\tilde  \CW_{[0, \xi_{\t_3}-1]}$ that has a  self-intersection $(\cdot, \cdot)_1$ with participation 
 of $\t'$. The subscript $1$
 indicates the fact that this simple self-intersection is not present in $ \CW_{[0, \xi_{\t_3}-1]}$.
 We  proceed as follows: we take any marked instant of time $\t''$ that is not involved into the self-intersections
 already present and consider the letter $\g''= \CW_{[0, \xi_{\t_3}-1]}(\xi_{\t''})$. If $\t''<\t'$, then we replace $\b_i$ by $\g''$.
 The fact that $\CW_{[0, \xi_{\t_3}-1]}$ performs this additional self-intersection $(\t'',\t')$ does not
 change the set $\Lambda(\a)$ because we reject the walks such that $(\t'',\t')$ is the open self-intersection.
 Therefore the edge $(\b_i,\a)$ still belongs to $\Lambda(\a)$. 
  If $\t'<\t''$, then we replace the letter $\g''$ by $\b_i$. Again, the fact that $(\b_i,\a)\in \Lambda(\a)$ remains true.
 The choice of $\b_i$ is bounded by $k_0$, we get not more than $(s-1)k_0$ different walks of this kind,
 $$
 \# \left\{ (\t_1,\t_2)_1 = \langle (\fo_1,\fo_2)\rangle_{\CW} ,  \CW \in\,  
 \sbW_{2s}\left(D; \fii^{(3)}\left(\fo_1, \fo_2, \left[ { \stackrel{\t_3}
 {{{\fu_3}}}} \right] \right)\right) 
 \right\} \le (s-1)k_0, \, \fu_3 = (\Lambda^{(k_0)}).
 $$
 
Let us consider the case when $\vk(\b_i)=2$ in the graph of the sub-walk $\CW_{[0, \xi_{\t_3}-1]}$. 
Then we can write that 
$$
 \# \left\{ (\t_1,\t_2)_1 = \langle (\fo_1,\fo_2)\rangle_{\CW} ,  \CW \in\,  
 \sbW_{2s}\left(D; \fii^{(3)}\left(\fo_1, \fo_2, \left[ { \stackrel{\t_3}
 {{{\fu_3}}}} \right] \right)\right) 
 \right\} \le k_0, \ \fu_3 = (\Lambda^{(k_0)}).
 $$
Gathering these two bounds, we obtain the estimate (5.4) with $\fu_3 = (\Lambda^{(k_0)})$. 
Lemma 5.1 is proved. $\Box$ 
\vs 
 
It is not hard to generalize these considerations to the case of walks
with the self-intersection of the form $\tilde \fii^{(k)}_2(\fo_1,\fo_2, \t_3,\dots,\t_k)$
and obtain here the same upper bounds  (5.4).


\subsubsection{General filtration procedure}

The general filtration procedure is as follows. 
Let us consider a diagram $\CG $ and recall that  $\langle \CG_\Diamond \rangle_s$ 
denotes a
realization of blue, green and black half-edges of $\CG$.
This procedure will be considered in more details  in the next sub-section.
Regarding the first vertex $v$ of $\CG$ that has at least one red $f$-edge attached, 
we consider the value $\t$  in the blue edge-window attached at $v$ as the value $\t_2$ or $\t_3$,
 in dependence whether 
$\vk(v)$ is equal to $2$ or $3$. Let us consider the  edges of $\CV(\CG)\setminus v$ such that
the integers in their windows  lie to the left of $\t$. They form a family of self-intersections 
that we regard as the family $\la \fI\ra_s$ introduced above. 

Now  we can use the filtration procedure of the previous sub-sections
and perform a chronological run $\fR_\CT(t)$ along $\CT_s= \CT(\th_{2s})$.
If the next in turn step $t'= \xi_{\t'}$ is marked and $\t'$ does not belong to $\la \fI\ra_s$
and to the realizations of the red edges seen by $\CW$, the walk $\CW$ creates a new,
next in turn, letter from the alphabet $\CA$. If $\t'$ belongs to $\la \fI\ra_s$, we 
follow  the rules of $\langle \fI\rangle_s$ and $\U$; we continue these actions  till the instant 
$\xi_{\t}-1$. Then we consider  all possible continuations of the sub-walk 
$\CW_{[0,\xi_{t}-1]}$ to the instant of time $\xi_t$,  choose one of them and continue the run
of the sub-walk till it meets the second blue window of the vertex that has red edges attached. 
Then the procedure repeats. 

\vs

It is clear that 
 the set of walks $\bW_{2s}^{[\theta]}(\rD;\CS,  \langle \CG_{\Diamond}\rangle_s)$ 
with given  $\CG(\CS) = \CG$ has a cardinality equal to the number of realizations
of all red edge-windows $\CG_\circ$ of $\CG$ and that 
$$
\vert \CG_\circ\vert  \le (2\th^*)^r\, \rD^{p}\, k_0^q \cdot (s(D+k_0))^{\mu_3'}.
\eqno (5.5)
$$
Relation (3.5b) follows from (5.5).

\subsubsection{Values in blue, green and black edge-windows}

Given  $\CG$, let us fill its  blue edge-windows  $\CE_\rb$ first. 
 We consider 
 the set of integers $[1,\dots,s]$ and  choose  the values for 
 $r$ groups of one
element, $p$ groups of one element, $q$ groups of one element, 
$\mu_2''$ groups of two elements, $\mu_3'$ groups of one element, $\mu_3''$ groups of three elements
and $\nu_k$ groups of $k$ elements, $k_0+1\le k\le s$.

The number of ways to choose these groups of subsets of  $[1, \dots, s]$  is given by the following expression,
$$
\vert \langle \CE_{\rb}\rangle_s\vert = {s!\over r!\, p!\, q!\ \mu_3'!\ \mu_2''! \,  (2!)^{\mu_2''} 
\  (3!)^{\mu_3''}\,  \mu_3''!  \ }\ 
\prod_{k=k_0+1}^s {1\over \nu_k!\ (k!)^{\nu_k}} \cdot {1\over (s- E)!},
$$
where $E = r+p+q + 2\mu_2'' + \mu_3' + 3\mu_3'' + \Vert \bar \nu \Vert$. Clearly, 
$$
\vert \langle \CE_\rb\rangle_s\vert \le {s^E\over 
r!\, p!\, q!\ \mu_3'!\ \mu_2''! \,  (2!)^{\mu_2''} 
\  (3!)^{\mu_3''}\,  \mu_3''! }\ 
\prod_{k=k_0+1}^s {1\over \nu_k!\ (k!)^{\nu_k}}.
\eqno (5.6) 
$$
The values in the green edge-windows $\CE_\rg$ 
are chosen from the set $[1,\dots, s]\setminus \langle \CE_\rb \rangle_s$.
Thus,
$$
\vert \langle \CE_\rg\rangle_s\vert \le s^{u_2+u_3}.
\eqno (5.7)
$$

The number of realizations  $\la \CG_\Diamond\ra_s$ is bounded by the product of the right-hand sides of  
(5.6) and (5.7).
This gives (3.5a). 
Lemma 3.2 is proved. $\Box$

\subsection{Proof of Lemma 3.3}

\subsubsection {The number of diagrams $\CG(\CS)$ }

As it is pointed out, given  $\CS = (r,p,q, \mu_2'', u_2; \mu_3',\mu_3'', u_3;\bar \nu)$,
the diagrams $\CG \in \bG(\CS)$ differ by the positions of the green edges attached to 
the vertices of $\bM_2' = \CV_2'(\CG)$ and $\bM_3= \CV_3(\CG)$. According to the last remark of the previous sub-section,
we can consider the vertices of $\CV_2'$ and $\CV_3$ as the ordered one. By construction,  
$\vert \CV_2'\vert = \mu_2'$ and $\vert \CV_3\vert  = \mu_3$.
So, we can consider $\CG$ as a union od three parts, $\CG = \CG_2\uplus\CG_3\uplus \CG_{(k_0+1)}$

We take $u_2$ green edges and distribute them over  vertices $\CV_2'$. 
We draw the green edges to the right of the blue edges. 

\vs
Let us denote by $u_2^{(i)}$, $1\le i\le k_0-2$  the number of vertices that have $i$ green edges attached, 
$\sum_{i=1}^{k_0-2} i u^{(i)}_2 = u_2$.  
Given $\mu_2'=r+p+q$ and $\bar u_2= (u^{(1)}_2,\dots, u^{(k_0-2)}_2)$, 
the number of all possible diagrams $\CG_2$ is equal to 
$$
 {\mu_2'\choose  u_2^{(1)}, \ \dots\ ,  u^{(k_0-2)}_2} =
  {\mu_2'!\over u_2^{(1)}! \ \cdots \ u_2^{(k_0-2)}! \ \ (\mu_2' - \vert \bar \u_2\vert)!}\ .
$$
Therefore the cardinality of the set   $\bG_2 = \{\CG_2\}$ is bounded as follows,
$$
\vert \bG_2\vert \le \  
 \prod_{i=1}^{k_0-2} \,  {(\mu_2')^{u_2^{(i)}}\over u_2^{(i)}!  }.
\eqno (5.8)
$$ 
Here we have used the following elementary bound, ${ a\choose b} \le a^b/b!$.
\vs
The $u_3$ green edges are  distributed over the  vertices of $\CV_3$ 
and are placed to the right of the blues edges at each vertex.
Assuming that the number of vertices that have $j$ green edges is given by $u^{(j)}_3$, $1\le j\le k_0-3$ with 
$\sum_{j=1}^{k_0-3} ju_3^{(j)} = u_3$, 
we see  that for  given
values of the parameters $\mu_3$ and $\bar u_3 = (u^{(1)}_3, \dots , u_3^{(k_0-3)})$
one obtains 
the following upper bound for the cardinality of the set $\bG_3 = \{ \CG_3\}$,
$$
\vert \bG_3\vert \le 
\  \prod_{j=1}^{k_0-3}  \, { ( \mu_3)^{u^{(j)}_3}\over u_3^{(j)}!}.
\eqno (5.9)
$$

\subsubsection{Sum over rules $\U$ and diagrams $\CG(\CS)$}

In this subsection we prove the Corollary of Lemma 3.2. 
Let us denote by $\bY = \bY(\CS)$ the family of all possible rules of continuation for the class
of walks with the set of parameters $\CS$, $\bY(\CS) = \{ \U(\CS)\}$.
To do this, we have consider a contribution of each vertex $\b$ such that 
the non-marked depart from $\b$ can be performed in a number of different ways. 
We denote by $\U_\b$ this local rule of passage. 
 Let $\xi_\t$ be the marked instant time of $i$-th arrival at a vertex $\b$ of self-intersection,
and the instant $\xi_\t+1$ is the non-marked one.
If the vertex $\b$ is closed at the instant of time $\xi_\t-1$, then there is only one possibility to continue
the run of the walk at the non-marked instant of time $\xi_\t+1$. 

\vs

Regarding a vertex of simple open self-intersection with the second arrival at the marked instant $\xi_\t$,
we see that the number of $\xi_\t$-open edges attached to $\b$ is bounded by 3. 
Then there are not more than $3$ possible continuations of the run of the walk at the 
non-marked instant of time. If  there are $r$ vertices of simple open self-intersection,
then  the contribution of these vertices to $\bY$ is bounded by $3^r$. 
The same concerns the vertices from $\bM_2'$ that have $p$-edges and $q$-edges 
attached. These can produce the open self-intersections too. 
Thus, the total contribution of the $o$, $p$ and $q$-arrivals at the vertices of $\bM_2'$
such that $\vk(\b)=2$
is bounded by $3^{r+p+q}$. 

\vs 
 
It is proved in \cite{KV} (see also \cite{K}) 
that any vertex $\b$
of self-intersection degree $k$ has $k$ non-marked departures from $\b$ and that
at any instant of time, there is not more than $2k$ open edges attached to $\b$.

\vs
Using this simple but important observation, we conclude that for any vertex $\b\in \bM_2'$ 
that has $u\ge 1$ green edges attached, the upper bound for the number of continuations
is given by 
$$
(2(2+u))^{2+u} \le (2k_0)^{2+u}\le (2k_0)^{3u}.
$$

Let us consider the vertex $\b\in \bM_3''$. If $\vk(\b)=3$, then the number of non-marked departures
from $\b$ with the choice of edges to close is not greater than 1 and the number of possible continuations
is bounded by $3$. This departure can be performed after the third marked arrival at $\b$. 

Regarding the vertex $\b$ of $\bM_3'$ such that $\vk(\b)= 3+u$ with $u\ge 1$, we get the following upper bound
for the total number of continuations at this vertex,
$$
(2(3+u))^{3+u} \le (2k_0)^{4u}.
$$
Any vertex $\b\in \bM_3''$ such that $\vk(\b)=3$ produces not more than 9 possible continuations. 

\vs
Finally, any $\nu$-vertex $\b$ such that $\vk(\b)=k$ contributes to the number of possible continuations 
with a factor 
bounded from above by  $(2k)^k$.

\vs 
Gathering the upper bounds obtained, we get the following inequality,
$$
\vert \bY(\CS)\vert \le 3^{r+p+q}\, 9^{\mu_3}\, (2k_0)^{3u_2+ 4u_3} 
\cdot \prod_{k=k_0+1}^s \, (2k)^{k\nu_k}.
\eqno (5.10)
$$

\vs
Gathering estimates (5.6)-(5.9) and (5.10) and combining them with the result of Lemma 3.2, we can 
estimate of number of walks for a class with given diagram $\CG(\CS)$ with 
$\CS= (\mu_2'',r,p,q,\bar u_2; \mu_3',\mu_3'', \bar u_3;\bar \nu)$ by the following
expression,
$$
{1\over \mu_2''!} \left( {s^2\over 2}\right)^{\mu_2''} \cdot 
{(6s\th^*)^r\over r!} \cdot {(3sD)^p\over p!} \cdot { (3sk_0)^q\over q!}\cdot
 {1\over \mu_3''!} \left( {9s^3\over 3!}\right)^{\mu_3''} \times   { (9s^2(D+k_0))^{\mu_3'}\over \mu_3'!}
 $$
 $$
 \times \ 
((2k_0)^3s)^{u_2}\cdot ((2k_0)^4 s)^{u_3} \cdot 
   \prod_{i=1}^{k_0-2} \,  {(\mu_2')^{u_2^{(i)}}\over u_2^{(i)}!  } \cdot  
   \prod_{j=1}^{k_0-3}  \, { ( \mu_3)^{u^{(j)}_3}\over u_3^{(j)}!}.
  \eqno (5.11)
$$
The sum of (5.11) over all possible values of $\bar u_2$ gives the following upper bound,
$$
((2k_0)^3s)^{u_2} 
\sum_{u_2^{(1)}+ \dots + u_2^{(k_0-2)} = u_2} \ \ 
  \prod_{i=1}^{k_0-2} \,  
  {(\mu_2')^{u_2^{(i)}}\over u_2^{(i)}!  }
  $$
  $$ 
  = 
{((2k_0)^3s\mu_2')^{u_2} \over u_2!}
\sum_{u_2^{(1)}+ \dots + u_2^{(k_0-2)} = u_2} \ \ 
 { u_2!\over u_2^{(1)}! \cdots u^{(k_0-2)}_2!}= 
 {\left( (2k_0)^3(k_0-2)s\mu_2'\right)^{u_2} \over u_2!}\, . 
 \eqno (5.12)
 $$ 
When deriving (5.12), we have used the multinomial theorem.

The sum of (5.11) over all possible values of $\bar u_3$ gives the following upper bound,
$$
((2k_0)^4 s)^{u_3}\sum_{u_3^{(1)}+ \dots + u_3^{(k_0-3)} = u_3} \  \    
 \prod_{j=1}^{k_0-3}  \, { ( \mu_3)^{u^{(j)}_3}\over u_3^{(j)}!} 
 \le
  { \left( (2k_0)^4(k_0-3) s\mu_3\right)^{u_3}\over u_3!}\, .
 \eqno (5.13)
$$
Combining relations (5.11), (5.12), (5.13) with the result of Lemma 3.2,  we obtain
inequality (3.6). Corollary of Lemma 3.2 is proved.

\vs
To complete the proof of Lemma 3.3, it remains to estimate the number of trajectories in the class of equivalence
$\CC_\CW$ (3.1). It is easy to see that given $\CW_{2s}$ of the class $\CG = \CG(\CS)$, we have the following equality,
 $$
 \vert \bU(\CI_{2s}; 2s)\vert = \vert \bV_g\vert = s- \s +1,
 \eqno (5.14)
 $$ 
where $\s = \mu_2 + 2\mu_3 + u_2+u_3 + \vert \bar \nu\vert_1$. 
Then \cite{SS1}
$$
\vert \CC_\CW\vert \ = \ \prod_{k=1}^{s-\s}\, \left(1 - {k\over n} \right) \ \le \ \exp\left\{ -{(s-\s)^2\over 2n}\right\}. 
\eqno (5.15) 
$$
This simple but important upper bound can be proved with the help of representation 
$1-k/n = \exp\{ \ln (1-k/n)\}$ and the use of the Taylor expansion.
Now, combining (5.15) with (3.6) and the result of Lemma 3.1, we get inequality (3.7). Lemma 3.3 is proved. $\Box$


\subsection{Walks with imported cells} 

As we have seen in Section 3, 
the proper and imported cells at the vertex of maximal exit degree $\bb$
are characterized by the
following  set of parameters $\CP_R = (\CQ_R, \CH_R)$ (4.12).
The aim of this sub-section is to  describe the general principles of the study of  the family of walks  
$\bW_{2s}^{(u)}(\rD, \bar d_R; \langle \CG^\star\rangle_s,\CH_R; \U)$ (4.16). In this subsection, we assume
that $I=0$ and there is no proper and mirror cells at $\bb$.

Let us consider the  walks that have only one imported cell generated by one BTS-instant 
determined by a couple $(\t_1, \phi)$, where
the marked instant $\t_1 $  is either $z_1$ or $ y_1$ and  denote it by 
$\bW_{2s} ^{(u)}( \rD, d_1, d_2; \la \CG^\star_{\t_1}\ra_s, \phi, \U)$.
 We choose  a tree of $\t_1$ edges $ \CT_{\t_1}$ 
and perform over it  a partial  chronological run $\fR_{\CT}^{(\t_1)}$, 
$$
\fR^{(\t_1)}_{\CT} = \left\{ \fR_\CT(t), \ 1\le t\le \xi_{\t_1}-1 \right\}.
\eqno (5.16)
$$ 
Going along $\fR_{\CT}^{(\t_1)}$,  we construct a sub-walk $\CW_{[0, \xi_{\t_1}-1]}$ according to  the self-intersections of 
$\langle \CG^\star  \rangle_s$. During this run, 
we  choose realizations of the values in the red windows $\CG_\circ$ 
with the help of the general filtration
procedure described in the proof of Lemma 3.2 (see subsections 5.1 and 5.2) 
and follow the rules $\U$ at the non-marked steps. 
Then we can write that 
$$
\CW_{[0,t_1]} =
\CW\left( \CT_{\t_1}, \langle \CG^\star_{\t_1}\rangle_s, 
\langle \CG_\circ\rangle_{[0, t_1]}, \U\right),
\eqno (5.17)
$$
where   $\langle \CG_\circ\rangle_{[0, t_1]}$ indicates  a realization of red edge-windows
of $\CG$ 
on the time interval  $[0, t_1] = [0,\xi_{\t_1}-1]$.

\subsubsection{Filtration of  values $\ell, \vp$ and $\psi$}
Let us assume 
that the instant $\t_1= y_1$ fills  the edge-window of the second arrival $\fa_2= \fa_2(\fii)$
of a simple self-intersection $\fii^{(2)} = \fii(\fo_1, y_1)$. 
As we have pointed out (see subsection 4.3), this edge-window can represent  either $o$-edge
or $p$-edge or $q$-edge of $\CG$. 

To construct a continuation of the sub-walk $\CW_{[0, t_1]}$ 
at  the marked instant of time $\xi_{\t_1}$,
we have to choose one of the vertices $\gamma_i$ of $g(\CW_{[0, t_1]})$ that  verify the condition that the
chosen  vertex $\gamma$  
is situated on the distance of $\phi$ non-marked steps from $\bb$, where 
$\phi $ denotes either $\ell$ or  $\psi_1$.  

\vs
Let us consider the case when the second arrival $\fa_2$ has to verify the $o$-property.
Then  the  collection  $\G  = \G^{(\ell, (o))}_{t_1}(\bb)$ of  admissible vertices $\g_i$ 
is such that 
$$ 
\bigsqcup_{\ell=1}^u \ \G^{(\ell, (o))}_{t_1}(\bb) = \bV^{(o)}_{t_1},
$$
 where $\bV^{(o)}_{t_1}$
is the set of $t_1$-open vertices. 
Clearly, 
$
\sum_{\ell=1}^u \, \vert \G^{(\ell, (o))}_{t_1}(\bb)\vert \le 2u.
$

\vs
Let us consider the case when $\fa_2$ verifies the $p$-condition. Then the admissible vertices
 $\g_i$ belong to the sub-set  $ \D^{(\ell)}(\a)\subseteq \D(\a)$, $\a =\CW(\xi_{\t_1}-1)$, of vertices
 that are situated on the distance of $\ell$ non-marked steps from $\bb$.
 Therefore
 $$
 \sum_{\ell=1}^u \ \vert \G^{(\ell, (\D))}_{t_1}(\bb)\vert  = \sum_{\ell=1}^u \vert \D^{(\ell)}(\a) \vert \le \rD.
 \eqno (5.18)
 $$
 The same reasoning can be applied to the case of $q$-condition imposed on $\fa_2$. 
 
 Summing up, we can write that 
   $$
\sum_{\ell=1}^u \ \vert \G^{(\ell, \fu_2)}_{t_1}(\bb)\vert \le 
\vert \G^{(\fu_2)}\vert = 
  \begin{cases}
 2u , &\text{if  $\fu_2 = (o)$}, \\
 \rD , &\text{if  $\fu_2 = (\D)$}, \\
k_0, &\text {if $\fu_2 = (\Lambda)$.}
\end{cases} 
 \eqno (5.19)
 $$

Let us consider the secondary imported cell generated by the remote BTS-instant $\t_1=y_1$.
In the graph $g(\CW_{[0, \xi_{y_1}-1]})$,  the way from $\bb$ to $\bb$ by $\psi_1$ non-marked steps is uniquely determined by
the rule $\U$. 
Therefore 
we can write that 
$$
\sum_{\psi_1=1}^u\,  \vert \G^{(\psi_1)}_{t_1}(\bb) \vert  \le 1.
\eqno (5.20)
$$
The same is true for  all subsequent secondary imported cells generated by $y_1$, and therefore (5.20)
holds for all values of $\psi_i$ with $1\le i\le f''_1$.

Now it is clear that in the case when the BTS-instant represented by  $\t_1 = z_1$ is the local one,
 then for all imported cells generated by $z_1$ the following  equality is valid,
 $$
\sum_{\vp_k=1}^u\  \vert \G^{(\vp_k)}_{t_1} (\bb)\vert \le 1, \quad 1\le k\le f'_1.
\eqno (5.21)
$$

\vs
\vs
Let us consider the case when the $y$-label of  $(y_1,\ell)$ is attributed to the edge-window the of the third arrival
$\fa _3$ at a vertex of $\bM_3'$ of $\CG$. 
Assume that $\fa_3$ verifies $\D$-condition. Repeating the proof of Lemma 5.1, attribute to $\CW(\xi_{y_1})$ 
a vertex $\g$ that belong to $\D^{(\ell)}(\a)$, $\a= \CW(\xi_{y_1}-1)$ and take a marked instant $\t'$
that determines $\g$. Taking into account the last inequality of (5.18), we see that 
after the sum over $\ell$, the total number of admissible values $\t'$ is bounded by $D$. 
The remaining part of the reasoning is the same as before. 
The case when $\fa_3$ verifies the $\Lambda$ condition can be studied by the similar argument.
Then (cf. (5.4))
$$
\sum_{\ell=1}^u 
 \# \left\{  \langle (\fo_1,\fo_2)\rangle_{\CW}, \ \CW \in 
 \sbW_{2s}\left(\fii^{(3)}\left(\fo_1, \fo_2, \left[ { \stackrel{(y_1,\ell)}{\fu_3}} \right] \right)\right)  \right\} 
 \le 
  \begin{cases}
 s \rD , &\text{if  $\fu_3 = (\D)$}, \\
s k_0, &\text {if $\fu_3 = (\Lambda)$.}
\end{cases}
 \eqno (5.22)
 $$
 It is clear that the sum over the values $\psi_i$ attributed to $y_1$ put into the edge-window of $\fa_3$ 
 give the upper bounds of the form (5.20). The same is true for the sums over variables $\phi_i$
 in the case of $\t_1 = z_1$.

\vs
Finally, let us consider the cases when the $y$-label is attributed to an edge-window $\fe$ of $\CG$
that represents  the third arrival $\fa_3$ to the corresponding vertex $v$. If $v\in \bM_2'$, then 
$\fe$ is the $u$-edge. If $v\in \bM''_3$, then $\fe$ represents the blue $\mu$-edge. If $v$ is a $\nu$-vertex,
then $\fe$ is the black $\nu$-edge. In all of these three cases, the vertex $\g = \CW(\xi_{y_1})$ is determined
uniquely by the sub-walk $\CW_{[0, t_1]}$ and therefore 
$$
\sum_{ \phi=1}^u \ \vert \G^{(\phi)}_{t_1} (\b) \vert \le 1.
\eqno (5.23)
$$
The same is true in the cases when $y_1$ is attributed to the edge-windows of the $k$-th arrivals $\fa_k$.  

\vs
Summing up, we see that after the summation over $\phi$, the number of realizations of the values
in the red edge-windows is bounded by  the same expressions 
(5.19) and (5.22)  as  in the ordinary case without $y$-labels considered in the previous 
sub-section.

\vs

Let us consider a walk that is constructed with the help of a diagram $\CG^\star$
that has two $y$-labels with the values to $(y_1, \ell_1)$ and $(y_2,\ell_2)$, respectively
that belong to the same vertex $v\in \CV(\CG^\star)$. 
We assume that $y_1<y_2$. 
In this case the filtering of the values of $\ell_1$ is performed as before, and the sum over $\ell_1$ gives
one of the estimates  (5.19)-(5.23) for the number of admissible vertices. 
When arrived at the instant of time $\xi_{y_2}-1$, the vertex $\g$ is determined by the construction 
of the sub-walk $\CW_{[0, \xi_{y_2}-1]}$, and $y_2$ is attributed  to the arrival $\a_k$ at $v$ 
with $\k\ge 3$. Then the sum over $\ell_2$ gives us the following upper bound, 
$$
\sum_{\ell_2=1}^u \ \vert \G^{(\ell_2)}_{t_2}(\bb)\vert \le 1.
\eqno (5.24)
$$

\vs
\vs

To get the final account on the sums over variables $\ell$, $\psi$ and $\vp$,
let us assume that the diagram $\CG_{\CQ}$ contains $\breve \mu_2'$ $y$-labels attributed to 
$\breve r $ 
$o$-edges, $\breve p$ $p$-edges and $\breve q$ $q$-edges of $\CG$, and $\breve \mu_3'$ $y$-labels 
attributed to the third arrivals at the vertices of $\bM_3'$. Then 
$$
\prod _{j=1}^J\, \sum_{\ell_j=1}^u \  \prod_{i=1}^{f''_j}\  \sum_{\psi_i=1}^u  \
\ \prod_{k=1}^K \ \prod_{l=1}^{f'_k} \ \sum_{\vp_l=1}^u \ 
\vert \G^{(\phi)}(\CG_\CQ)\vert \le 
\, (2u)^{\breve r}\, \rD^{\breve p} \, k_0^{\breve q}\ (2s(\rD+k_0))^{\breve \mu_3'} = \breve \fF(\CG),
\eqno (5.25)
$$
where we denoted by $\vert \G^{(\phi)}(\CG_\CQ)\vert$ the product over all cardinalities of the sets of admissible vertices
presented. 
 
 
\subsubsection{Underlying Dyck paths and trees}

Regarding a walk $\CW_{2s} \in \bW_{2s} ^{(u)}( \rD, d_1, d_2; \la \CG^\star_{\t_1}\ra_s, \phi, \U)$, let us denote
by $t_1$ and $t_2$ the instants of time such $\CW_{2s} (t_i)=\bb$, $\t_1 = \xi_{\t_1}$ and $t_2$ represents
the arrival $\fa'$ at $\bb$ by the corresponding imported cell. According to the definition of 
 variables $(\t_1,\phi)$ that the sub-walk 
 $\CW_{[t_1+1, t_2-1]}$ is such that it has 
 $\phi$ non-marked
steps and after each of such step it has a tree-type sub-walks $\tilde \CW^{(k)}$ 
that are reduced by $\hat \CR$ to the empty walks. This means that the Dyck structure of $\CW_{2s}$
is such that the nest cell $\varpi' = \varpi(\t_1, \phi)$ that corresponds to $\fa'$ is obtained after $\phi$ descending steps
are performed in $\CT_{\t_1}$ from the vertex $\u = \fR_{\CT_{\t_1}}(\xi_{\t_1})$. Then 
the vertex $\u'$ of $\CT_s$ of $\CW_{2s}$ is determined, where 
the exit sub-cluster of $d_2$ edges is to be placed.  

Therefore $\CT_s$ is to be of the following structure: consider  a tree $\CT_{\t_1}^{[l]}$ that contains $\t_1$ edges and has the descending
part of the length $l\ge \phi$ (see Figure 1). We will say that the vertex $\u_1 = \fR_\CT(\xi_{\t_1})$ is on the distance $l$ from the 
root of the tree. 
On the $l$ vertices of the descending part, construct a 
  realization of $l$ sub-trees of the total number of $s-\t_1$ edges 
such that at the nest cell $\varpi'$ there exists a tree with the root sub-cluster of $d_2$ edges
(see Section 6 for more details). We denote such a family of trees by $\CT^{\{l\}}_{s-\t_1}(d_2,\t_1, \phi)$. 

  We see that the family of trees that corresponds to the Dyck paths generated by the elements of 
  $\bW_{2s}^{(u)}(d_2, \langle \CG^\star\rangle_s, \phi, \U)$ is given by the following expression,
  
  $$
  \stackrel{ \displaystyle \bigsqcup_{l=1}^{\t_1} \left\{ \underbrace{  \CT_{\t_1}^{[l]} \otimes
   \CT_{s-\t_1}^{\{l\}} (d_2, \t_1,\phi)  }\right\}
 \displaystyle  = \bT_s^*,
  }{ u\hbox{-condition}}
  $$
  where we denoted
  $$
  \bT_s^* = \bT^{(u)}_s(d_2,\t_1, \phi).
 \eqno (5.26)
  $$
  The under-brace 
 with $u$-condition means that we construct the sub-trees  $\CT_{\t_1}^{[l]}$ and
 $\CT_{s-\t_1}^{\{l\}} (d_2, \t_1,\phi)$ in the way that the height of the common tree obtained $\CT_{s}$ 
 attains $u$. In Section 6 we describe in details the set of such trees.

\vs

Ignoring 
the condition that the vertex $\bb = \CW_{2s}(x_1)$ has the first proper cell $x_1$ with $ d_1$ edges of the exit sub-cluster,
 we can write  that 
 $$
   \vert \bW_{2s}^{(u)}( d_2, \langle \CG^\star\rangle_s, \phi, \U)\vert =
 \sum_{l=1}^{\t_1} \ \sum_{\CT_{\t_1}^{[l]}}   \ 
 \  \sum_{\langle \CG_\circ \rangle_{[0, \xi_{\t_1}-1]}} \ \ 
 \vert \G^{(\phi)}_{\t_1}(\bb)\vert 
 $$
 $$
 \times
 \sum_{ \CT_{s-\t_1}^{\{l\}}(d_2, \t_1, \phi)} 
 \ \  \sum_{ \langle \CG_\circ \rangle^{(\t_1, \phi)}_{[\xi_{\t_1}+1, 2s]}} 1.
 \eqno (5.27)
 $$
  In the last sum of (5.27), we have denoted
by 
$\langle \CG^{{\tiny{\hbox{(r)}}}}\rangle^{(\t_1, \phi)}_{[\xi_{\t_1}+1, 2s]} $  a realization of the values
in the red edge-windows of $\CG$ on the time interval $[\xi_{\t_1}+1, 2s]$; this realization also depends 
on $\CT_{\t_1}^{[l]}$,
 and $\CT_{s-\t_1}^{\{l\}}(d_2, \t_1, \phi)$. We denote by $r_1, p_1, q_1, 2\mu_3'(1)$ and $r_2, p_2, q_2, 2\mu'_3(2)$ the number of red windows
 to be determined during the time intervals $[0, \xi_{\t_1}-1]$ and  $[\xi_{\t_1}+1, 2s]$, respectively.
 It is clear that 
 $r_1=r_2 +1 =r$. The same equalities are verified by other red edge-windows.

  \vs 
  
  Let us forget for the moment that the walks of the left-hand side of (5.26) are such that their Dyck paths
  have the height $\th^*=u$. 
  Then it is not hard to show that the following upper bound holds for any given value of $\phi$, 
 $$
 \sum_{ \CT_{s-\t_1}^{\{l\}}(d_2, \t_1, \phi)
 }\   1 \ \ Ê\le \  
 e^{- \eta d_2} \,  \sum_{ \CT_{s-\t_1}^{\{l\}}
}\  
1,
\eqno (5.28)
$$
where $\eta = \ln(4/3)$ (see Lemma 6.1 of the next section). 

\vs

Taking into account that  the upper bound (cf. (5.5)) 
$$
\vert \left\{ \langle \CG_\circ\rangle^{(\t_1, \phi)}_{[\xi_{\t_1}+1,2s]}\right\} \vert 
\le  \vert \CG_\circ^{(2)}\vert =  
(2u)^{r_2} D^{p_2} k_0^{q_2}
(s(D+k_0))^{\mu_3'(2)}
\eqno (5.29)
$$
 is  uniform  with respect to $\phi$, 
 we use (5.19) and deduce from (5.27) that 
   $$
\sum_{\phi=1}^u\,   \vert \bW_{2s}^{(u)}(d_2, \langle \CG^\star\rangle_s, \phi, \U)\vert \le 
  \sum_{l=1}^{\t_1} \ \sum_{\CT_{\t_1}^{[l]} }  \ \  \sum_{\langle \CG_\circ\rangle_{[0, \xi_{\t_1}-1]}} \ 
  \sum_{ \CT_{s-\t_1}^{\{l\}}} e^{-\eta d_2} \cdot  
  \vert \CG_\circ^{(2)}\vert
 \cdot   
\sum_{\phi=1}^u\  
\vert \G^{(\phi)}_{\t_1}(\bb)\vert \, .
 $$
 Taking into account that 
 $$
  \sum_{l=1}^{\t_1} \
   \sum_{\CT_{ \t_1}^{[l]} } \ 
  \sum_{ \CT_{s-\t_1}^{\{l\}} } \ 1\ = \rt_s,
  $$
  we finally get   the following upper bound,
  $$
\sum_{\phi=1}^u\,  \vert \bW_{2s}^{(u)}(d_2, \langle \CG^\star \rangle_s, \phi , \U)\vert
\le \ e^{- \eta d_2}\,  \rt_s
\  \vert \CG_\circ\vert ,
\eqno (5.30)
$$
where  $\vert \CG_\circ \vert $  is  given by 
  (5.5). 

\vs
The $u$-condition of (5.26) makes the use of (5.28) more complicated. 
The  proof
of the exponential estimates of the form of (5.30)  is presented in Section 6.




\section{Catalan trees and D-lemma}

\subsection{Exponential bound for Catalan trees}

The following statement slightly improves corresponding results of \cite{K01} and \cite{K}.
\vs 

{\bf Lemma 6.1.} {\it Consider the family  of Catalan trees constructed with the help of $s$ edges
and such that the root vertex $\varrho$ has $d$ edges attached to it and denote by $\tilde \rt_{s}{(d)}$ its cardinality.
Then the  upper bound 
$$
\tilde \rt_ {s}{(d)}   \le  e^{- \eta d}  \, \rt_s , \quad \eta = \ln (4/3)
\eqno (6.1)
$$
is true for any given integers $d$ and $s$ such that $1\le d\le s$. }
\vskip 0.3cm
{\it Proof.}  By definition,
$$
\tilde \rt_{s}{(d)}= \sum_{u_1+\cdots+u_{d-1} +u_d=s-d} \rt_{u_1}\ \cdots \  \rt_{u_{d-1}}\, \rt_{u_d},
\eqno (6.2)
$$
where the sum runs over integers $u_j\ge 0$.
We will say that (6.2) represents the  number of Catalan trees of $s$ edges that have 
a {\it sub-cluster} attached to the root $\varrho$ that contains  $d$ edges.

Using the fundamental recurrence relation
$$
\rt_{s+1} = \sum_{j=0}^s \, \rt_j\, \rt_{k-j}, \quad s\ge 0, \quad \rt_0=1,
\eqno (6.3)
$$
we can rewrite (6.2) in the following form,
$$
\tilde \rt_{s}{(d)}= \sum_{v=0}^{s-d}\  \ \sum_{u_1+\cdots +u_{d-2}+v=s-d} \rt_{u_1}\ \cdots \  \rt_{u_{d-2}}\
\left( \sum_{u_{d-1}+u_d = v}\rt_{u_{d-1}}\, \rt_{u_d}\right)
$$
$$
=  \sum_{u_1+\cdots +u_{d-2}+u_{d-1}=s-d+1} \rt_{u_1}\ \cdots \
\rt_{u_{d-2}}\  \rt_{u_{d-1}}
-
\sum_{u_1+\cdots +u_{d-2}=s-d+1} \rt_{u_1}\ \cdots \
\rt_{u_{d-2}}.
\eqno (6.4)
$$
Relation (6.4) implies  that
$$
\tilde \rt_{s}{(d)} = \begin{cases}
\tilde \rt_{s}{(d-1)} - \tilde \rt_{s-1}{(d-2)}, & \text{for all  $3\le d\le s$,}\\
\rt_{s-1}, & \text{for $d=2$ and $s\ge 2$,}\\
\rt_{s-1}, & \text{for $d=1$ and $s\ge 1$,}\\
\end{cases}
\eqno (6.5)
$$
where the last two relations are easy to obtain directly.
It follows from  (6.5)  that
$$
\tilde \rt_{s}{(d)}\, \le\,  t_{s-1},\quad {\hbox {for all  }} 1\le d\le s.
\eqno (6.6)
$$

Let us return to  (6.4) and rewrite it in the form
$$
\tilde \rt_{s}{(d)}\,  = \, \sum_{v=0}^{s-d} \  \tilde \rt_{s-v-1}{(d-1)}\cdot \rt_{v}, \quad 2\le d\le s.
\eqno (6.7)
$$
Assuming that $d\ge 3$ and applying (6.6) to the right-hand side of (6.7), we get inequality
$$
\tilde \rt_s{(d)} \, \le \, \sum_{v=0}^{s-d}\   \rt_{s-2-v}\, \rt_v\,  \le \, \rt_{s-1} - \rt_{s-2}.
\eqno (6.8)
$$
Using the explicit expression for the Catalan numbers (3.2), it is easy to show that
$$
2\rt_s\le \rt_{s+1}\le 4 \rt_s,\quad s\ge 1.
\eqno (6.9)
$$
Then  we deduce from (6.8) that 
$$
\tilde \rt_s{(d)} \le \left( {3\over 4}\right) \, \rt_{s-1}, \quad d\ge 3.
\eqno (6.10)
$$
We see that (6.1) holds for  $d=3$, $s\ge 3$.
The standard reasoning by recurrence based on (6.7) proves the bound
$$
\tilde \rt_{s}{(d)} \le \left( {3\over 4}\right)^{d-2}\, \rt_{s-1}, \quad 3\le d\le s.
\eqno (6.11)
$$
Remembering (6.5), we see that  (6.11) is true in the case of $d=2, s\ge 2$ also. 
Using 
the first inequality of (6.9), we  deduce from (6.11) the upper bound  (6.1). \mbox{Lemma 6.1} is proved. $\Box$


\subsection{Tree-type walks with  multiple edges}

Given a Catalan tree $\CT_s$, let us  denote by $\RN^{(2)}(\CT_s)$ the number of choices
of two edges of $\CT_s$ that have the same parent vertex.
Clearly, the sum $\RN^{(2)}_s= \sum_{\CT_s\in \bT_s} \RN^{(2)}(\CT_s)$ represents the number of
even closed tree-type  walks of $2s$ steps
whose graphs have exactly one $p$-edge passed four times and $s-2$ grey edges  passed two times.

\vskip 0.3cm
{\bf Lemma 6.2.}  {\it
For any given $s\ge 2$, the following relations hold, 
$$
\RN^{(2)}_s=  { (2s)!\over (s-2)!\, (s+2)!}= \left( s - {3s\over s+2}\right) \rt_s ,
\eqno (6.12)
$$
and therefore $\RN^{(2)}_s\le s \rt_s$;  the   lower bound
$\RN^{(2)}_s \ge (s \rt_s)/ 2  $ is true for all $s\ge 4$.
}

\vskip 0.3cm
{\it Proof.}
It is not hard to see  that
$$
\RN^{(2)}_s = \sum_{u+v_1+v_2+v_3=s-2} (2u+1)\, \rt_u \rt_{v_1} \rt_{v_2} \rt_{v_3}, \quad s\ge 2,
\eqno (6.13)
$$
where the sum runs over all integers $u\ge 0$ and $v_i\ge 0$.
Then the generating function $\Phi^{(2)}(\vsi)= \sum_{k\ge 0} \RN^{(2)}_k \vsi^k$ with $\RN^{(2)}_0=\RN^{(2)}_1=0$
is given by relation
$$
\Phi^{(2)}(\vsi) =  2\vsi^3  \ \rf'(\vsi)\,  \rf^3(\vsi) + \vsi^2\, \rf^4(\vsi),
\eqno (6.14)
$$
where
$\rf(\varsigma) = \sum_{s=0}^\infty \rt_s\, \varsigma^s$ is the generating function of the Catalan numbers.
It follows from (6.3)  that $\rf(\vsi)$ verifies  equation
$$
\rf(\vsi) = 1 + \vsi\,  \rf ^2(\vsi),
\eqno (6.15)
$$
and then
$$
\rf(\varsigma) = {1-\sqrt{1-4\varsigma} \over 2\varsigma} \ .
\eqno (6.16)
$$
It follows from (6.15) that 
$$
\rf'(\vsi) = \rf^2(\vsi) + 2 \vsi\, \rf'(\vsi)\, \rf(\vsi).
\eqno (6.17)
$$
Using (6.15) and (6.17) and taking into account that
$$
\rf'(\vsi) = \sum_{k=0}^\infty\,  (k+1)\rt_{k+1}\, \vsi^k,
\eqno (6.18)
$$
one can easily derive from (6.14) relation (6.12). Lemma 6.2 is proved.

\vskip 0.5cm
Let us consider a general  case of the tree-type walks such that their graphs have
exactly  one edge passed  $2l$ times and other $s-l$ edges passed two times.
The  number of such walks  $\RN^{(l)}_s$ is given by the
 the total number of possibilities to mark $l$ edges that have the same parent vertex at the Catalan trees. 
 Similarly to (6.13), we can write that
$$
\RN^{(l)}_s = \sum_{u+v_1+\dots + v_{2l-1}= s-l} \ (2u+1) \, \rt_u \rt_{v_1} \rt _{v_2} \dots  \rt_{v_{2l-1}}.
\eqno (6.19)
$$
The corresponding generating function $\Phi^{(l)}(\vsi)$ is given by relation
$$
\Phi^{(l)}(\vsi) 
= 2\vsi^{l+1}\,  \rf'(\vsi) \, \rf^{2l-1}(\vsi) + \vsi^l\,  \rf^{2l}(\vsi)
\eqno (6.20)
$$
and therefore
$$
\Phi^{(l)}(\vsi)
= \vsi^l\, \rf^{2l-1}(\vsi) \, \left( {2\over \sqrt{1-4\vsi}} - \rf(\vsi)\right).
\eqno (6.21)
$$
Using (6.15) and (6.18), it is not hard to   deduce from (6.20) that
$$
\RN^{(l)}_s \le 2^l s \,  \rt_s, \quad 2\le l\le s.
\eqno (6.22)
$$
This inequality means that the constant $\eta = \ln (4/3)$
of the exponential estimate (6.1) can be considerably increased for large values of $s$ and $d$.

\vs 
Similarly to (6.12),  
it is not hard to show that
$$
\RN^{(3)}_s = \, {(2s)!\over (s-3)!\, (s+3)!} \, =\  \rt_s \left( s-8 - {36s +48\over s^2 + 5s + 6}\right)
\eqno (6.23)
$$
and therefore $\RN^{(3)}_s \le s \rt_s$. 
Regarding $\RN^{(1)}_s$ as a number of trees with one marked edge, we see that 
$$
\RN^{(1)}_s = s\rt_s = {(2s)!\over (s-1)!\, (s+1)!}.
\eqno (6.24)
$$ 
Relation (6.24) indicates a natural connection  between expressions (3.2) and  (6.12), (6.23). 
It is natural to assume that the following equality 
$$
\RN^{(l)}_s = {(2s)!\over (s-l)!\, (s+l)!}
 \eqno (6.25)
$$
is true for all values of $l\in [1,\dots, s]$.
However, relation (6.21)  seems to be not so convenient  to prove  (6.24). 
It would be useful to find  representations of  $\RN^{(l)}_s$  different from 
(6.19) and (6.20).
Let us note that 
 (6.25) would imply a useful upper bound $\RN^{(l)}_s\le s\rt_s$ for all $s$ 
and $l\le s$ that is more strong than (6.20). Clearly, the last upper bound is in complete accordance
with the Galton-Watson view of Catalan trees.









\subsection{$D$-lemma}  

Our aim  is to  prove the exponential-type estimate of the form (4.18). For simplicity,
we consider the case when the non-trivial tree-type sub-clusters with $d_i>0$
correspond either to the proper or to the imported cells at $\bb$. 
 The case of mirror cells will be considered at the end of the present subsection. 
\vs

{\bf Lemma 6.3}
{ \it Consider a family of walks 
$\bW_{2s}^{(u)}(\rD, \bar d_R; \langle \CG^\star_R\rangle_s, \CH_R, \U)$ (4.16). Then for any integer $m\ge 0$}
$$
\sum_{u=1}^s \ u^m \, \ \sum_{\CH_R} \, \vert \bW_{2s}^{(u)}( \rD, \bar d_R;  \langle \CG^\star_R\rangle_s, \CH_R, \U)\vert
\le A_s^{(m)}
\cdot 4^R \, (D_R+1)\, e^{-\eta D_R}\, \rt_s,
\eqno (6.26)
$$
{\it where we denoted } 
$$
A_s^{(m)}= A_s^{(m)}(\CG, \rD,k_0)
\,=
 \, s^{(m+r)/2} \  B_{m+r} \  2^r \,  \rD^p\, k_0^q \, (s(\rD+k_0))^{\mu_3'},
\eqno (6.27)
$$
{\it with }  
$$
 B_{m+r} = \sup_{s\ge 1}\ B_{m+r}^{(s)}, \quad B_r^{(s)}=  {1\over \rt_s} \, \sum_{u=1}^s \
\left({u\over \sqrt s} \right)^{r} \cdot  \vert \bT^{(u)}_s \vert
\eqno (6.28)
$$
{\it and }
$$
\bT^{(u)}_s = \left\{ \CT_s: \ \th^*(\CT_s)=u\right\}.
\eqno (6.29)
$$

We prove Lemma 6.3 by recurrence. Let us introduce the following denotations related with (6.29).
Given natural $a$ and $u$, let us denote by $\dot \bT^{(u)}_a = \bT^{(u)}_a$ a family of Catalan trees $\CT_a$ of $a$ edges 
such that for any such a tree the corresponding Dyck path has the height $u$, $\th^*(\CT_a)=u$. In this case we simply say that 
$\CT_a$ has the height $u$. 
Let $\ddot \bT^{(u)}_a$ be a family of Catalan trees such that $\th^*(\CT_a)\le u$.
We  denote
 the cardinalities of these sets of trees by $\dT^{(u)}_a$ and $\ddT^{(u)}_a$, respectively, 
$$
\dT^{(u)}_a = \vert \left\{ \CT_a\in \bT_a: \ \th^*(\CT_a)= u\right\}\vert
\quad \hbox{and}\quad 
\ddT^{(u)}_a =  \vert \left\{ \CT_a\in \bT_a: \ \th^*(\CT_a)\le u\right\}\vert \, . 
\eqno (6.30)
$$
We assume that   $\dT^{(u)}_a = 0$ when $a<u$,  $\dT^{(0)}_a = \ddT^{(0)}_a=0$ for any $a\ge 1$  and
 $\dT^{(0)}_0 = \ddT^{(0)}_0=1$.

We also denote by 
$\dot \bT^{(u, [l])}_a$ the family of trees of the height $u$ that have the descending part of length $l$.
The families $\dot \bT^{(u,\{l\})}_a$, $\ddot \bT^{(u, [l])}_a$ and 
$\ddot  \bT^{(u,\{l\})}_a$ as well as the families  of trees that have a sub-cluster at one of the nest cell  
are determined in obvious manner. This will be clarified in the computations that follow. 









\subsubsection{The initial step of recurrence}

In this subsection, we study a family  of walks 
$\bW^{(u)}_{2s} ( d, \langle \CG^\star_{\t_1} \rangle_s, \phi, \U)$
given by (4.16) with $R=1$
such that the first non-trivial sub-cluster of the tree-type part $\tilde W$ with $d$
edges is attached to the nest cell  $\varpi_1 = (\t_1, \phi_1)$  of the corresponding tree
(see also  (5.26)).
The variable $\t_1$ denotes  one of the three possible values that are  $x_1$, $y_1$, or $z_1$
and $\phi_1=\phi $ is equal to  $0$, $\ell_1$ or $\vp_1$, respectively. 
If $\t_1 = y_1$, then
$\phi$ can also be equal to $\psi_1$. 

According to the definitions of (6.30), the subset of trees (5.26) can be represented as follows,
$$
\bT^{(u)}_s (d, \t_1, \phi) = 
$$
$$
\bigsqcup_{l=1}^{\t_1} \left\{ \dot \CT_{\t_1}^{(u, [l])} \otimes 
\ddot \CT_{s-\t_1}^{(u, \{l\})}(d,\t_1,\phi)\right\}
\sqcup \ \bigsqcup_{l=1}^{\t_1} 
 \left\{ \ddot \CT_{\t_1}^{(u-1, [l])} \otimes \dot \CT_{s-\t_1}^{(u, \{l\})}(d,\t_1,\phi)\right\} \, . 
\eqno (6.31)
$$
The accurate version  of  (5.27) is given  then by the following  two sums,
$$
   \vert \bW_{2s}^{(u)}( d, \langle \CG^*\rangle_s, \phi, \U)\vert
   $$
   $$
    =\ 
 \sum_{l=1}^{\t_1} \ \sum_{\dot \CT_{\t_1}^{(u,[l])}}   \ \  
 \sum_{\langle \CG_\circ \rangle_{[0, t_1]}} \ \ 
 \vert \G^{(\phi,\fu)}_{t_1}(\bb)\vert \, 
 \sum_{ \ddot \CT_{s-\t_1}^{(u,\{l\})}(d, \t_1, \phi)} 
 \ \  \sum_{ \langle \CG_\circ\rangle^{(\t_1, \phi)}_{[t_1+2, 2s]}} 1
 $$
$$
 + \ \sum_{l=1}^{\t_1} \ \sum_{\ddot \CT_{\t_1}^{(u-1,[l])}}   \ \  
 \sum_{\langle \CG_\circ \rangle_{[0, t_1]}} \ \ 
 \vert \G^{(\phi,\fu)}_{t_1}(\bb)\vert \, 
 \sum_{ \dot \CT_{s-\t_1}^{(u,\{l\})}(d, \t_1, \phi)} 
 \ \  \sum_{ \langle \CG_\circ \rangle^{(\t_1, \phi)}_{[t_1+2, 2s]}} 1,
 \eqno (6.32)
 $$
 where we denoted $t_1 = \xi_{\t_1}-1$. 
 It should be noted that  
 the sum $ \  \sum_{ \langle \CG_\circ\rangle^{(\t_1, \phi)}_{[t_1+2, 2s]}} 1$
is bounded from above by  $\vert \CG_\circ^{(2)}\vert $ (5.29)  uniformly with respect to $\phi$. 
 
 \vs

Let us estimate the cardinality of the family $\ddot \bT_{s-\t_1}^{(u,\{l\})}(d, \t_1, \phi)$, 
$$
 \vert \ddot \bT_{s-\t_1}^{(u,\{l\})}(d, \t_1, \phi)\vert \ = \ 
 \sum_{ \ddot \CT_{s-\t_1}^{(u,\{l\})}(d, \t_1, \phi)} 1\  .  
 $$
Using (6.30), we can write that
$$
  \vert \ddot \bT_{s-\t_1}^{(u,\{l\})}(d, \t_1, \phi)\vert \ = \ 
\ \sum_{ \stackrel{   \bar b_d, \bar c_{l},} 
{ \vert \bar b_d\vert + \vert \bar c_{l} \vert  = s-\t_1} 
}\ \ \ddt^{(u-l+\phi-1)}_{\{ d, \bar b_d\}} \ 
\ddt^{(u)}_{[l+1, \bar c_l]}\ ,
\eqno (6.33)
$$
where $\bar b_d = (b_1, \dots, b_d)$, $\vert \bar b_d\vert = b_1+\dots+ b_d$, 
$\bar c_l = (c_0, c_1, \dots, c_l)$, $\vert \bar c_l \vert = c_0+c_1+ \dots+ c_l$, 
$$
 \ddt^{(u-l+\phi-1)}_{\{ d, \bar \b_d\}}= \prod_{k=1}^d \, \ddT^{(u-l+\phi-1)}_{b_k}
$$
and 
 $$
\ddt^{(u)}_{[l+1, \bar c_l]} = \prod_{i=0}^{l} \ \ddT^{(u-l+i)}_{c_i}.
 $$
 Regarding (6.34), we can use the following obvious inequalities,
 $$
 \ddT^{(u-l+\phi-1)}_{b_k} \le \ddT^{(u-l+\phi)}_{b_k}\le \rt_{b_k}
 \eqno (6.34)
 $$
 and write that for a given $b\ge d$, due to Lemma 6.1, 
 $$
 \sum_{\bar b_d:\, \vert \bar b_d\vert = b-d}\ddt^{(u-l+\phi-1)}_{\{ d, \bar b_d\}}\le
  \sum_{\bar b_d:\, \vert \bar b_d\vert = b-d} \prod_{k=1}^d \, \rt_{b_k}= \t_b(d) \le e^{-\eta d}\, \rt_b.
  \eqno (6.35)
$$ 
Then
$$ 
 \vert \ddot \bT_{s-\t_1}^{(u,\{l\})}(d, \t_1, \phi)\vert \ \le \ 
 \sum_{b=d}^{s-\max\{\t_1, 1\}}
 \ e^{-\eta d}\, \rt_b\ 
 \ \sum_{ \stackrel{   \bar c_{l},} 
{  \vert \bar c_{l} \vert  = s-\t_1-b} 
}\ \ 
\ddt^{(u)}_{[l+1, \bar c_l]}\ .
\eqno (6.36)
$$
Now we can perform the sum over $\phi$ in the right-hand side of (6.32) and get
with the help of (5.19), the following upper bound (cf. (5.5)),
$$
\sum_{ \phi=1}^u\ \ \sum_{ \la\CG_\circ\ra_{[0, t_1]} } \ 
\vert \G^{(\phi, \fu)}_{t_1}(\bb)\vert \cdot  \vert \CG_\circ^{(2)}\vert \le
 (2u)^r\, \rD^p\, k_0^q\, (s(\rD+k_0))^{\mu_3'} = \vert \CG_\circ\vert .
$$
The sum over $\phi$ being  performed,   the last expression of (6.36) can be
 inserted into the first term of the right-hand side of (6.32). This   gives  the following sum,
$$
 \sum_{l=1}^{\t_1} \ \sum_{\dot \CT_{\t_1}^{(u,[l])}}   \ \sum_{   \bar c_{l},
  \vert \bar c_{l} \vert  = s-\t_1-b} 
\ \ 
\ddt^{(u)}_{[l+1, \bar c_l]}\ = \vert \grave \bT^{(u, \t_1)}_{s-b}\vert ,
\eqno (6.37)
$$
where we have  denoted by $\grave \bT^{(u, \t_1)}_{s-b}$ the family of all trees  $\CT_{s-b}$ such that the height
$\th^*(\CT_{s-b})=u$ is attaint for the first time during the time interval $[0, t_1]= [0, \xi_{\t_1}-1]$.

\vs
Let us consider the second term of the right-hand side of (6.32) and estimate the cardinality
$$
\vert \dot \bT^{(u,\{l\})}_{s-\t_1}(d,\t_1,\phi)\vert =
 \sum_{ \dot \CT_{s-\t_1}^{(u,\{l\})}(d, \t_1, \phi)} \  1.
$$
Using the first inequality of (6.35), we can write that
$$
\vert \dot \bT^{(u,\{l\})}_{s-\t_1}(d,\t_1,\phi)\vert \ 
\le
\ \sum_{ \stackrel{   \bar b_d, \bar c_{l},} 
{ \vert \bar b_d\vert + \vert \bar c_{l} \vert  = s-\t_1} 
}\ \left( 
\ddt^{(u-l+\phi)}_{\{d,\bar b_d\}} \ \dt^{(u)}_{[l+1, \bar c_l]} \ + \ 
\dt^{(u-l+\phi)}_{\{d,\bar b_d\}} \ \ddt^{(u)}_{[l+1, \bar c_l]} \right),
\eqno (6.38)
$$
where 
$$
\dt^{(u-l+\phi)}_{\{d,\bar b_d\}} = 
\sum_{k=1}^{d} \, \ddT^{(u-l+\phi-2)}_{b_1}\cdots  \ddT^{(u-l+\phi-2)}_{b_{k-1}}\, \dT^{(u-l+\phi-1)}_{b_k}
\, \ddT^{(u-l+\phi-1)}_{b_{k+1}}\cdots \ddT^{(u-l+\phi-1)}_{b_d}.
\eqno (6.39)
$$

Let us consider the first term of the right-hand side of (6.38). 
The same computation as before shows that for any given $b\ge d$, 
$$
 \sum_{\bar b_d:\, \vert \bar b_d\vert = b-d}\ddt^{(u-l+\phi)}_{\{ d, \bar b_d\}}\ \le \ 
 e^{-\eta d}\, \rt_b.
 \eqno (6.39)
$$
Similarly to (6.37), we observe that
$$
 \sum_{l=1}^{\t_1} \ \sum_{\ddot \CT_{\t_1}^{(u,[l])}}   
 \ \sum_{   \bar c_{l},
  \vert \bar c_{l} \vert  = s-\t_1-b} 
\ \ 
\dt^{(u)}_{[l+1, \bar c_l]}\ = 
\vert \acute \bT^{(u, \t_1)}_{s-b}\vert ,
\eqno (6.40)
$$
where we $ \acute \bT^{(u, \t_1)}_{s-b}$ denotes the family of trees $\CT_{s-b}$ 
such that the height $\th^*(\CT_{s-b})=u$ is attaint for the first time during the time interval
$[\xi_{\t_1},2s]$.

\vs

Let us consider the second term of the right-hand side of (6.38). 
Applying (6.34) to all the factors  $\ddT$ of (6.39), we can write that
$$
\dt^{(u-l+\phi)}_{\{d,\bar b_d\}}\ \le \   
 \sum_{k=1}^{d} \, \rt_{b_1}\cdots  \rt_{b_{k-1}}\, \dT^{(u-l+\phi-1)}_{b_k}
\, \rt_{b_{k+1}}\cdots \rt_{b_d}.
$$
Then for any given $b\ge d$,
$$
 \sum_{\bar b_d:\, \vert \bar b_d\vert = b-d}\dt^{(u-l+\phi)}_{\{d,\bar b_d\}}\ \le \ 
 d \, e^{-\eta(d-1)} \ \sum_{b_1= u-l+\phi-1}^{b-d} \ \dT^{(u-l+\phi-1)}_{b_1}\ \rt_{b-b_1-1}.
\eqno (6.41)
 $$
It is useful to note that the factor 
$\dT^{(u-l+\phi-1)}_{b_1}\, \ddt^{(u)}_{[l+1, \bar c_l]}$ being substituted into the second term of (6.32)
produces  an expression 
$$
 \ \sum_{l=1}^{\t_1} \ \sum_{\ddot \CT_{\t_1}^{(u-1,[l])}}  \ \dT^{(u-l+\phi-1)}_{b_1}\, 
  \ \sum_{ \stackrel{  \bar c_{l},}{
  \vert \bar c_{l} \vert  = s-\t_1-(b-b_1-1)}  }  \ddt^{(u)}_{[l+1, \bar c_l]} 
\  = \ \vert \check \bT_{s-(b-b_1-1)}^{(u, \t_1,\phi, 1)}\vert ,
\eqno (6.42)
 $$
 where we have denoted by $ \check \bT_{s-b'}^{(u, \t_1,\phi, 1)}$,  $ b' = b-b_1-1$ the family of trees $\CT_{s-b'}$ 
 such that the height $\th^*(\CT_{s-b'})=u$ is attaint for the first time 
 during the chronological run over a sub-tree attached by exactly one edge to the nest cell $(\t_1, \phi)$ of $\CT_{s-b'}$. 
 This definition is self-explained by the left-hand side of (6.42). 
It is clear that 
$$
\vert \check  \bT_{s-b'}^{(u, \t_1,\phi, 1)}\vert \, \le \, 
\vert \dot \bT^{(u)}_{s-b'}\vert \quad {\hbox{and}} \quad \vert \grave \bT^{(u, \t_1)}_{s-b}\vert \, + \, \vert \acute \bT^{(u, \t_1)}_{s-b}\vert  \, = \, \vert \dot \bT^{(u)}_{s-b}\vert.
\eqno (6.43)
$$ 
\vs

Remembering that all of the upper bounds (6.36), (6.39) and (6.41) are valid for any given values of $\phi$, we
turn back to (6.32)
and get the following upper bound,
$$
   \vert \bW_{2s}^{(u)}( d, \langle \CG^*\rangle_s, \phi, \U)\vert
   \  \le
   \   \vert \CG_\circ \vert \ e^{-\eta d} 
   \left( 
    \sum_{b=d}^{s-1}  \ \rt_b \ \vert \dot \bT^{(u)}_{s-b}\vert
   + {4d\over 3}  \, \sum_{b'=d-1}^{s-1} \rt_{b'} \ \vert \dot \bT^{(u)}_{s-b'}\vert\right).
   \eqno (6.44)
   $$
  Extracting the factor $u^r$ from  $\vert   \CG_\circ \vert$ (5.5), 
  we can write that
   $$
   \sum_{u=1}^{s-b} \, u ^{m+r} \ \vert \dot \bT^{(u)}_{s-b}\vert\  
   = (\sqrt s)
 ^{m+r} \ \sum_{ u=1}^{s-b}\, \left( { u\over \sqrt s} \right)^{m+r}\, \vert \Theta_{2s-2b}^{(u)}\vert   
   \le\  s^{(m+r)/2} \, \rt_{s-b}\,  B_{m+r}\, . 
   \eqno (6.45)
   $$
Taking into account that $d\ge 1$ and using the following elementary relations based on (3.2) and (6.3),
$$
\sum_{b=1}^{s-1} \, \rt_b \, \rt_{s-b} = \rt_{s+1}- 2\rt_s \le 2\rt_s
$$
and 
$$
{4\over 3} \sum_{b'=0}^{s-1} \, \rt_{b'} \, \rt_{s-b'} = {4\over 3} (\rt_{s+1}- \rt_s) \le 4\rt_s,
$$
we deduce from (6.44) and (6.45) that
$$
\sum_{u=1}^s \ u^m\ \sum_{\phi=1}^u\ 
 \vert \bW_{2s}^{(u)}( d, \langle \CG^*\rangle_s, \phi, \U)\vert  
 \le 
 (2+4d)\,  e^{-\eta d} \, \rt_s\,   
 $$
 $$
 \times \,   s^{(m+r)/2}\,   B_{m+r}\,  2^r\, D^p\, k_0^q \, (s(D+k_0))^{\mu_3'}  .
 \eqno (6.46)
 $$
 This proves relations (6.26), (6.27)  with  $R=1$.



\subsubsection{Recurrent estimates}

Let us denote by $\t_{R+1}$ the marked instant that corresponds to  the imported cell $(\t_{R+1},\phi_{R+1})$ and write down 
the following expression
(cf. (6.32)),
$$
\vert \bW_{2s}^{(u)}( \rD, \bar d_{R+1};  \langle \CG^\star_{R+1}\rangle_s, (\CH_{R},\phi_{R+1}), \U)\vert 
= \S^{(1)}(\phi_{R+1}) + \S^{(2)}(\phi_{R+1}),
\eqno (6.47)
$$
where 
$$
\S^{(1)}(\phi_{R+1})= 
\sum_{l=1}^{\t_{R+1}} 
\ \  \sum_{\dot  \CT_{\t_{R+1}}^{(u, [l]) }}\ \ \  
\sum_{ \la  \CG_\circ ^{(1)}\ra }  \ \vert \G^{(\phi_{R+1},\fu)}_{t_{R+1}}(\bb)\vert 
\ 
\sum_{\ddot \CT^{(u, \{l\})}_{s-\t_{R+1}}(d_{R+1}, \t_{R+1}, \phi_{R+1})} \ \ 
\sum_{ \la \CG_\circ^{(2)} \ra} 1
 $$
 and 
 $$
 \S^{(2)} (\phi_{R+1}) 
 = \sum_{l=1}^{\t_{R+1}} 
\ \  \sum_{\ddot  \CT_{\t_{R+1}}^{(u, [l]) }}\ \ \  
\sum_{ \la  \CG_\circ ^{(1)}\ra }  \ \vert \G^{(\phi_{R+1},\fu)}_{t_{R+1}}(\bb)\vert 
\ 
\sum_{\dot \CT^{(u, \{l\})}_{s-\t_{R+1}}(d_{R+1}, \t_{R+1}, \phi_{R+1})} \ \ 
\sum_{ \la \CG_\circ^{(2)} \ra} 1.
$$
In these relations,  we have denoted by $\langle  \CG_\circ^{(1)}\ra $ and $\langle  \CG_\circ ^{(2)}\ra$   
realizations of values of the red edge-windows of $\langle\CG_{\CQ_R}\rangle_s$
such that the marked instants of the corresponding  blue edge-windows 
are strictly less than $\t_{R+1}$ and are great or equal to $\t_{R+1}$, respectively. 

\vs 
Taking into account the uniform with respect to $\phi_{\t_{R+1}}$ estimate of $\langle  \CG_\circ^{(2)}\ra $ (5.29) 
and denoting 
$\t' = \t_{R+1}$, we can write 
with the help of (6.37) that the sum $\S^{(1)} = \sum_{\phi_{R+1}=1}^u \S^{(1)}(\phi_{R+1})  $
is bounded from above as follows, 
$$
\S^{(1)}  \le \ e^{-\eta d_{R+1}}\ 
\sum_{b=d_{R+1}}^{s-1-\t'}\ \rt_b \ 
 \left( \sum_{ \CT\in \grave \bT_{s-b}^{(u,\t') }} \  \sum_{ \langle  \CG_\circ^{(1)}\ra } 1 \right) \cdot 
\vert \CG_\circ ^{(2)}\vert \cdot \vert \G^{(\fu_{R+1})}\vert .
\eqno (6.48)
$$
 It should be noted that the 
expression standing in the parenthesis of (6.48) represents the family of walks
$$
\grave \bW_{2s-2b}^{(u, \t')}(\rD, \bar d_R; \langle \CG^{(\star, 1)}_{R}\rangle_{s-b}, \CH_R, \U) \ 
= \ 
\sum_{ \CT\in \grave \bT_{s-b}^{(u,\t') }} \  \sum_{ \langle \CG_\circ^{(1)}\ra  }\,  1
\eqno (6.49)
$$
such that the conditions of Lemma 6.3 are verified. Here we have denoted by 
$
\langle \CG^{(\star, 1)}_R\rangle_{s-b}$ 
the part of the realization of the diagram $\langle \CG^\star_R\rangle_s$ that takes into account the 
instants of self-intersections that are strictly less than $\t' = \t_{R+1}$.

\vs 
Regarding the sum $\S^{(2)}  = \sum_{\phi_{R+1}=1}^u \S^{(2)}(\phi_{R+1}) $, we use representation (6.38) and write that 
$$
\S^{(2)} \le \acute \S^{(2)} + \dot \S^{(2)},
\eqno (6.50)
$$
where 
$$
\acute \S^{(2)} \le 
\ e^{-\eta d_{R+1}}\ 
\sum_{b=d_{R+1}}^{s-1-\t'}\ \rt_b \ 
 \left( \sum_{ \CT\in \acute \bT_{s-b}^{(u,\t') }} \  \sum_{ \langle  \CG_\circ ^{(1)}\ra } 1 \right) 
\ \vert \CG_\circ^{(2)}\vert  \cdot \vert \G^{(\fu_{R+1})}\vert
\eqno (6.51)
$$
and 
$$
\dot \S^{(2)}(\phi') \le d_{R+1} \, e^{-\eta(d_{R+1}-1)}\, 
\sum_{b=d_{R+1}-1}^{s-1} \, \rt_b \ 
\cdot \vert \CG_\circ ^{(2)}\vert \ 
$$
$$
\times \ \left(\sum_{l=1}^{\t'}\   \sum_{\ddot \CT^{(u, [l])}_{\t'} }\   
\sum_{ \langle \CG_\circ^{(1)}\ra } \ 
\sum_{\vert \bar c_l\vert = s-\t' - b} \ \ddT^{(u-1)}_{[l+1, \bar c_l]}\   \dT^{(u-l+\phi'-1)}_{b-1}\right)
 \cdot \vert \G^{(\fu_{R+1})}\vert.
\eqno (6.52)
$$

\noindent 
It is not hard to see that 
$$
\S^{(1)} + \acute \S^{(2)} \le \rD^{p_2}\, k_0^{q_2} \ (2s(\rD+k_0))^{\mu_3'(2)}  \cdot \vert \G^{(\fu_{R+1})}\vert
$$
$$
\times e^{-\eta d_{R+1}}\, \sum_{b=1}^{s-1} \ \rt_b \cdot 
 (2u)^{r_2}\,  \cdot 
 \vert \bW_{2s-2b}^{(u, \t')}(\rD, \bar d_R; \langle \CG^{(\star, 1)}_{R}\rangle_{s-b}, \CH_R, \U) \vert.
 \eqno (6.53)
$$
Regarding the sum
$$
\S_{R+1}'= \sum_{u=1}^s\ u^m\,  \sum_{\CH_R} \
\left(\S^{(1)} + \acute \S^{(2)} \right),
$$
we  apply to the right-hand side of (6.53) the main proposition of lemma  (6.27) and get inequality
$$
\S_{R+1}'\le 
2 \cdot 4^R\,  D_R\ e^{-\eta D_{R+1} } \, \rt_s \cdot  B_{m+r}\ 2^r \, \rD^p \, k_0^q \ (s(\rD+k_0))^{2\mu_3'}.
\eqno (6.54)
$$

\vs

Let us consider the right-hand side of (6.52). It is not hard to observe that the expression 
standing in the parenthesis of (6.52) is bounded from above by  the number of walks $\CW_{2s-2b}$ 
that on the time interval $[0, \xi_{\t_{R+1}}-1]$ verify conditions of Lemma 6.3 and
such that the height $u$ of the corresponding Dyck path $\th(\CW_{2s-2b})$  is attaint for the first time
above the edge attached to the nest cell $(\t',\phi')$.  Let us denote this set of walks by
$\dot \bW_{2s-2b}^{(u, \t',\phi')}(\rD, \bar d_R; \langle \CG^{(\star,1)}_{R}\rangle_{s-b}, \CH_R, \U)$ (cf. (6.49)). 
To simplify the proof, it is useful to observe that 
$$
\vert 
\dot \bW_{2s-2b}^{(u, \t',\phi')}(\rD, \bar d_R; \langle \CG^{(\star,1)}_{\CQ_R}\rangle_{s-b}, \CH_R, \U)
\vert
 \le 
\vert 
\stackrel{\ldots}{ \bW}_{2s-2b}^{(u,\D_i^c)}(\rD, \bar d_R; \langle \CG^{(\star,1)}_{R}\rangle_{s-b}, \CH_R, \U)
\vert,
$$
where $\stackrel{\ldots}{ \bW}_{2s-2b}^{(u,\D_i^c)}(\rD, \bar d_R; \langle \CG^{(\star,1)}_{R}\rangle_{s-b}, \CH_R, \U)$
is the family of walks with the same properties as before and   the only difference
that the height $\th^*(\CW_{2s})=u$ is attaint somewhere excepting those parts of $\th$ that lie 
over the exit sub-clusters $\D_i$, $1\le i\le d_R$. 

\vs 

We are going to show that
$$
\sum_{u=1}^s u^m\,   \sum_{\CH_R}\  \, \vert  \stackrel{\ldots}{ \bW}_{2s-2b}^{(u, \Delta_i^c)}
(\rD, \bar d_R; \langle \CG^{(\star,1)}_{R}\rangle_{s-b}, \CH_R, \U)\vert 
$$
$$
\le \ 2^R \, e^{-\eta D_R} \, \rt_{s-b} \  B_{m+r_1}\, 2^{r_1}\,  \rD^{p_1}\, k_0^{q_1} \ (s(\rD+k_0))^{\mu_3'(1)}\, .
\eqno (6.55)
$$

\vs 
One can prove (6.55) by recurrence. We consider here  the initial case of  $R=1$ only. It is not hard  to see that
$$
\vert \stackrel{\ldots}{ \bW}_{2s}^{(u, \D_i^c)}(\rD, d; \langle \CG_{\t_1}^\star \rangle_s, \phi_1, \U)\vert =
\sum_{l=1}^{\t_1} \left\{ \sum_{\ddot \CT_{\t_1}^{(u-1, [l])}}  \sum_{\dot \CT_{s-\t_1 - b}^{(u, \{l\})}} +
 \sum_{\dot \CT_{\t_1}^{(u, [l])} }
 \sum_{\ddot \CT_{s-\t_1 - b}^{(u, \{l\})}} \right\}\ \sum_{\langle \CG_\circ\rangle_{[0, t_1]}} \ 1
$$
$$
\times \ \vert \G^{(\phi_1,\fu_1)}_{t_1} (\bb)\vert   \ \ \sum_{\ddot \CT_b^{(u-l+\phi_1 -1)}(d)} 
\ \ \sum_{\langle  \CG_\circ\rangle_{[t_1+2, 2s]}} \ \ 1,
\eqno (6.56)
$$
where $t_1 = \xi_{\t_1}-1$.

Using the upper bound (5.5) and inequality (6.35), we deduce from 
(6.56) the following inequality,
$$
 \sum_{u=1}^s\ u^m\,   \sum_{\phi_1 =1}^u \ 
 \vert \stackrel{\ldots}{ \bW}_{2s}^{(u, \D_i^c)}(\rD, d; \langle \CG_{\t_1}^*\rangle_s, \phi_1, \U)\vert 
 $$
 $$
 \le \sum_{b=d_1}^{s-1} \ \sum_{u=1}^s \ 
 u^{m+r}\,  \vert \dot \bT_{s-b}^{(u)}\vert \cdot e^{-\eta d_1} \, \rt_b
 \ 2^{r}\, \rD^p\, k_0^q \ (2s(\rD+k_0))^{\mu_3'}.
 \eqno (6.57)
 $$
Standard computations show that (6.57) proves (6.55) in the case of $R=1$. 

\vs 
With the help of (6.55), we can deduce from (6.52) the following inequality,
$$
\S''_{R+1} = \sum_{u=1}^s\ u^m\,  \sum_{\CH_R}\   \sum_{\phi'=1}^u\  
\dot \S^{(2)}(\phi') \le {4d_{R+1}\over 3}\, 2^{R+1} \, e^{-\eta D_{R+1}} \
 \rt_s 
$$
$$
 B_{m+r}\ 2^r\, \rD^p\, k_0^q\, (s(\rD+k_0))^{2\mu_3'}.
$$
Remembering (6.54), we obtain that
$$
\S'_{R+1} + \S''_{R+1} \le 
\left(2\cdot 4^R \, D_R+  {4d_{R+1}\over 3}\, 2^{R+1}\right)  \, e^{-\eta D_{R+1}} \
 \rt_s \ 
 $$
 $$
 \times 
B_{m+r}\ 2^r\, \rD^p\, k_0^q\, (s(\rD+k_0))^{2\mu_3'}.
$$
It is clear that the last inequality implies (6.27) with $R$ replaced by $R+1$. Lemma 6.3 is proved. $\Box$

\subsubsection{Walks with mirror cells at $\bb$}

Let us consider a family of walks $\bW_{2s}^{(u)}(\rD, \bar d_R; \langle \CG^\star_R\rangle_s, \CH_R, \U)$ (4.16)
with given set $(\bar x, \bar m)_I$ and assume that $x_1$ is attributed by a number $m_1>0$. 
This means that the walk arrives at $\bb$ at at the marked instant $x_1$ and then performs a tree-type sub-walk 
$\check \CW = \sqcup_{i=1}^{m_1} \check \CW^{(i)}
$
such that during this sub-walk it arrives $m_1$ times by non-marked steps at $\bb$. Moreover, 
each of the $m_1$ tree-type sub-walks $\check \CW^{(i)}$ has a number of marked instants  $x^{(i)}_j$
such that $\check \CW^{(i)}(\xi_{x^{(i)}_j})=\bb$. These marked instants being determined, 
let us denote the maximal one by $x'_1 = \max\{ x^{(i)}_j\}$. 

It is not hard to see 
that the construction of the sub-trees and the nest cell is as follows: we consider a tree $\CT_{x_1}^{[l_1]}$
such that the vertex $\u_1$ is on the distance of $l_1$ from the tree root $\varrho$. Then 
we add $l_2$ edges to the vertex $\u_1$ and get the vertex $\u_2$. 
On $l_2$ roots obtained, we construct sub-trees with the help of $x_1'-x_1$ edges
and get the tree $\CT_{x_1'}^{[l_1+l_2]}$. The mirror cell we consider will correspond to the arrival at $\u_2$
by $\l_2$ steps from  $\u_2$. Then the ordinary procedure of constructing the trees with $u$-condition like (5.26)
can be used. 

In the proof of Lemma 6.3 by recurrence, we needed the construction of the last proper 
imported cell only. Therefore it is clear that the construction of the mirror
cell presented above fits completely this scheme. 
We do not present the detailed arguments here.

 
 
\section{Estimates from below}

Let us consider random matrices $H^{(n,\r)}$ (2.1) with random variables $a_{ij}$ that  have
all moments finite. Then the following statement for the moments $\rM_n^{(n,\r)}$ (2.5) is true.

\vskip 0.3cm
{\bf Theorem 7.1}. {\it Let $s_n = \chi n^{2/3}$ and $ \r_n = \z n^{2/3}$ with given $\chi, \z>0$. Then
$$
\liminf_{n\to\infty} \rM_{2s_n}^{(n,\r_n)}\  \ge  \ 
{4V_4 \over \z  ( \pi \chi)^{1/2} } e^{-e\chi^3} (1+o(1))\ ,
\eqno (7.1)
$$
where $V_4 = \E \vert a_{ij}\vert^4$ and  $\E \vert a_{ij}\vert^2 = v^2=1/4$.}
\vskip 0.3cm
{\it Proof.}

Let us consider a Dyck path $\theta_{2s}$ and corresponding Catalan tree $\CT_s$.
Regarding the chronological run $\fR_\CT$, let us determine marked instants $\t_1$ and $\t_2$
such that  the corresponding vertices of $\CT_s$ have the same parent. 
We denote such a pair by $(\t_1,\t_2)_p$. 
Regarding the example 
tree $\CT_8$ 
given on Figure 1, we can take $(\t_1,\t_2)_p= (3,4)$ or, say $(\t'_1,\t'_2)_p=(6,8)$. 

It is clear that  the  tree-type walk $\CW_{2s} = \CW_{2s}^{[\th]}(\t_1,\t_2)$ 
that has the Dyck structure $\th_s= \th(\CT_s)$ and one simple self-intersection
$(\t_1,\t_2)_p$ is a walk with one $p$-edge. It is clear also that the family of all possible such walks
$$
\bW_{2s}^{[1,p]} = \bigsqcup_{\CT_s} \ \bigsqcup_{(\t_1,\t_2)_p} \CW^{[\th]}_{2s}(\t_1,\t_2) 
$$
has the cardinality $\vert \bW_{2s}^{[1,p]}\vert = \RN_s^{(2)}$ (6.12).

\vs

Let us denote the elements of  $ \bW_{2s}^{[1,p]}$  by $w= w_{2s}$. 
We are going  to construct  a family of walks $\bW_{2s}(w;\mu_2)$ by introducing $\mu_2$ additional simple self-intersections
in it, $0\le \mu_2\le M$. 
The graph of the walk $w_{2s}$ has $s$ vertices and therefore the cardinality of this family is bounded from below as follows,
$$
\vert \bW_{2s}(w;\mu_2)\vert = {s!\over 2^{\mu_2}\, \mu_2! (s-2\mu_2)!} 
\ge {1\over \mu_2!} \left( { (s-2M)^2\over 2}\right)^{\mu_2}.
\eqno (7.2)
$$
It is clear that the weight of any trajectory $\CI_{2s}$ (2.8) such that $\CW(\CI_{2s}) \in \bW(w;\mu_2)$ 
admits the following  lower bound, 
$$
\Pi_a(\CI_{2s}) \, \Pi_b(\CI_{2s})\ \ge \ {V_2^{s-1}\, V_4\over n^{s-2}\, \rho}.
\eqno (7.3)
$$
Remembering (3.1), we conclude that the number of trajectories in the class $\CC(\CW_{2s})$, $\CW_{2s} \in \bW_{2s}(w;\mu_2)$
is bounded as follows,
$$
\vert \CC(\CW_{2s})\vert  =  \prod_{i=0}^{s-\mu_2-1} (n-i)= 
n^{s-\mu_2} \ \prod_{k=1}^{s-\mu_2-1} \left( 1 - {k\over n}\right)
\ge n^{s-\mu_2} \exp\left\{ -{s^2\over 2n}\right\}.
\eqno (7.4)
$$
The last inequality is obtained by the same argument as the upper bound (5.15).
\vs 
Taking $M=\lfloor c n^{1/3}\rfloor+1$ with  $c>0$, we deduce from relations (7.2), (7.3) and (7.4)
that
$$
\rM_{2s}^{(n,\rho)}\,  \ge \ \sum_{w \in \bW_{2s}^{[1,p]}} \ \sum_{\mu_2=0}^M \ \ \sum_{\CI_{2s} \in \CC(\CW)}
\ \sum_{\CW \in \bW_{2s}(w;\mu_2)} {V_2^{s-1}\, V_4\over n^{s-1}\, \rho}
$$
$$
\ge \,   n\, \RN^{(2)}_s \, \exp\left\{ -{s^2\over 2n}\right\}\  \sum_{\mu_2=0}^M  \,  
{1\over \mu_2!} \left( { (s-M)^2\over 2n}\right)^{\mu_2}
\cdot {V_2^{s-2}\, V_4\over \rho} .
\eqno (7.5)
$$
Elementary use of the Stirling formula shows that
$$
\sum_{\mu_2 = M+1}^{\infty} {1\over \mu_2!} \left( { (s-M)^2\over 2n}\right)^{\mu_2} \ge 
{1\over \sqrt{2\pi cn^{1/3}}} \ \sum_{\mu_2 = M+1}^{\infty} \left( {e\chi^2 \over 2c}\right)^{\mu_2} = o(1),
$$
for  $c = e\chi^2$ and therefore with this choice of $c$, we have 
$$
\exp\left\{ -{s^2\over 2n}\right\}\  \sum_{\mu_2=0}^M  \,  
{1\over \mu_2!} \left( { (s-M)^2\over 2n}\right)^{\mu_2} \ \ge\  {1\over 2} \exp\{- e\chi^3/2\}.
$$
Remembering the lower bound $\RN^{(2)}_s \ge (s \rt_s)/2$ (see Lemma 6.2), 
we  deduce from (7.5) the following inequality,
$$
\rM_{2s}^{(n,\rho)}\,  
\ge\ n \rt_s \, V_2^s \ {s\, V_4\over 4V_2^2  \rho} \exp\{- e\chi^3\}.
$$
Then (7.1) follows. Theorem 7.1 is proved. $\Box$


\section{Discussion}

We have studied the asymptotic properties 
of the probability distribution of the spectral norm of large dilute random matrices.
We have shown that the probability distribution of the maximal eigenvalue
of dilute  Wigner random matrices $H^{(n,\r_n)}$, when regarded at the scale $n^{-2/3}$,
admits a universal upper bound 
in the limit of infinite $n,\r_n$ such that $n^{2/3(1+\vep)}  \le \r_n\le n$, $\vep>0$ and $s_n= \chi n^{2/3}$, $\chi>0$.
This result is a consequence of the
existence of a universal upper bound $\fL(\chi)$ of the moments 
$\tilde \rM_{2s_n}^{(n,\r_n)}$, $s_n = \chi n^{2/3}$
of  $\tilde H^{(n,\r_n)}$ (2.6) and,
in more general situation, of 
the moments $\hat \rM_{2s_n}^{(n,\r_n)}$ of corresponding random matrices with truncated elements.
\vs

According to the general scheme developed in papers \cite{SS2,S} in the case of Wigner ensemble of random matrices, 
this kind of asymptotic behavior
of the moments $\rM_{2s_n}= \E \rL_{2s_n} $,  can be regarded as the strong evidence
of the universality of the probability distribution of one maximal eigenvalue of $H^{(n,\r_n)}$, 
or its several consecutive neighbors (see also \cite{FS,So-2}).
Indeed, as it is described in \cite{S}, 
the study of the correlation functions of $\rL_{2s'_n}$ and $ \rL_{2s''_n}$  can be reduced, in a major part,
to the study of the related moment $\rM_{2s'_n+2s''_n-2}$ whose behavior can be shown to be universal 
(see however, \cite{K2}).
  
Therefore one can expect that the dilute Wigner random matrices in the limit of the {\it weak dilution},
i.e. for the values of $\r_n$ such that $n^{2/3}\ll \r_n\le n$, $n\to\infty$ belong to the
class of universality determined by the Gaussian Orthogonal Ensemble of random matrices (GOE) in 
the case of real symmetric
 $H^{(n,\r_n)}$,
or to the class of  GUE in  the  hermitian case \cite{M}.

\vskip 0.1cm

  Theorem 7.1 shows that
in the asymptotic regime $n,\r_n\to\infty$ such that $\r_n= \z n^{2/3}$,
the estimate from below of
 $\rM_{2s_n}^{(n, \r_n)}$ involves the factor $V_4$.
 Therefore the upper bound $\limsup_{n\to\infty} \rM_{2s_n}^{(n,\r_n)}\le \fM(\chi)$ (2.6) 
 cannot be true in this asymptotic regime.
This means that in the asymptotic regimes
 of the {\it moderate and strong} dilution, when $\r_n = \z \n^{2/3}$ or $\r_n= o(n^{2/3})$, respectively,
 the limiting  probability distribution of the maximal eigenvalue 
 cannot be universal fails in the sense that the limiting expressions
should  depend on the moments  higher than $V_2$  of the random variables $a_{ij}$.

\vskip 0.1cm

Moreover, inequality (7.1) shows  that in the asymptotic regime $n,\r_n\to\infty $ when
$\r_n = \s n^{\ep}$ with $0<\ep< 2/3$,
to get the finite upper bound for  the moment $\rM_{2s_n}^{(n,\r_n)}$ of  the order $s_n$,
one should restrict the growth of $s_n$ and consider the case when  $s_n$ is proportional to $\r_n$
but not to $n^{2/3}$ as before. According to the general considerations based on the inequalities of the form (4.26),
 one can  conclude that the  scale at the border of the limiting spectrum of $H^{(n,\r_n)}$ should be also changed
to be  proportional to $\r_n$ and not to $n^{2/3}$ as it is in the case of $n^{2/3(1+\vep)} \le \r_n$.

\vs
Therefore we can put forward a conjecture  that the rate $\r_n= n^{2/3}$ represents the critical
point where the eigenvalue distribution at the edge of the spectrum changes its properties,
such as the scale and the dependence on the probability distribution of the matrix elements $a_{ij}$.

\vs
Another important observation concerns the subsequent terms of the estimate from below given by (7.1).
Repeating the computations  of the  proof of Theorem 7.1 and using (6.23), we observe that 
the moments $\rM_{2s}^{(n,\rho)}$ (2.5) admit the following asymptotic expansion, 
$$
\rM_{2s}^{(n,\rho)} \simeq {n\rt_s\over 4^s}\, \left( c^{(1)} +  \sum_{k\ge 1} c^{(2)}_k \left( {s\,  V_4\over \r}\right)^k 
+ \sum_{l\ge 1}
c^{(3)}_l \left({s\, V_6\over \r^2}\right)^l + 
o(s/\r^2) \right),
\eqno (8.1)
$$
where $s= \chi n^{2/3}, \ \ \r= \z n^{2/3}$ and $c^{(i)}>0$ depend on $\chi$ and $\z$ but do not depend on $n$. We see that in this case,
the terms with $V_4$ are present in the asymptotic development of $\rM_{2s}^{(n,\r)}$, but the higher 
moments $V_6, V_8, \dots$ disappear from it. Therefore we can formulate 
a conjecture 
 that 
the regimes of moderate and strong dilutions exhibit a new kind of universality, say 
{\it $V_4$-universality} at the border of the limiting spectra.

\vs 

The next observation is that the asymptotic expansion of the form
$$
{4^s\over n \rt_s} \rM_{2s}^{(n,\rho)}   \simeq c^{(1)} +  \sum_{k\ge 1} c^{(2)}_k \left( {s\,  V_4\over \r}\right)^k,
\quad s= \chi n^{2/3}, \ \ \r= \z n^{2/3}
\eqno (8.2)
$$
is related with the corresponding limit of the moments of sparse random matrices
studied in \cite{KSV}. The fact that the right-hand side of (8.2) depend on the terms with $V_4$ only
could essentially simplify the recurrent relations obtained in \cite{KSV}. Moreover, it is natural to assume
that the coefficients $c^{(2)}_k$ will be related with the corresponding terms of
$
\lim_{s/\r = \chi/\zeta, \ s\to\infty} m_{s}^{(\r)},
$
where $m_{s}^{(\r)}$ are determined by the following recurrent relation, with obvious initial conditions, 
$$
m_s^{(\r)}= v^2 \sum_{a_1+a_2 = s-1} m_{a_1}^{(\r)} \, m_{a_2}^{(\r)}  \, +  \, {V_4\over \r}
\sum_{b_1+\dots + b_4 = s-2} m_{b_1}^{(\r)} \, 
m_{b_2}^{(\r)} \,  m_{b_3}^{(\r)}  \,  m_{b_4}^{(\r)}.
\eqno (8.3)
$$
One can show that the asymptotic expression of the numbers $m_{s}^{(\r)}$ (8.3) should be related with
the generating functions of the numbers of ternary trees. We postpone the study of (8.3) to subsequent publications.

\vs
Our last remark is related with  the difference between the ensembles of real symmetric and hermitian matrices. 
As it is mentioned in Section 2, the upper bounds $\fM_{\hbox{\tiny{GOE}}}(\chi)$
and $\fM_{\hbox{\tiny{GUE}}}(\chi)$ are slightly different (see also relations (4.35) and (4.36)). 
This difference is due to the contribution 
of walks with simple open self-intersections and the breaks of the tree structure performed by the walk at them. 
The contribution of these walks should vanish in the asymptotic regime of the strong dilution, when
$\r_n\le n^{2/3-\vep}$ and $s_n\le  n^{2/3-\vep}$ and only the tree-type walks give the non-vanishing contribution. 
This means that  the difference between the spectral properties of real symmetric random matrices and their hermitian
analogs could disappear in the asymptotic regime of strong dilution. It would be interesting
to study this phenomenon in more details. It should be noted that the moments $\rM_{2s}^{n,\rho)}$ of random matrices
$H^{(n,\rho)}$ (2.1) with $\rho = O(1)$ as $n\to\infty$ have been studied in \cite{KSV} in the case of $s= O(1)$. 
The explicit expressions obtained there  as well as the technique developed in  \cite{K01} 
could be useful in the studies of  more complicated asymptotic regime described above, 
 when 
$s_n\to\infty$ as $n\to\infty$. 

\vs 
{\bf Acknowledgement.} The author is grateful for the anonymous referee for the careful reading of the manuscript.

\end{document}